\documentclass[preprint,12pt]{elsarticle}

\usepackage{lineno}

\usepackage{amssymb}
\usepackage{amsmath}
\usepackage{subcaption}
\usepackage{multirow}
\usepackage{tikz}
\usepackage[hidelinks]{hyperref}
\usepackage{breqn}
\usepackage{booktabs}
\usepackage{float}
\biboptions{square,numbers,sort&compress}

\journal{Applied Mathematical Modelling}

\begin{document}

\begin{frontmatter}

\title{Modeling the Hydrodynamics in the Oslofjord using an Advanced Circulation Model}

\author[inst1]{Matthew Scarborough\corref{cor1}}
\ead{matthew.scarborough@nmbu.no}
\cortext[cor1]{Corresponding author}

\author[inst3]{Kai H\aa{}kon Christensen}

\author[inst4]{Albert Cerrone}

\author[inst3]{Nils Melsom Kristensen}

\author[inst5,inst6]{Eirik Valseth}


\affiliation[inst1]{organization={The Department of Data Science, The Norwegian University of Life Sciences},
            addressline={Dr\o{}bakveien 31},
            city={\AA{}s},
            postcode={1433},
            country={Norway}}
\affiliation[inst3]{organization={Norwegian Meteorological Institute},
            addressline={Postboks 43 Blindern},
            city={Oslo},
            postcode={0313},
            country={Norway}}
\affiliation[inst4]{organization={The Department of Civil Engineering, University of Notre Dame},
            addressline={156 Fitzpatrick Hall of Engineering},
            city={South Bend, Indiana},
            postcode={46556},
            country={USA}}
\affiliation[inst5]{organization={The Department of Mechanical Engineering and Technology Management, The Norwegian University of Life Sciences},
            addressline={Dr\o{}bakveien 31},
            city={\AA{}s},
            postcode={1433},
            country={Norway}}
\affiliation[inst6]{organization={Department of Scientific Computing and Numerical Analysis, Simula Research Laboratory},
            addressline={Kristian Augusts gate 23},
            city={Oslo},
            postcode={0164},
            country={Norway}}

\begin{abstract}
This study introduces a new unstructured computational mesh for hydrodynamic simulations of the Oslofjord. The mesh was created with global bathymetry and shoreline data, using OceanMesh2D. It contains 70,410 nodes, with a resolution at the coastline of 50 meters. We use the new mesh to create an advanced circulation model of the fjord. The model is run for four time periods with different characteristics, and validated against the current state of the art and elevation gauges in the fjord.

Initial analyses show that the model effectively propagates tides through the narrow channels of the fjord, while requiring minimal computation time. Three different combinations of tidal constituents are used to force the model, and we analyze the cost and benefits of using additional constituents. In order to assess the model in a realistic scenario, the water surface elevation output from a coarse global model is used to force the model an extreme weather event. The results demonstrate the mesh's ability to reproduce high-resolution total water levels in the fjord, with a maximum root mean square error of just over 10 centimeters.
\end{abstract}



\end{frontmatter}


\clearpage 
\section{Introduction}

The most defining geological feature of Norway is undoubtedly its complex coastline, a labyrinth of fjords created by millenia of slow glacial flow. Because of these intricacies, Norway has more than 100,000 kilometers of coastline, despite its mainland spanning fewer than 2000 kilometers \citep{fjord}. Accordingly, the Norwegian economy heavily depends on maritime industries, which employed over 90,000 people and generated 219 billion NOK in 2024 \citep{shipowners}.

Because of Norway's largely steep, rocky coast and its upward vertical land motion, sea level rise may not seem like an imminent threat. However, sea level rise is accelerating worldwide, and it is essential to understand how this will affect coastal communities. Projections based on different levels of worldwide emissions range from 13 to 46 centimeters of average relative sea level rise in Norway -- and a rise of only 10 centimeters could triple the flood risk in many areas \cite{sealevel}. In addition, extreme weather events are likely to become much more common with the rising sea levels. Flooding in Norway is often localized, due to the generally steep nature of its topography, but because of its long complex coast, quite a large area is potentially vulnerable \cite{sealevel}. In particular, three types of low-lying coastline are at risk: the moraine coast in the southwest, glaciofluvial deltas at the heads of fjords, and the fringes of lowland and shallow sea areas, often at the mouth of fjords \cite{aunan}.

This initial effort focuses on the Oslofjord, as much prior research has centered around it, which will be helpful for validation. The fjord is about 100 kilometers long, and is divided into ``inner'' and ``outer'' fjords by the shallow, narrow Dr\o{}bak Sill \cite{fjordos}. The sill is a major obstruction for the water exchange in the fjord; because it is so shallow and narrow, its tidal currents often exceed 1 m/s \cite{fjordos}.
The inner fjord consequently has a lower tidal influence. It is further divided into two natural basins, the Vestfjorden and Bunnefjorden, separated by a 50-meter plateau. The seabed in the inner fjord is characterized by hundreds of pockmarks, which are crater-like depressions between 16 and 100 meters in diameter \citep{Webb09}. This leads us to desire a model that will encompass details of this scale.

The Oslofjord is characterized by its high degree of complexity. It contains many islands of various sizes, which are often very close together, resulting in several narrow sounds and channels in the fjord. The total number of islands and smaller skerries is estimated to be above 1000. This includes 40 larger islands in the inner harbor of Oslo and the extensive Hvaler archipelago in the south, which alone comprises 833 islands and skerries~\cite{fjord,geonorge}.
The narrow areas are very hard to resolve with numerical methods relying on rectangular grids, and a more adaptable resolution mesh would provide a way to resolve these narrow areas more accurately in the model.
Near the entrance to the fjord is a 400-meter deep basin, and there are others further in as deep as 150--200 meters \cite{fjordos}, see Figure \ref{fig:bathymetry} below. Finally, we note that at least 37 rivers discharge into the Oslofjord, including two of Norway's largest: Glomma and Drammenselva, which together have a mean discharge of over 1000 $\text{m}^{3}$/s \cite{milliman}.

Another peculiarity of the Oslofjord is that some very large ferries produce mini-tsunamis when they pass significant changes in water depth, for example near the Dr\o{}bak Sill. These waves can be up to a kilometer long and have been recorded as high as 1.4 meters \cite{tsunamis}. They travel ahead of the ship at twice the ship's velocity, and they can induce currents of about 1 m/s. All of these factors contribute to a new, significant type of erosion in the area. This phenomenon reinforces the need for enhanced modeling capabilities in the Oslofjord, as the shallow, narrow sill could produce similar underpredicted weather phenomena further up in the fjord during major weather events.

Large floods in southern Norway, including the Oslofjord, are usually caused by anticyclonic westerlies \citep{floods}. While they rarely cause fatalities, many floods cause damage to vital infrastructure, as well as industries and private property located near the coast.
One such flood occurred in October of 1987, when the remnants of Hurricane Floyd crossed the Atlantic and caused heavy winds and 255 cm of storm surge in the Oslofjord \cite{engen:1988, andresen:etal:1987}. This was accompanied by several centimeters of rain throughout southern Norway, and large areas of Oslo and Drammen were inundated. One woman was killed, and the total damages amounted to more than 650 million NOK \cite{floods}.

More recently, 2023's Storm Hans also caused major damage to southeastern Norway, forcing thousands of people to evacuate due to extreme rainfall over a three-day period, up to 189.2 mm in Liarvatn \cite{hansmet}. Parts of the Braskereidfoss dam collapsed, as did the Randklev Bridge in Ringebu. Including the individual property damage, this amounts to over 2 billion NOK (about €180 million) \cite{hansmet}. Fifteen of the Norwegian Water Resources and Energy Directorate's measuring stations just north of Oslo recorded 50-year floods, and the Drammens waterways leading into the Oslofjord experienced a watercourse rise of 1--3 meters over the period.
Dozens of roads were closed, and over 200 people in the area were evacuated \cite{hansmet}.
Similarly damaging was Storm Amy, in October 2025. Amy caused elevated water levels all around the south of Norway, and was characterized by very high winds and precipitation. Over 11,000 insurance claims were reported, with an estimated total cost of 1.8 billion NOK \cite{amy,amymet}. Gusts up to 29 m/s were observed at Tryvannsh\o{}gda just outside Oslo, with 20 m/s gusts in the city \cite{amymet}. The area around Oslofjord also experienced high water levels, with waves over 1.5 meters in the inner fjord \cite{sehavniva}, and reported waves up to 6.5 meters further out \cite{amymet}.

The most devastating storms cause flooding via many different factors, including high winds, rainfall, and storm surge.  The Norwegian Directorate for Civil Protection evaluates the risk of such an extreme flood in the Oslofjord, categorizing the risk as moderate, with large economic consequences and moderate amounts of personal injuries and long-term environmental destruction \cite{dsb}. The likelihood of another smaller-scale flood in one of the vulnerable towns along the Oslofjord is assessed to be almost 100\% \cite{dsb}.

As a result, coastal flooding due to storm surge is a vital challenge that must be properly addressed in Norway, and the Oslofjord especially. This study attempts to enhance the storm surge modeling capabilities in Norway by developing and validating a new computational mesh compatible with The ADvanced CIRCulation (ADCIRC) model, a high fidelity finite element solver for the shallow water equations (SWE)~\citep{vreugdenhil1994numerical,adcirc}. As a finite element solver, the naturally built-in unstructured mesh capabilities allow us to incorporate the irregular shapes seen in the Oslofjord. Additionally, as ADCIRC is a parallelized 2D SWE solver, we seek to develop a model that can provide rapid predictions with minimal computational efforts. 
There are presently several ocean models that incorporates water level variability due to weather effects that cover the Norwegian coast. These models are run both as operational forecast models, and in hindcast mode to create historical archives. Section \ref{sec:lit} elaborates on some of these models.

In the following, we present the development of an ADCIRC model for the Oslofjord and examine its performance using several historical storm scenarios. First, we address the existing models from the Norwegian Meteorological Institute (MET Norway) and elsewhere in Section \ref{sec:lit}. In Section \ref{sec:model}, we describe the numerical model and generate an unstructured mesh of the Oslofjord. Tidal analyses are discussed for a variety of simulated time periods in Section \ref{sec:results}, including some additional analysis using different sets of tidal constituents. Full realistic forcing is then used to simulate a storm in Section \ref{sec:full_forcing}, and the new Oslofjord mesh is compared to two existing models. Finally, conclusions and ideas for future works are expounded upon in Section \ref{sec:conclusions}.

The main novel contribution of this work is the Oslofjord mesh, which is analyzed and evaluated in a number of conditions. The mesh may be used in a variety of hydrodynamic applications, e.g. water level forecasting, compound flooding models, and transport or dredging studies.

\section{Overview of Relevant Models Covering the Oslofjord} \label{sec:lit}

There are several existing models for storm surge and hydrodynamics that include the Oslofjord, most notably those operated by MET Norway. At the core of these models is the Regional Ocean Modeling System (ROMS). ROMS uses a finite difference method to solve the Reynolds-averaged Navier Stokes (RANS) equations, with hydrostatic and Boussinesq assumptions \citep{roms}. Thus it is a baroclinic model that may optionally be run barotropically. It uses a split-explicit time-stepping scheme, in which the free surface and 2-D momentum equations use a much smaller time step, while the full 3-D model uses a larger timestep \cite{roms}.

\subsection{Norkyst}

Norkyst is one of MET Norway's premier models, developed in close collaboration with the Institute of Marine Research. It covers the entire coast of mainland Norway, from the North Sea to the Barents Sea,  and is based on ROMS, with a horizontal resolution of 800 meters \citep{norkyst}. In addition to tidal and atmospheric inputs, it uses forcing from Norway's 1760 main rivers. However, Norkyst was not specifically designed to model the behavior of water inside the fjords; rather, its intention is to provide a broader model of the mesoscale phenomena just outside of the fjords \citep{fjordos}. Norkyst can be run in barotropic or baroclinic modes.

\subsection{FjordOs} \label{sec:fjordos}

MET Norway also had a very high resolution model covering the entire Oslofjord called the FjordOs model. This model was discontinued in 2024, and replaced by a two-way nested child domain inside the Norkyst model, covering only the inner parts of the fjord. FjordOs \citep{fjordos} had a variable resolution between 50 and 300 meters. It used nine tidal constituents, enumerated in Table \ref{tab:constituents}, and also solved the 3-D RANS equations. Like Norkyst, FjordOs also included river discharge to force the model, using the 37 main rivers that flow into the Oslofjord. FjordOs aimed to resolve the features in the Oslofjord that Norkyst is unable to resolve adequately. To that aim, it took advantage of ROMS's optional curvilinear coordinate system, which allows for a nonuniform grid.

\subsection{Operational Water Level Forecasting} \label{sec:forecasting}

MET Norway's operational model for storm surge and water level forecasting, also based on ROMS, creates nowcasts and forecasts several times a day \citep{kristensen23}. The storm surge model operates on an even larger domain, covering the North, Norwegian, and Barents Seas. It has a 4-kilometer horizontal resolution, 10-second timestep, and quadratic bottom friction parameter of $2.5 \times 10^{-3}$. Its atmospheric forcing comes from the European Centre for Medium-Range Weather Forecasts, and contains 16-kilometer spatial and 3-hour temporal resolutions. Unlike Norkyst and FjordOs, however, this model utilizes ROMS's barotropic (2-D) mode, requiring much less computation time as a result \citep{kristensen23}. A ROMS benchmark, for reference, compares a full baroclinic model with a simplified barotropic one, using 64 CPU cores in the inner Oslofjord. The baroclinic run takes over 16 minutes, whereas the barotropic run completes in less than one minute for an hour of simulation time \citep{beiser}.

In addition to the deterministic model, MET Norway also employs an ensemble prediction system (EPS) for uncertainty quantification \citep{kristensen23}. Another EPS called GPU Ocean utilizes the computational framework of the barotropic ocean model with a reduced-gravity model for the upper ocean layer, essentially combining simplified barotropic and baroclinic models \citep{beiser}. Using GPU acceleration, this becomes a lightweight tool useful when the full baroclinic model is infeasibly computationally expensive, speeding up the baroclinic model by 350 times \cite{beiser}. 

The Norwegian Mapping Authorities (NMA) produce tidal predictions for a set of permanent stations based on harmonic analysis \cite{kristensen23}. After the operational forecasting model computes the storm surge, these tidal predictions are used as astronomical corrections to increase the accuracy of the predicted TWL.

\subsection{Global Models}

\begin{figure}[h!]
    \centering
    \includegraphics[width=0.85\linewidth]{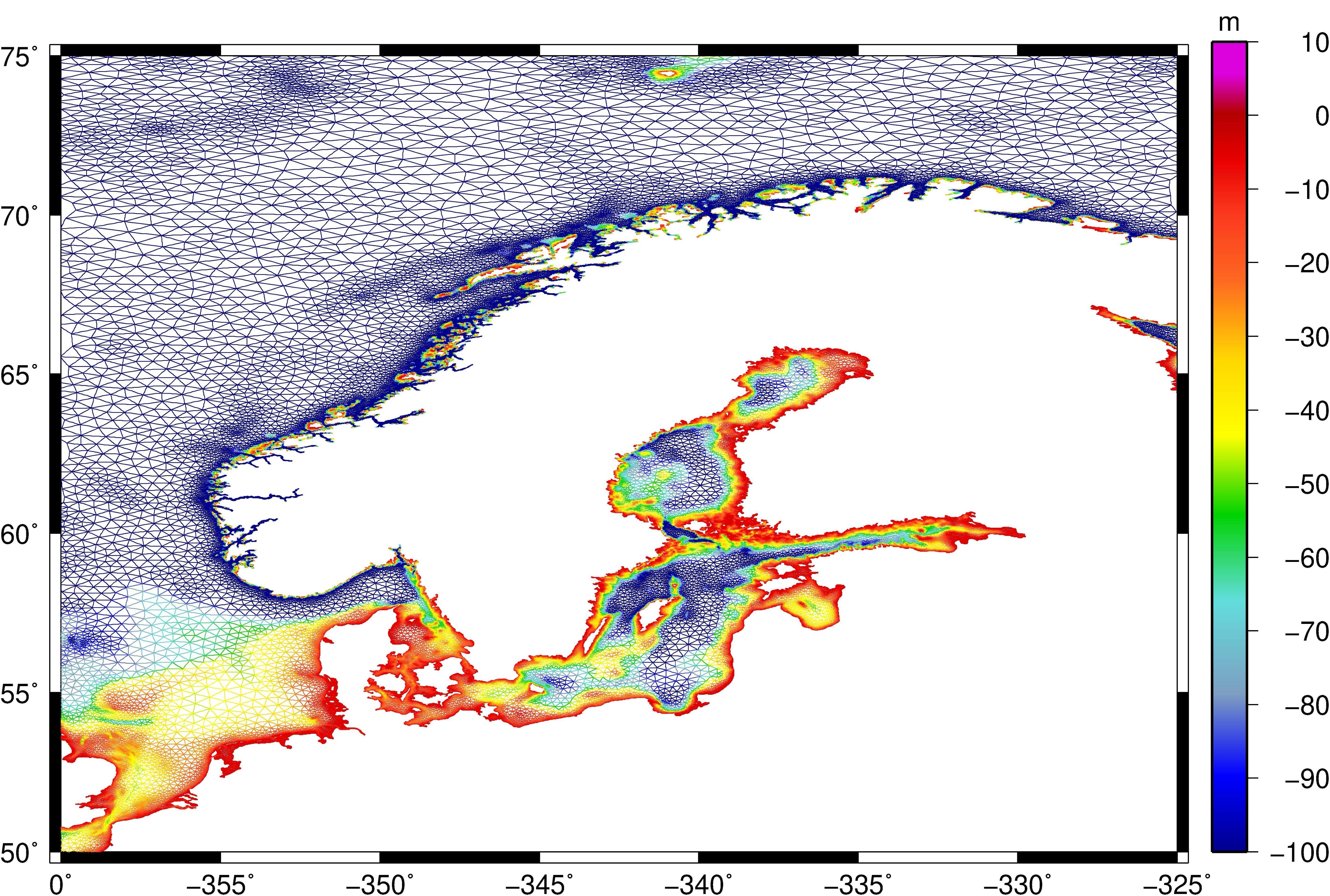}
    \caption{\textbf{STOFS-2D-Global mesh around the Norwegian mainland}.}
    \label{fig:stofs}
\end{figure}

Whereas the aforementioned efforts all use ROMS for ocean modeling, there is one coastal ocean model nearer to the extent of the present study. It is an ongoing project, led by the U.S.'s National Oceanic and Atmospheric Administration (NOAA), to model storm surge on a global level, called the Global Surge and Tide Operational Forecast System (STOFS) 2-D \citep{stofs}. Unlike the Norwegian models mentioned above, STOFS is 2D SWE-based model, and its mesh contains more than 24 million elements. It produces nowcasts and forecasts multiple times per day for use with weather forecasting, navigation, and disaster mitigation. It is refined up to 80 meter resolution on the U.S. west coast, and up to 120 meters on the U.S. east coast and for Pacific islands, but most of the world's coastlines are not nearly so refined. The maximum coastal resolution is 1.5 kilometers, including around Norway; see Figure~\ref{fig:stofs} for an overview of the mesh.

With this new Oslofjord mesh, we create a more refined ADCIRC model, denoted ADCIRC-OF, in pursuit of a middle ground between the highly scalable STOFS and the more computationally intensive 3D models. 

\section{Model Formulation} \label{sec:model}

\subsection{Numerical Model}

Our primary goal in this study is the development of a model based on a 2D unstructured mesh of the Oslofjord. 
While the Oslofjord exhibits complex seasonal baroclinic stratification, for our focus on resolving water surface elevations, e.g., storm surges, a 2D barotropic formulation is appropriate. 

The central objective is to develop a robust numerical framework capable of forecasting extreme sea-level events and coastal flooding. Recent events, such as Storm Hans in 2023 and the storm Amy in 2025, have demonstrated the vulnerability of the Oslofjord's densely populated coastline to significant flooding and infrastructure damage. By exploiting the inherent unstructured mesh refinement of a finite element method, our model can resolve the intricate geometry of the fjord, including the critical ``choke point'' at the Dr\o{}bak Sill. This allows us to accurately capture the fluxes through such narrow straits and channels, which are critical to 2D accurate flood simulations. 

In order to maintain the validity of the numerics and accurately capture the process of inundation and recession of water in low-lying areas, ADCIRC uses an advanced wetting and drying algorithm, which can turn elements on and off, thereby capturing the critical process of inundation of dry land \cite{luettich99, dietrich05}.

The ADCIRC model used here is based on the two-dimensional shallow water equations (SWE), which consist of the conservative depth-averaged equations of mass conservation as well as $x$ and $y$ momentum conservation \cite{vreugdenhil1994numerical,tan1992shallow}. This set of equations is stated as follows in Equation \ref{eq:SWE}, with a schematic of the shallow water depths in Figure \ref{fig:elevation_def}.

{Find }  $(\zeta, \mathbf{u})$  { such that:}  
\begingroup\makeatletter\def\f@size{11}\check@mathfonts
\begin{align}
\begin{split}
    \frac{\partial  \zeta}{\partial t} + \nabla \cdot (H{\mathbf{u}})  &= 0, \text{ in } \Omega, \\
    \frac{\partial (Hu_x)}{\partial t} + \nabla \cdot \left( Hu_x^2 + \frac{g}{2}(H^2-h_b^2), Hu_xu_y \right) - g\zeta \frac{\partial h_b}{\partial x} + \kappa u_x &= F_x, \text{ in } \Omega, \\ 
    \frac{\partial (Hu_y)}{\partial t} + \nabla \cdot \left( Hu_xu_y, Hu_y^2 + \frac{g}{2}(H^2-h_b^2) \right) - g\zeta \frac{\partial h_b}{\partial y} + \kappa u_y &= F_y, \text{ in } \Omega,
\end{split} \tag{1} \label{eq:SWE}
\end{align}
\endgroup
where $\zeta$ is the free surface elevation (positive upwards from the geoid),  $h_b$ the bathymetry (positive downwards from the geoid), $H$ is the total water column, $\mathbf{u} = \{ u_x,u_y\}^{\text{T}}$ is the depth-averaged velocity field, $\kappa$ is the bottom friction factor, and the source terms $F_x$ and $F_y$ represent other forces. These can include Coriolis force, tidal potential forces, wind stresses, and wave radiation stresses, and vertically integrated lateral stresses \citep{luettich2004formulation}. 
The bottom friction terms can assume a linear or nonlinear form, or a mix of both. For the linear formulation used in this study, $\kappa$ is constant.
Coriolis force is expressed as $fHu_y$ and $fHu_x$, where $f = 2 \times 7.29212 \times 10^{-5} \times \sin{\phi}$ and $\phi$ is the latitude. Wind stress is included and contributes through the pressure gradient terms and the input wind velocities, which are converted to wind stresses via the Garratt drag formulation \cite{garratt}.
The computational domain is denoted by $\Omega$ and its boundary $\Gamma$ is typically specified by three distinctive sections $\Gamma = \Gamma_{ocean}\cup\Gamma_{land}\cup\Gamma_{river}$. On these sections, the boundary conditions include specified elevation conditions, zero normal flow, and specified normal flow, respectively.

\begin{figure}[h!]
    \centering
    \begin{tikzpicture}[scale=1.4]

    \definecolor{waterblue}{RGB}{73,215,240}
    \definecolor{sand}{RGB}{145,98,23}
    
    \draw[black, very thick] (-4,0) -- (4,0);
    
    \draw[brown, thick] plot[smooth, tension=.7] coordinates {(-4,-2.5) (-2,-1.8) (0,-2.3) (2,-2.0) (4,-2.4)};
    \fill[sand] plot[smooth, tension=.7] coordinates {(-4.5,-3) (-4, -2.5) (-2,-1.8) (0,-2.3) (2,-2.0) (4,-2.4) (4.5,-3)} -- cycle;

    \draw[blue, thick] plot[smooth, tension=.7] coordinates {(-4,1.2) (-2,0.9) (0,1.2) (2,0.8) (4,1.2)};
    \fill[waterblue] plot[smooth, tension=.7] coordinates {(-4,1.2) (-2,0.9) (0,1.2) (2,0.8) (4,1.2)} -- plot[smooth, tension=.7] coordinates { (4,-2.4) (2,-2.0) (0,-2.3) (-2,-1.8) (-4,-2.5) } -- cycle;
    
    \node[right] at (1, -0.6) {$\qquad H = \zeta + h_b$};
    
    \node[right] at (4.5, 0.450) {$\quad \zeta $};
    \node[right] at (4.5, -1.2) {$\quad h_b$};

    \draw[<->] (1.3, 0.9) -- (1.3, -2.1);

    \draw[<->] (4.5, 1.2) -- (4.5, 0);
    \node[above] at (4.5, 1.3) { };

    \draw[<->] (4.5, 0) -- (4.5, -2.4);
    \node[below] at (4.5, -2.4) { };
    
    \draw[black , thick] (-4,0) -- (4,0);

\end{tikzpicture}
    \caption{\textbf{Definition of shallow water elevations}. The horizontal line is the geoid, where $\zeta = h_b = 0$.}
    \label{fig:elevation_def}
\end{figure}

At the core of our model is the two dimensional depth-integrated (2DDI) version of ADCIRC. ADCIRC does not solve the aforementioned SWE in the primitive form, but rather the Generalized Wave Continuity Equation (GWCE) form, which provides enhanced numerical stability and efficiency \cite{lynch1979wave,adcirc}. The 2DDI GWCE combines the time-differentiated primitive continuity equation with the spatially-differentiated primitive momentum equation, and it is shown in Cartesian coordinates in Equation \ref{eq:gwce}.

\begingroup\makeatletter\def\f@size{9}\check@mathfonts
\begin{align}
\begin{split}
    &\frac{\partial^{2} \zeta}{\partial t^{2}} + \tau_{0} \frac{\partial \zeta}{\partial t} \\
    &\qquad + \frac{\partial}{\partial x} \Bigg\{ u_{x} \frac{\partial \zeta}{\partial t} - Hu_{x} \frac{\partial u_{x}}{\partial x}  - Hu_{y} \frac{\partial u_{x}}{\partial x} + fHu_{y} \\
    &\qquad \qquad - H \frac{\partial}{\partial x} \left[ \frac{p_{s}}{\rho_{0}} + g (\zeta - \alpha \eta) \right] - E_{h_{2}}^{M} \frac{\partial^{2} \zeta}{\partial x \partial t} + \frac{\tau_{sx}}{\rho_{0}} + (H \tau_{0} - \kappa ) u_{x} \Bigg\} \\
    &\qquad + \frac{\partial}{\partial y} \Bigg\{ u_{y} \frac{\partial \zeta}{\partial t} - Hu_{x} \frac{\partial u_{y}} {\partial x}  - H u_{y} \frac{\partial u_{y}}{\partial y} - fHu_{x} \\
    &\qquad \qquad - H \frac{\partial}{\partial y} \left[ \frac{p_{s}}{\rho_{0}} + g (\zeta - \alpha \eta) \right] - E_{h_{2}}^{M} \frac{\partial^{2} \zeta}{\partial y \partial t} + \frac{\tau_{sy}}{\rho_{0}} + (H\tau_{0} - \kappa ) u_{y}  \Bigg\} = 0,
\end{split} \tag{2} \label{eq:gwce}
\end{align} \endgroup
where $\tau_{0}$ is a user-specified constant affecting the numerical diffusion. Here, $p_{s}$ is the atmospheric pressure at the surface, $\alpha$ is the effective Earth elasticity factor, $\eta$ is the Newtonian equilibrium tidal potential, $E_{h_{2}}^{M}$ is the horizontal eddy diffusion/dispersion coefficient, and $\tau_{sx}$ and $\tau_{sy}$ are wind stresses at the surface. We refer to \cite{adcirc} for further details on these functions and parameters. Note: since $\cos\gamma=1$ for the 2DDI case, it was removed for simplicity.

\subsection{Weak Formulation and Discretization}

ADCIRC uses the continuous Bubnov-Galerkin finite element method to solve the SWE, which are discretized in space using linear triangular elements, and we describe this discretization scheme in brief here (c.f. \cite{adcirc}).

We seek globally continuous approximations $\zeta^{h}$, $Hu_{x}^{h}$, and $Hu_{y}^{h} \in V_{h} = P^{1}(\Omega)$, where $V_{h}$ is the space of piecewise continuous linear polynomials. To obtain a weighted residual statement for the GWCE, we weight it by arbitrary test function $v \in V_{h}$. First, the terms involving gradients of the pressure, tidal potential, and surface elevation may be rewritten as:
\begingroup\makeatletter\def\f@size{9}\check@mathfonts
\begin{align*}
    H \frac{\partial}{\partial x} \left[ \frac{p_{s}}{\rho_{0}} + g (\zeta - \alpha \eta) \right] &= gh_{b} \frac{\partial \zeta}{\partial x} + \frac{g}{2} \frac{\partial \zeta^{2}}{\partial x^{2}} + gH \frac{\partial}{\partial x} \left(\frac{p_{s}}{\rho_{0}g} - \alpha \eta \right) \\
    H \frac{\partial}{\partial y} \left[ \frac{p_{s}}{\rho_{0}} + g (\zeta - \alpha \eta) \right] &= gh_{b} \frac{\partial \zeta}{\partial y} + \frac{g}{2} \frac{\partial \zeta^{2}}{\partial y^{2}} + gH \frac{\partial}{\partial y} \left(\frac{p_{s}}{\rho_{0}g} - \alpha \eta \right),
\end{align*}
\endgroup
Then, after integrating over the domain $\Omega$, applying Gauss's divergence theorem, and simplifying, the GWCE's weak formulation is written as:
\begingroup\makeatletter\def\f@size{9}\check@mathfonts
\begin{align}
\begin{split}
    &\left( \frac{\partial^{2} \zeta}{\partial t^{2}}, v \right)_{\Omega} +
    \left( \tau_{0} \frac{\partial \zeta}{\partial t}, v \right)_{\Omega} +
    \left( gh \frac{\partial \zeta}{\partial x}, \frac{\partial v}{\partial x} \right)_{\Omega}  +
    \left( gh \frac{\partial \zeta}{\partial y}, \frac{\partial v}{\partial y} \right)_{\Omega} = \\
    &\qquad \left( u_{x} \frac{\partial \zeta}{\partial t}, \frac{\partial v}{\partial x} \right)_{\Omega} +
    \left( u_{y} \frac{\partial \zeta}{\partial t}, \frac{\partial v}{\partial y} \right)_{\Omega} + 
    \left( W_{x}, \frac{\partial v}{\partial x} \right)_{\Omega} + 
    \left( W_{y}, \frac{\partial v}{\partial y} \right)_{\Omega} \\
    &\qquad \qquad - \int_{\Gamma_{Q}} \left( \frac{\partial Q_{n*}}{\partial y} + \tau_{0} Q_{n*} \right) v \mathrm{d}\Gamma, \label{eq:weakgwce}
\end{split} \tag{3}
\end{align}
\endgroup
where $Q_{n*}$ is defined to be the specified normal flux across the boundary $\Gamma$, and
\begingroup\makeatletter\def\f@size{9}\check@mathfonts
\begin{align*}
    W_{x} &\equiv -Hu_{x} \frac{\partial u_{x}}{\partial x} - Hu_{y} \frac{\partial u_{x}}{\partial y} + H f u_{y} - \frac{g}{2} \frac{\partial \zeta^{2}}{\partial x^{2}} - gH \frac{\partial}{\partial x} \left( \frac{p_{s}}{\rho_{0}g} - \alpha \eta \right),  \\
    &\qquad \qquad + \frac{\tau_{sx}}{\rho_{0}} + \left( H \tau_{0} - \kappa \right) u_{x} \\
    W_{y} &\equiv -Hu_{x} \frac{\partial u_{y}}{\partial x} - Hu_{y} \frac{\partial u_{y}}{\partial y} + H f u_{x} - \frac{g}{2} \frac{\partial \zeta^{2}}{\partial y^{2}} - gH \frac{\partial}{\partial y} \left( \frac{p_{s}}{\rho_{0}g} - \alpha \eta \right) \\
    &\qquad \qquad + \frac{\tau_{sy}}{\rho_{0}} + \left( H \tau_{0} - \kappa \right) u_{y}. 
\end{align*}
\endgroup

Similarly, for the momentum equations in Equation \eqref{eq:SWE}, we multiply by $v$, then integrate over $\Omega$ and apply the divergence theorem to obtain their weak forms:
\begingroup\makeatletter\def\f@size{9}\check@mathfonts
\begin{align}
\begin{split}
    &\left( \frac{\partial u_{x}}{\partial t}, \phi_{i} \right)_{\Omega}
    - \left( fu_{y}, \phi_{i} \right)_{\Omega}
    + E_{h_{2}}^{M} \left( \frac{\partial Hu_{x}}{\partial x}, \frac{\partial}{\partial x} \left( \frac{\phi_{i}}{H} \right) \right)_{\Omega}
    + E_{h_{2}}^{M} \left( \frac{\partial Hu_{x}}{\partial y}, \frac{\partial}{\partial y} \left( \frac{\phi_{i}}{H} \right) \right)_{\Omega} =\\
    &\qquad - \left( \frac{\partial}{\partial x} \left[ \frac{p_{s}}{\rho_{0}} + f \left( \zeta - \alpha \eta \right) \right] - \frac{\tau_{sx}}{\rho_{0}H} + \frac{\kappa}{H} u_{x}, \phi_{i} \right)_{\Omega}
    - \left(  u_{x} \frac{\partial u_{x}}{\partial x} + u_{y} \frac{\partial u_{x}}{\partial y}, \phi_{i} \right)_{\Omega}, \label{eq:weakxmomentum}
\end{split}  \tag{4}
\end{align}
\begin{align}
\begin{split}
    &\left( \frac{\partial u_{y}}{\partial t}, \phi_{i} \right)_{\Omega}
    - \left( fu_{x}, \phi_{i} \right)_{\Omega}
    + E_{h_{2}}^{M} \left( \frac{\partial Hu_{y}}{\partial x}, \frac{\partial}{\partial x} \left( \frac{\phi_{i}}{H} \right) \right)_{\Omega}
    + E_{h_{2}}^{M} \left( \frac{\partial Hu_{y}}{\partial y}, \frac{\partial}{\partial y} \left( \frac{\phi_{i}}{H} \right) \right)_{\Omega} = \\
    &\qquad - \left( \frac{\partial}{\partial y} \left[ \frac{p_{s}}{\rho_{0}} + f \left( \zeta - \alpha \eta \right) \right] - \frac{\tau_{sy}}{\rho_{0}H} + \frac{\kappa}{H} u_{y}, \phi_{i} \right)_{\Omega}
    - \left(  u_{x} \frac{\partial u_{y}}{\partial x} + u_{y} \frac{\partial u_{y}}{\partial y}, \phi_{i} \right)_{\Omega}. \label{eq:weakymomentum}
\end{split}  \tag{5}
\end{align}
\endgroup

We represent the unknowns $\zeta^{h}$, $u_{x}^{h}$, and $u_{y}^{h}$ as linear combinations of basis functions $\phi_{i}(x,y) \in V_{h}$, piecewise linear functions over each element $e \subset \Omega$:
\begingroup\makeatletter\def\f@size{9}\check@mathfonts
\begin{align}
    \zeta^{h} &= \sum_{j=1}^{3} \zeta^{(j)}(t) \phi_{j}(x,y), \nonumber \\
    u_{x}^{h} &= \sum_{j=1}^{3} u_{x}^{(j)}(t) \phi_{j}(x,y), \tag{6} \label{eq:basis} \\
    u_{y}^{h} &= \sum_{j=1}^{3} u_{y}^{(j)}(t) \phi_{j}(x,y), \nonumber 
\end{align} 
\endgroup
where $\zeta^{(j)}(t)$, $u_{x}^{(j)}(t)$, and $u_{y}^{(j)}(t)$ are unknown coefficients.

Substituting \eqref{eq:basis} into the weak formulations \eqref{eq:weakgwce}, \eqref{eq:weakxmomentum}, and \eqref{eq:weakymomentum} yields a system of ordinary differential equations in time. For the GWCE, a three-level semi-implicit time-stepping scheme is used, with the nonlinear terms $W_{x}$ and $W_{y}$ being treated explicitly. Similarly, to solve the momentum equations, a two-level implicit Crank-Nicolson scheme is used for the linear terms, and an explicit scheme is used for the nonlinear terms. This scheme is detailed in \cite{adcirc}.

\subsection{Mesh Generation}

The computational mesh was generated with OceanMesh2D~\citep{oceanmesh}, a MATLAB-based tool for generating unstructured meshes for coastal models. There are several required elements to create the mesh, including shoreline data, a digital elevation model (DEM), a few resolution parameters, and a definition for the open boundary. 

The DEM for the mesh bathymetry was created using data from a global collection of satellite bathymetry and topography data, SRTM15+ \citep{srtm}, which contains shipboard sounding and satellite altimetry data around the globe at a 15 arc-second spatial interval, which corresponds to about 250--460 meters at the latitudes of the Oslofjord. The data from Geonorge, the Norwegian Mapping Authority's site for location data, does have higher resolution available than SRTM15+; however, many gaps exist in the publicly-available data for the Oslofjord domain in particular, resulting in an uneven bathymetry. 

The mesh's shoreline is from the Global Self-consistent, Hierarchical, High-resolution Geography Database (GSHHG), which is an amalgamation of three public domain datasets \citep{gshhg}. Its main source for ocean shorelines is the World Vector Shorelines, which has an approximate precision of 50 meters.

\begin{figure}
    \centering
    \begin{subfigure}[t]{0.45\textwidth}
        \includegraphics[width=\linewidth]{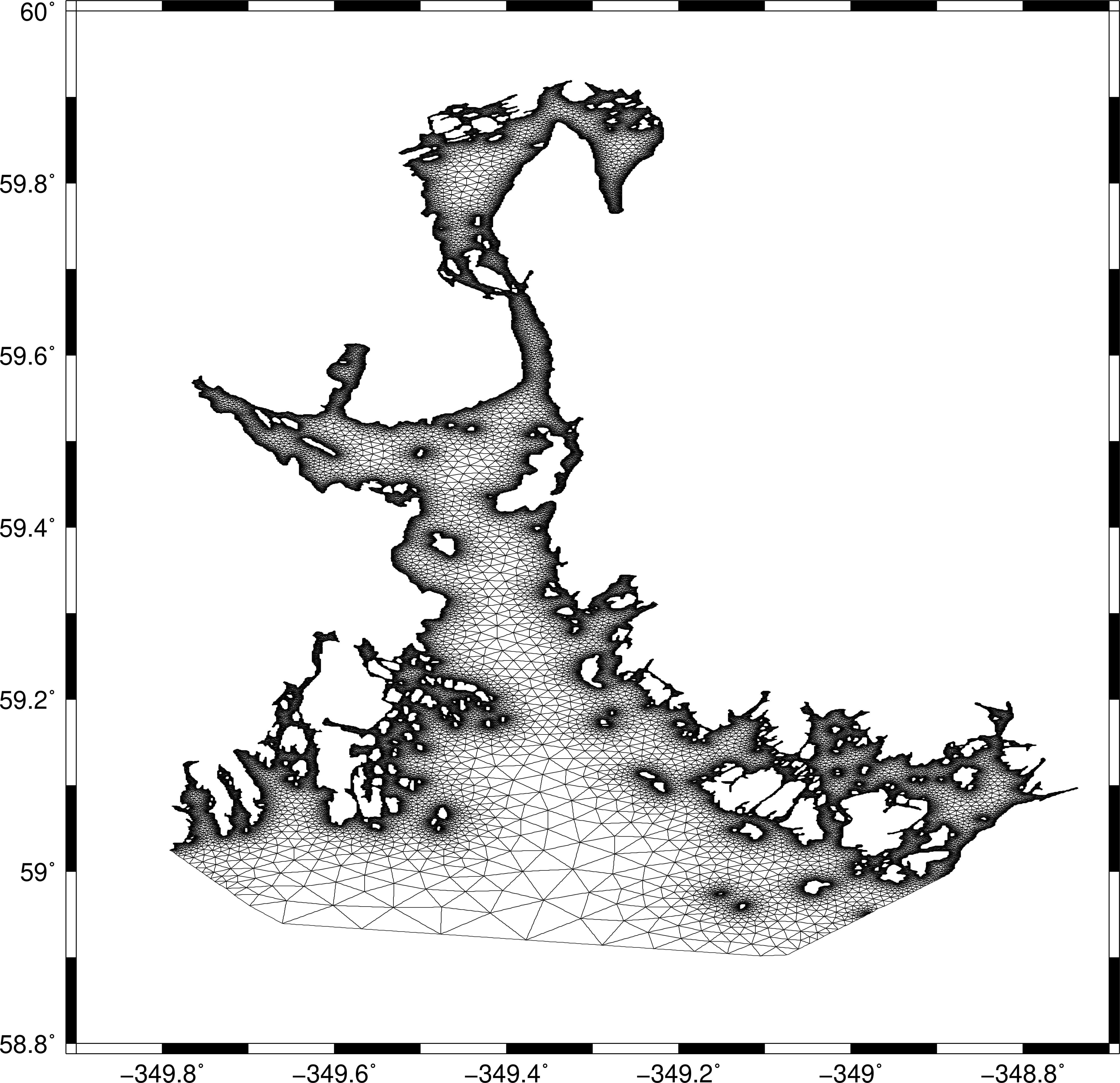}
        \caption{Oslofjord mesh grid, with 50-m resolution in the finest areas}
        \label{fig:mesh}
    \end{subfigure}
    \begin{subfigure}[t]{0.5\textwidth}
        \includegraphics[width=\linewidth]{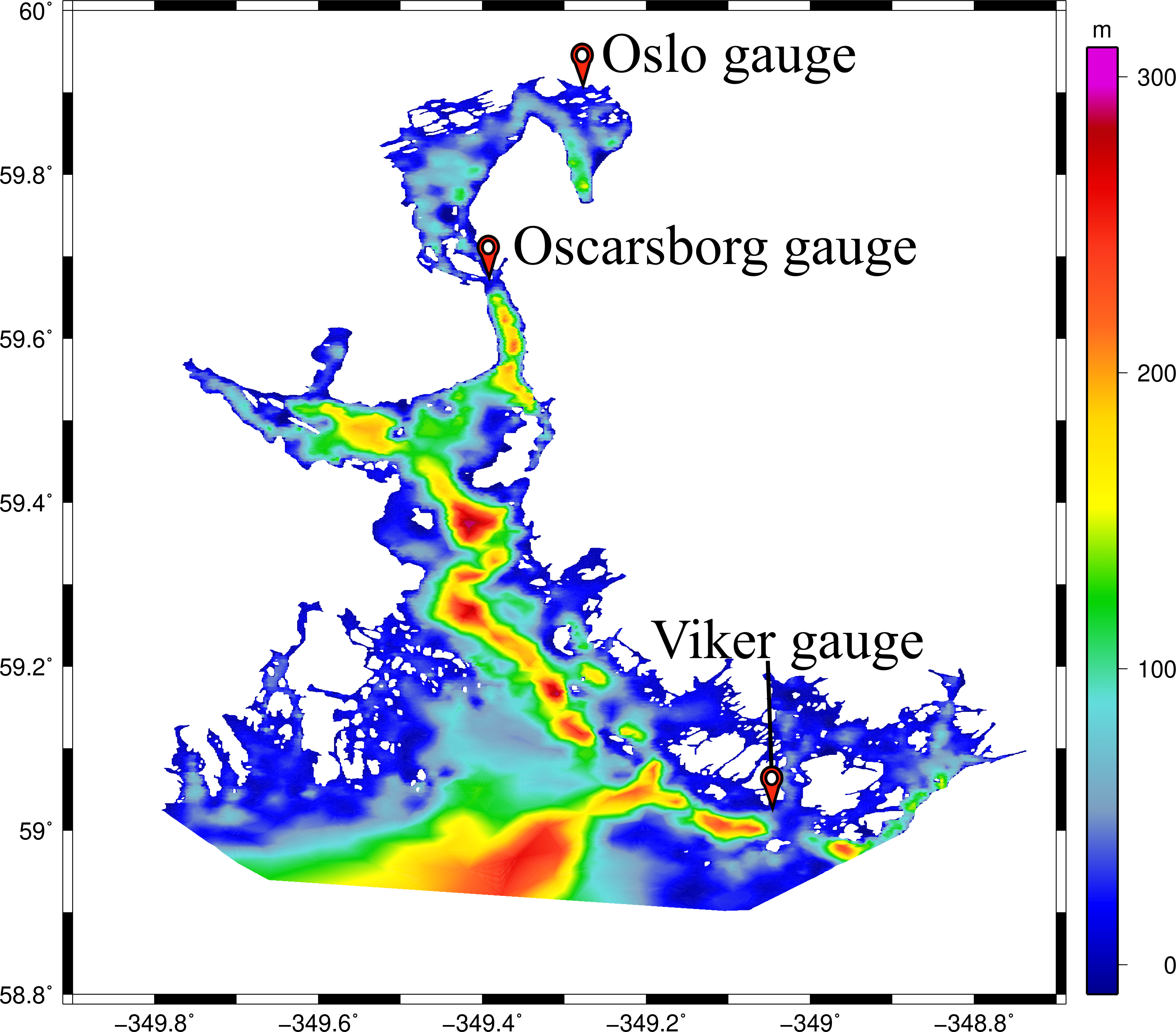}
        \caption{Mesh bathymetry, showing the locations of ocean surface elevation gauges}
        \label{fig:bathymetry}
    \end{subfigure}
    \caption{\textbf{The mesh and bathymetry of the Oslofjord used in ADCIRC-OF}.}
    \label{fig:mesh_and_bathymetry}
\end{figure}

The resulting Oslofjord mesh is defined at 50-meter resolution along the shoreline, varying between thirty meters in the tightest channels and several kilometers further out in the fjord. The element size for the full mesh is shown in \ref{appendix:size}.
The resolution is much finer around islands and in areas with high bathymetry gradients, in order to more effectively resolve submarine features important for dissipative and reflective effects \cite{oceanmesh}. The mesh contains a total of 120,781 elements and 70,410 nodes. The mesh and its bathymetry are shown in Figure \ref{fig:mesh_and_bathymetry}.

The open boundary of the mesh was created manually, as a bounding box along the southern edge, to capture tidal effects at the wide mouth of the fjord. There are also 250 inner and outer boundaries with no normal flow, representing the mainland and islands. The complex geography of the fjord and its many islands results in several narrow channels in the mesh.

\begin{figure}
    \centering
    \begin{subfigure}[t]{0.45\textwidth}
        \includegraphics[width=\linewidth]{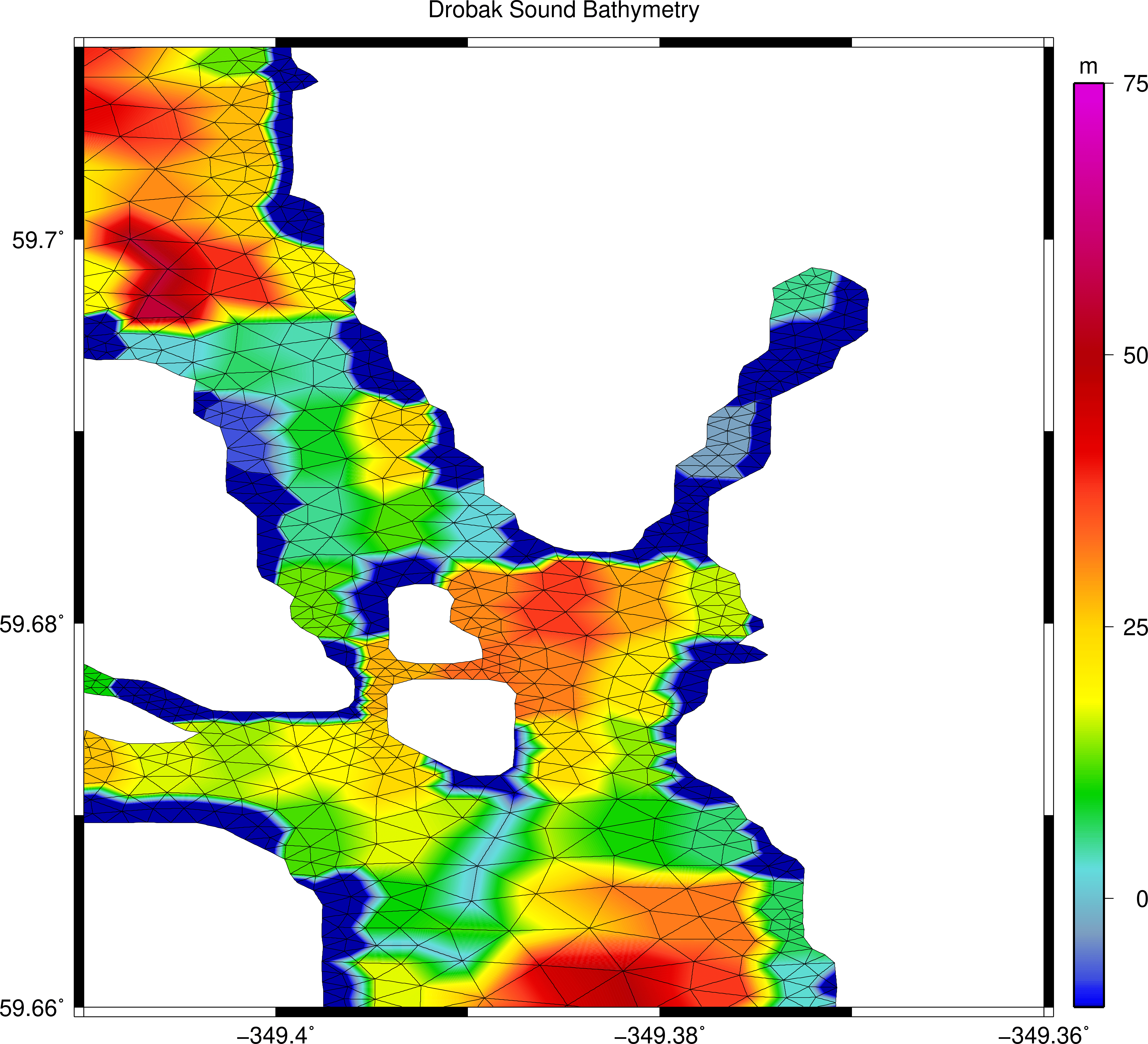}
        \caption{Bathymetry}
        \label{fig:drobakbathymetry}
    \end{subfigure}
    \begin{subfigure}[t]{0.45\textwidth}
        \includegraphics[width=\linewidth]{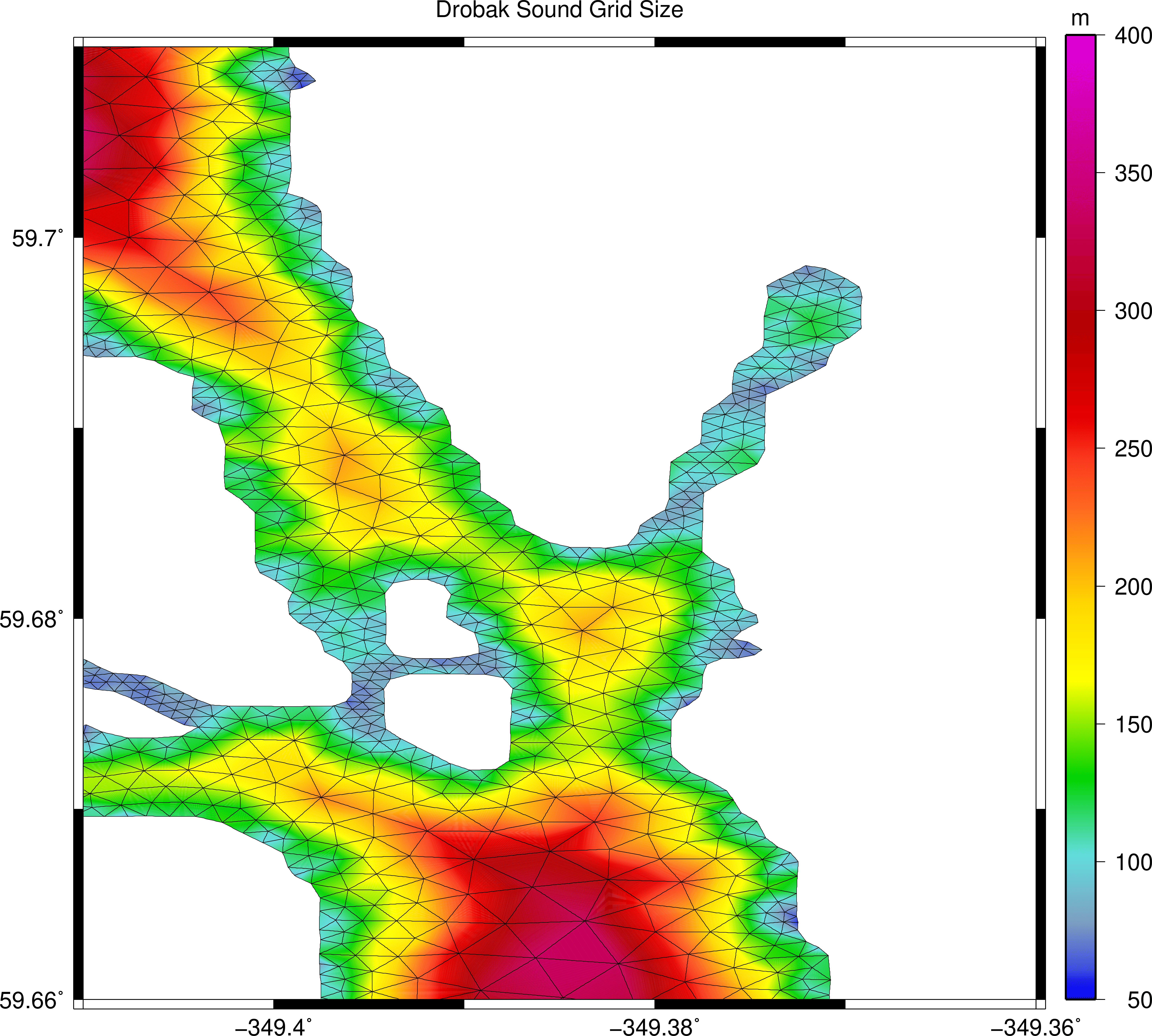}
        \caption{Grid size}
        \label{fig:drobakmeshsize}
    \end{subfigure}
    \caption{\textbf{The Oslofjord mesh bathymetry and grid size near Dr\o{}bak sound and the Oscarsborg gauge}. The Oscarsborg Fortress jetty is visible in the lower left.}
    \label{fig:drobak}
\end{figure}

\begin{figure}
    \centering
    \begin{subfigure}[t]{0.45\textwidth}
        \includegraphics[width=\linewidth]{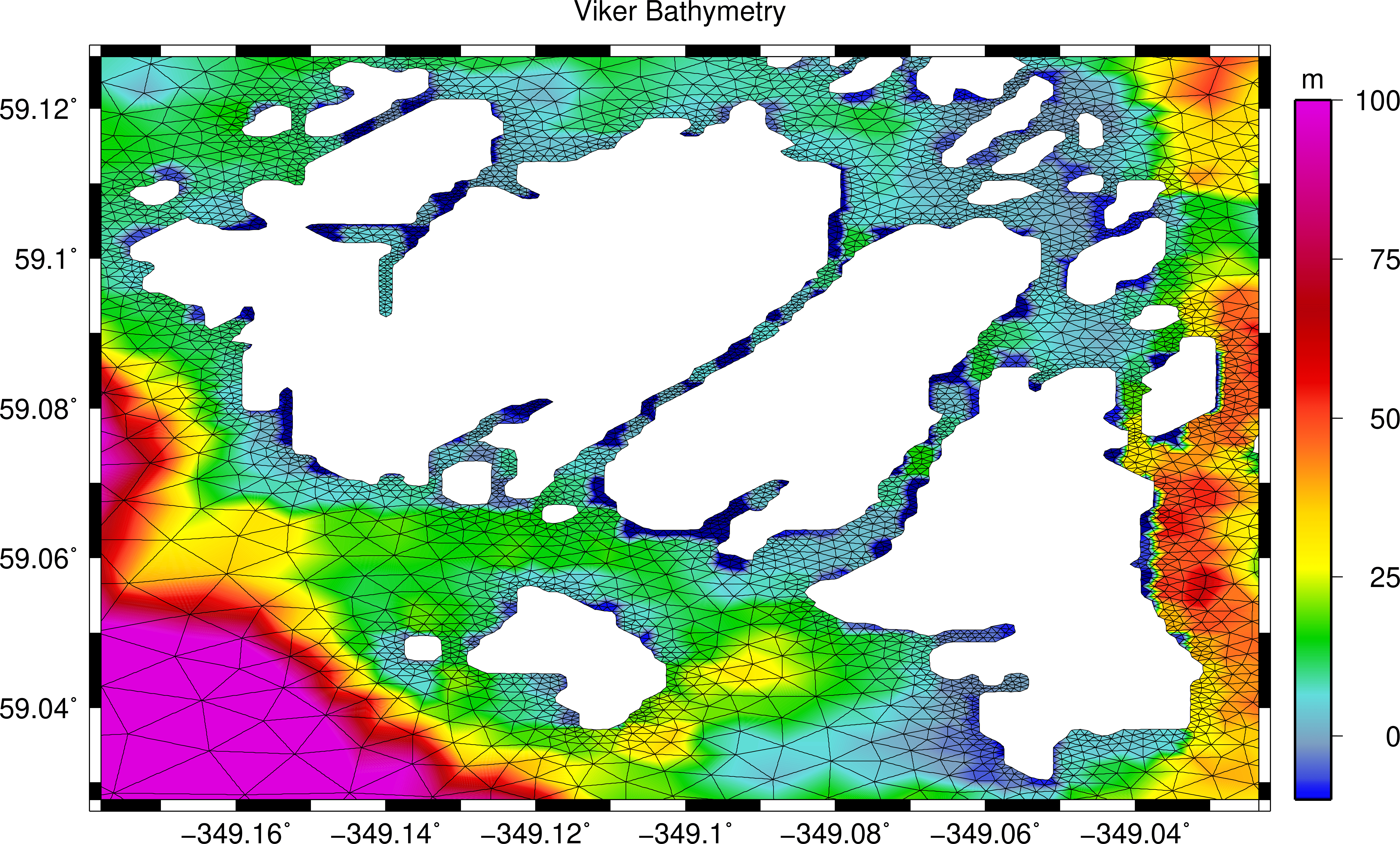}
        \caption{Bathymetry}
        \label{fig:vikerbathymetry}
    \end{subfigure}
    \begin{subfigure}[t]{0.45\textwidth}
        \includegraphics[width=\linewidth]{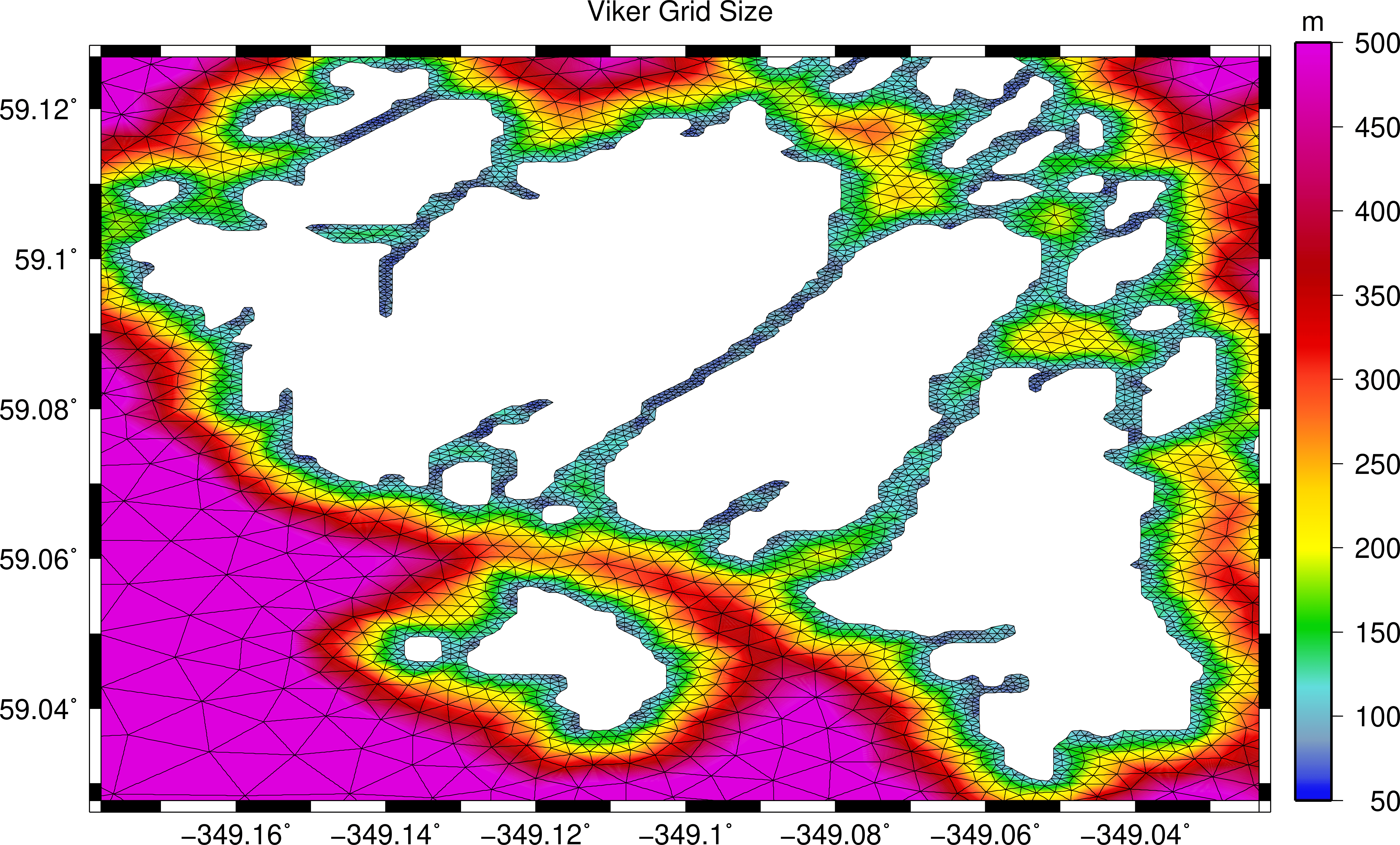}
        \caption{Grid size} 
        \label{fig:vikermeshsize}
    \end{subfigure}
    \caption{\textbf{The Oslofjord mesh bathymetry and grid size near Ytre Hvaler National Park and the Viker gauge}. The smallest elements in the channels between islands are of 30--50 meter diameter.}
    \label{fig:viker}
\end{figure}

Some examples of these channels are shown in Figures \ref{fig:drobakmeshsize} and \ref{fig:vikermeshsize}.
A few areas, including part of the Drammensfjord, contained channels that were too thin even for our 50-m resolution, introducing local instabilities into the model. Those smallest channels were removed from the initial mesh.

The bathymetry was adjusted for two nodestrings in Dr\o{}bak Sound, in order to account for the Oscarsborg Fortress's underwater wall. The man-made jetty extends south from S\o{}ndre Kaholmen to Sm\aa{}skj\ae{}r, and then west to the coast. The water depth along the jetty has an average depth of 0.5 meters, and there is a 140-m ship channel in the south, which has a depth of about 10.4 meters \cite{oscarsborg_jetty}. An approximation of this feature adapted to the Oslofjord mesh is clearly shown in Figure \ref{fig:drobakbathymetry}.

\subsection{Model Setup and Forcings} \label{sec:methods}

To develop the ADCIRC-OF model, we add boundary conditions and forcing data to the Oslofjord mesh for a given period. We use semi-implicit timestepping to solve the SWE, in order to improve the stability and accuracy of the model. Due to the many fine, narrow channels in the mesh, we neglect advective terms in the model equations, and a timestep of 0.5 seconds is selected.

The mesh includes many elements around the coastline that are initially ``dry,'' with local water level less than 10 centimeters. ADCIRC's wetting-drying algorithm ensures that inundation is possible if water in surrounding elements would flow into the dry ones \cite{dietrich05}.

To mirror the MET Norway prediction system for the Oslofjord which we compare against~\cite{kristensen23}, a quadratic bottom friction formulation was used, with constant friction factor of 0.0025. While a spatially varying bottom friction value would better describe the fjord, using the same constant friction factor will provide a baseline for future study.

Wind and pressure data were collected from the Global Forecast System (GFS) \citep{gfs} using MetGet \citep{metget}. GFS has a temporal resolution of one hour and a spatial resolution of 0.25 $\times$ 0.25 degrees, corresponding to about 15 $\times$ 28 kilometers in the Oslofjord. The data are interpolated in space onto the ADCIRC mesh, and in time to synchronize with the model timestep. 

The tidal constituents were extracted from the TPXO9 global barotropic tide model \citep{tpxo}. Varying sets of these constituents were used, as will be expounded upon in Section \ref{sec:tides}, but the primary model was forced with the major 8 constituents.

\section{Validation and Tidal Analysis} \label{sec:results}

In this section, we present results from the developed model, beginning with an analysis of the model's tidal forcing capability over four distinct time periods, including an extraordinary wind event. We investigate the impacts of using additional tidal constituents to force the developed model and construct harmonic analyses for each simulation.

The technical details of the simulation setup are given in Table \ref{tab:run_details}. For a 0.5-second timestep, the models average around 12 seconds of runtime per hour simulated.
    
\begin{table}
    \centering
    \scriptsize
    \resizebox{\textwidth}{!}{
    \begin{tabular}{c|c|c|c|c|c|c}
        \toprule
        \textbf{Version} & \textbf{Computing System} & \textbf{Processors} & \textbf{Cores} & \textbf{Period} & \textbf{Runtime (min.)} & \textbf{dt} (s) \\
        \midrule
        \multirow{4}*{ADCIRC v55.01} & \multirow{4}*{TACC Frontera} && \multirow{4}*{40} & 5--15 Sep 2022 & 48 & 0.5 \\
        && Intel 8280 && 1--15 Jan 2025 & 68 & 0.5 \\
        && ``Cascade Lake'' && 1--15 Jul 2025 & 66 & 0.5 \\
        &&&& Storm Hans (7--10 Aug 2023) & 69 & 0.1 \\
        \bottomrule
    \end{tabular}}
    \caption{\textbf{ADCIRC-OF run details}.}
    \label{tab:run_details}
\end{table}


The Norwegian Mapping Authority (NMA) has 24 permanent water elevation gauges around the country, three of which are in our Oslofjord domain. These gauges provide elevation data every ten minutes at three key points in the mesh: Viker in the southeast, the Oscarsborg fortress in Dr\o{}bak Sound, and Oslo Harbor. Data from these gauges, along with NMA predicted tides, are collected from Se Havniv\aa{} \cite{sehavniva}.

The tidal constituents considered in this study are outlined in Table \ref{tab:constituents} They are divided into groups based on the sets of constituents used to force ADCIRC-OF. In order to reconstruct the tidal signals, we consider the Rayleigh criterion for minimum time series span to discern between two constituents:
\begin{align}
    T \geq \frac{R}{\Delta f}, \label{eq:rayleigh} \tag{7}
\end{align}
where $\Delta f$ is the difference between the constituents' frequencies and $R$ is the chosen Rayleigh number. To determine which constituents we can reconstruct, we examine their periods, also shown in Table \ref{tab:constituents} \cite{barker}. $\text{K}_{2}$ and $\text{M}_{2}$, for example, only differ in period by a few minutes, thus requiring a time span $T$ of over six months for the conventional choice of $R = 1$. Because the time spans simulated in this study are relatively short, we use a Rayleigh number $R = 0.5$.

Accordingly, the tidal reconstructions in this study only use four constituents: $\text{M}_{2}$, $\text{S}_{2}$, $\text{K}_{1}$, and $\text{O}_{1}$. Other similar constituents may be inferred via their astronomical equilibrium relationships, but for clarity we omit these \cite{doodson}. The shortest period studied only consists of three days, and hence for that period we can only distinguish two constituents for reconstruction and analysis: $\text{M}_{2}$ and $\text{K}_{1}$.

\begin{table}
    \centering
    \scriptsize
    \begin{tabular}{c|c|c|c|c|c}
        \toprule
        \textbf{Name} & \textbf{Period} $f$ (h) & \textbf{Description} & \textbf{Major 8} & \textbf{FjordOs} & \textbf{All} \\
        \midrule
        $M_{2}$ & 12.42  & Principal lunar semidiurnal constituent & \checkmark & \checkmark & \checkmark \\
        $S_{2}$ & 12     & Principal solar semidiurnal constituent & \checkmark & \checkmark & \checkmark \\
        $S_{2}$ & 12.67  & Larger lunar elliptic semidiurnal constituent & \checkmark & \checkmark & \checkmark \\
        $K_{2}$ & 11.97  & Lunisolar semidiurnal constituent & \checkmark & & \checkmark \\
        $K_{1}$ & 23.93  & Lunar diurnal constituent & \checkmark & \checkmark & \checkmark \\
        $O_{1}$ & 25.82  & Lunar diurnal constituent & \checkmark & \checkmark & \checkmark \\
        $P_{1}$ & 24.07  & Solar diurnal constituent & \checkmark && \checkmark \\
        $Q_{1}$ & 26.87  & Larger lunar elliptic diurnal constituent & \checkmark & \checkmark & \checkmark \\
        $M_{f}$ & 327.86 & Lunisolar fortnightly constituent &&& \checkmark \\
        $M_{m}$ & 661.31 & Lunar monthly constituent &&& \checkmark \\
        $M_{4}$ & 6.21   & Shallow water overtides of principal lunar constituent && \checkmark & \checkmark \\
        $MN_{4}$ & 6.27  & Shallow water quarter diurnal constituent && \checkmark & \checkmark \\
        $MS_{4}$ & 6.10  & Shallow water quarter diurnal constituent && \checkmark & \checkmark \\
        $2N_{2}$ & 12.91 & Lunar elliptical semidiurnal second-order constituent &&& \checkmark \\
        $S_{1}$ & 24     & Solar diurnal constituent &&& \checkmark \\
        \bottomrule
    \end{tabular}
    \caption{\textbf{List of the tidal constituents, and their periods, that are used in the ADCIRC-OF experiments.}}
    \label{tab:constituents}
\end{table}

Using these reconstructed tides, along with the ADCIRC-OF results and observed gauge data, we calculate the residuals between TWL and tides. This is often referred to as the storm surge, or weather effect contribution to the TWL. Since the tides and weather effect interact nonlinearly, called tide-surge interactions, it is not possible to fully separate them, and hence the residual is the common way to refer to this part of the TWL. This separation into a tidal and non-tidal residual part is very helpful when we assess the skill of the new mesh at reproducing the observed total water levels, since we can evalute these two parts individually. The ``observed residuals'' are calculated by subtracting the NMA tidal predictions from the observed TWL. In order to calculate the ``modeled residuals,'' we reconstruct the tidal signal from the ADCIRC-OF model with atmospheric and tidal forcing. Then, similarly, we subtract these reconstructed tides from ADCIRC-OF's simulated TWL.

\subsection{Major 8 Constituents}\label{results_main}
First, the model was run with the standard major 8 tidal constituents, and using four time periods under different weather conditions. For the first three periods we examine, a timestep of 0.5 seconds is used, while the last, most extreme weather event required a smaller 0.1-second timestep. We can broadly quantify the error between ADCIRC-OF and the NMA tidal predictions~\cite{kristensen23} using the root mean square error (RMSE) and Pearson correlation coefficients. These quantities are also calculated between the observed and modeled residuals. These values are given in Table \ref{tab:rmse_corr_tides} for each simulated period.

\begin{table}
    \centering
    \small
    \begin{tabular}{c|c|c|c|c|c}
        \toprule
        \textbf{Period} & \textbf{Gauge} & \multicolumn{2}{l|}{\textbf{Tides}} & \multicolumn{2}{l}{\textbf{Residuals}} \\
        && $\text{RMSE}_{\text{tides}}$ [cm] & $\rho_{\text{tides}}$ & $\text{RMSE}_{\text{resid}}$ [cm] & $\rho_{\text{resid}}$ \\
        \midrule
        & Viker & 4 & 0.95 & 19 & -0.04 \\
        Sep 2022 & Oscarsborg & 5 & 0.95 & 22 & 0.16 \\
        & Oslo & 10 & 0.74 & 23 & 0.30 \\
        \midrule
        & Viker & 5 & 0.94 & 17 & 0.19 \\
        Jan 2025 & Oscarsborg & 6 & 0.93 & 17 & 0.46 \\
        & Oslo & 10 & 0.69 & 17 & 0.51 \\
        \midrule
        & Viker & 5 & 0.89 & 10 & -0.08 \\
        Jul 2025 & Oscarsborg & 6 & 0.86 & 10 & -0.08 \\
        & Oslo & 6 & 0.89 & 10 & 0.02 \\
        \midrule
        & Viker & 5 & 0.95 & 43 & 0.00 \\
        Storm Hans & Oscarsborg & 7 & 0.94 & 43 & -0.04 \\
        & Oslo & 10 & 0.84 & 42 & -0.06 \\
        \bottomrule
    \end{tabular}
    \caption{\textbf{RMSE and correlation between the tidal ADCIRC-OF model and NMA tidal predictions, and between modeled and observed residuals}. The time series for these results are shown in Figures \ref{fig:sept22}--\ref{fig:hans_tides}.}
    \label{tab:rmse_corr_tides}
\end{table}

We note that $\text{RMSE}_{\text{tides}}$ values peak at 10 cm at the Oslo gauge, and decrease further out, towards Viker, to 4 or 5 centimeters for each simulated period. Similarly, $\rho_{\text{tides}}$ is generally lowest at Oslo and increasing further south. However, for the July 2025 case, the correlations for all three gauges have only negligible differences. We also note for this case, that the correlation is lower than the other cases for the Viker and Oscarsborg stations. Visual inspection of the modeled residual in Figure \ref{fig:jul25}, could suggest some ``tidal contamination'' of the residual, but this has not been investigated further.

The errors in the modeled residuals are considerably higher than those for the tides. $\text{RMSE}_{\text{resid}}$ is highest for the Storm Hans simulation, at about 40 cm. It is smallest for the July 2025 case, at about 10 cm. This is likely due to the combination of the fact that the model is forced solely by tides at the open boundary, and that this is the period with the least weather contribution of the four. We can only speculate that this could result in improved $\text{RMSE}_{\text{tides}}$ for Oslo due to minimal effect of tide-surge interaction.
However, for each case, the residual correlations are very low, showing a clear lack of skill for modeling the residual weather effects with only these tidal inputs, and hence the TWL. 

In order to isolate the effect of tides on the observed TWL, we apply a Doodson $X_{0}$ filter to the raw gauge data \cite{doodson}. This simple symmetric filter is designed to damp out the main tidal frequencies, and it produces a weighted average of 39 hourly values. We reconstruct the tidal signal from these ``de-surged'' TWL at each weather station, and compare these with the ADCIRC-OF tidal reconstructions, allowing us to examine how well the ADCIRC-OF model reproduces the tidal signal through the fjord.

We elaborate on results from each specific simulation period in the following subsections.

\subsubsection{September 2022}

\begin{figure}
    \centering
    \includegraphics[width=\linewidth]{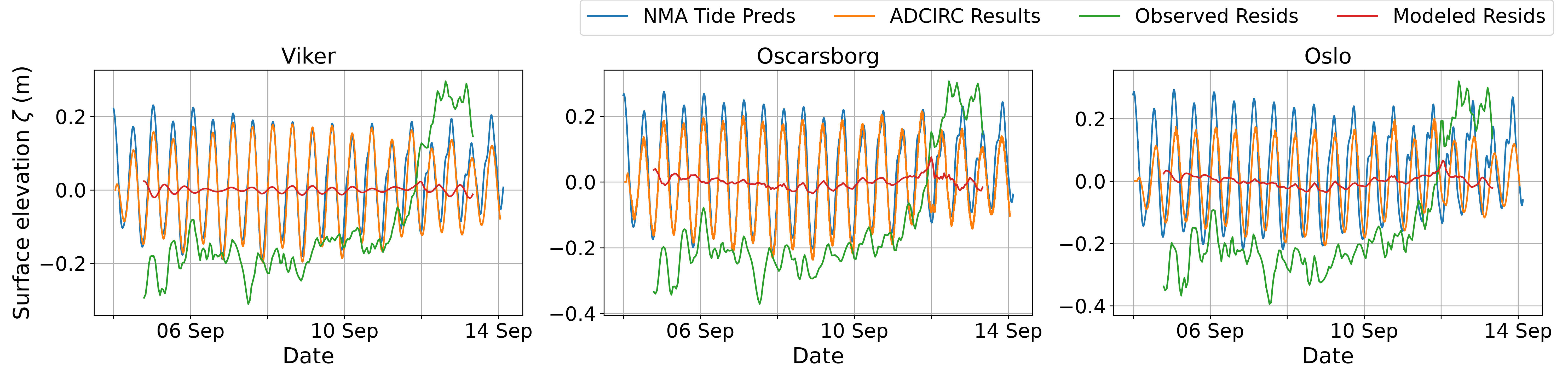}
    \caption{\textbf{Tidal ADCIRC-OF model results for September 2022}. Water surface elevation $\zeta$ given by ADCIRC-OF and NMA tidal predictions, as well as the modeled and observed residuals.}
    \label{fig:sept22}
\end{figure}

\begin{table}[htbp]
  \centering
  \small
  \caption{\textbf{Reconstructed tidal constituents for ADCIRC-OF results (with major 8 forcing) and observations for September 2022}.
  Only constituents satisfying the Rayleigh criterion (Eq. \eqref{eq:rayleigh}) over the 10-day period.
  $a$ denotes amplitude (cm), and $G$ is Greenwich phase lag ($^\circ$). Astronomical predictions reconstructed from NMA tide predictions.}
  \label{tab:constituents_sept22}
  \resizebox{\textwidth}{!}{
  \begin{tabular}{l|l|rr|rr|rr|rr}
    \toprule
    \textbf{Station} & \textbf{Const.} & \multicolumn{2}{c|}{\textbf{Observed}} & \multicolumn{2}{c|}{\textbf{Astronomical}} & \multicolumn{2}{c|}{\textbf{ADCIRC-OF}} & \multicolumn{2}{c}{\textbf{Model Error}} \\
    & & \multicolumn{2}{c|}{\textbf{(De-surged)}} & \multicolumn{2}{c|}{\textbf{Predictions}} & \multicolumn{2}{c|}{\textbf{Model}} & \multicolumn{2}{c}{\textbf{(vs Obs)}} \\
    \cmidrule(lr){3-4} \cmidrule(lr){5-6} \cmidrule(lr){7-8} \cmidrule(lr){9-10}
    & & $a$ & $G$ & $a$ & $G$ & $a$ & $G$ & $\Delta a$ & $\Delta G$ \\
    \midrule
    \textbf{Viker} & & & & & & & & & \\
    & $\text{M}_{2}$ & 14 & 101 & 13 & 93 & 15 & 98 & 1 & -3 \\
    & $\text{S}_{2}$ & 5 & 14 & 4 & 24 & 2 & 36 & -3 & 22 \\
    & $\text{K}_{1}$ & 1 & 215 & 1 & 203 & 0 & 217 & 0 & 2 \\
    & $\text{O}_{1}$ & 3 & 242 & 2 & 269 & 2 & 271 & -1 & 29 \\
    \midrule
    \textbf{Oscarsborg} & & & & & & & & & \\
    & $\text{M}_{2}$ & 17 & 120 & 16 & 109 & 16 & 103 & 0 & -17 \\
    & $\text{S}_{2}$ & 6 & 33 & 5 & 44 & 3 & 39 & -4 & 6 \\
    & $\text{K}_{1}$ & 1 & 193 & 1 & 190 & 0 & 161 & 0 & -32 \\
    & $\text{O}_{1}$ & 3 & 239 & 3 & 276 & 2 & 273 & -1 & 34 \\
    \midrule
    \textbf{Oslo} & & & & & & & & & \\
    & $\text{M}_{2}$ & 18 & 126 & 17 & 115 & 14 & 150 & -3 & 24 \\
    & $\text{S}_{2}$ & 7 & 36 & 5 & 50 & 2 & 89 & -5 & 53 \\
    & $\text{K}_{1}$ & 1 & 192 & 1 & 184 & 1 & 171 & 0 & -21 \\
    & $\text{O}_{1}$ & 3 & 238 & 3 & 276 & 2 & 293 & -1 & 55 \\
    \bottomrule
  \end{tabular}}
\end{table}

The first model run was of a 10-day period in September of 2022. We observe very close similarity between ADCIRC-OF and the NMA tidal predictions for the outer two gauges, with $\rho_{\text{tides}}$ about 95\%. At the Oslo gauge, the correlation drops to 74\%, and $\text{RMSE}_{\text{tides}}$ doubles. This difference is also shown in the time series in Figure \ref{fig:sept22}, with the tide predictions peaking higher than ADCIRC-OF at Oslo. We note that in this and subsequent Figures, the residuals presented are shorter than the predictions because of the Doodson filter applied to the gauge data.

However, the modeled residuals are quite low relative to the observed residuals. This indicates that ADCIRC-OF is primarily forced by the tidal input, and the atmospheric forcing has a very small effect on the solution. The observed residuals for this case are low for the first several days, between 10 and 30 cm below the geoid, and they jump up to over 20 cm above the geoid in the last few days. This demonstrates that weather effects, even in this relatively unremarkable period, have a significant effect on the total water levels in the fjord, and these effects are not captured by the ADCIRC-OF model with tidal forcing. The $\text{RMSE}_{\text{resid}}$ is 19 centimeters at Viker, but increases further inland, towards Oslo. This pattern makes sense, as errors propagate through the narrow channels of the fjord. 

The reconstructed tidal constituents used to compute these residuals are shown in Table \ref{tab:constituents_sept22}. As we observe in the fourth column, the amplitudes of each constituent only differ by a few centimeters between ADCIRC-OF and the observed tides. These amplitude errors increase slightly towards Oslo, peaking at $-5$ cm. The phases have errors less than $30^{\circ}$ at Viker, and increasing to a peak of $55^{\circ}$ at Oslo.

\subsubsection{January 2025}

\begin{figure}
    \centering
    \includegraphics[width=\linewidth]{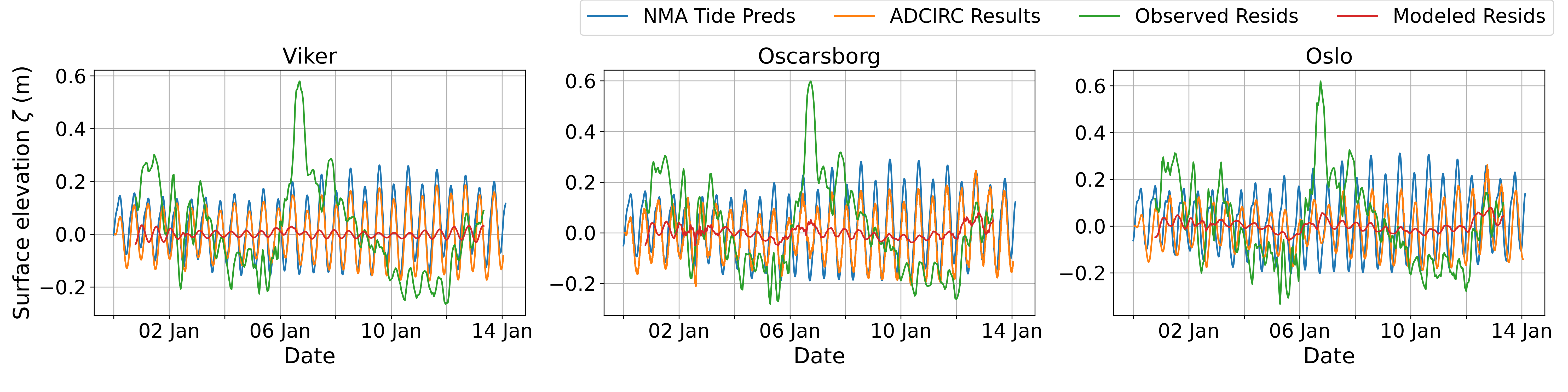}
    \caption{\textbf{Tidal ADCIRC-OF model results for January 2025}. Water surface elevation $\zeta$ given by ADCIRC-OF and NMA tidal predictions, as well as the modeled and observed residuals between TWL and tides.}
    \label{fig:jan25}
\end{figure}

\begin{table}[htbp]
  \centering
  \small
  \caption{\textbf{Reconstructed tidal constituents for ADCIRC-OF results (with major 8 forcing) and observations for January 2025}.
  Only constituents satisfying the Rayleigh criterion (Eq. \eqref{eq:rayleigh}) over the 14-day period.
  $a$ denotes amplitude (cm), and $G$ is Greenwich phase lag ($^\circ$).}
  \label{tab:constituents_jan25}
  \resizebox{\textwidth}{!}{
  \begin{tabular}{l|l|rr|rr|rr|rr}
    \toprule
    \textbf{Station} & \textbf{Const.} & \multicolumn{2}{c|}{\textbf{Observed}} & \multicolumn{2}{c|}{\textbf{Astronomical}} & \multicolumn{2}{c|}{\textbf{ADCIRC-OF}} & \multicolumn{2}{c}{\textbf{Model Error}} \\
    & & \multicolumn{2}{c|}{\textbf{(De-surged)}} & \multicolumn{2}{c|}{\textbf{Predictions}} & \multicolumn{2}{c|}{\textbf{Model}} & \multicolumn{2}{c}{\textbf{(vs Obs)}} \\
    \cmidrule(lr){3-4} \cmidrule(lr){5-6} \cmidrule(lr){7-8} \cmidrule(lr){9-10}
    & & $a$ & $G$ & $a$ & $G$ & $a$ & $G$ & $\Delta a$ & $\Delta G$ \\
    \midrule
    \textbf{Viker} & & & & & & & & & \\
    & $\text{M}_{2}$ & 14 & 84 & 14 & 85 & 13 & 92 & -1 & 8 \\
    & $\text{S}_{2}$ & 6 & 80 & 3 & 61 & 4 & 76 & -2 & -3 \\
    & $\text{K}_{1}$ & 2 & 263 & 1 & 275 & 0 & 182 & -1 & -81 \\
    & $\text{O}_{1}$ & 2 & 294 & 3 & 267 & 2 & 266 & 0 & -27 \\
    \midrule
    \textbf{Oscarsborg} & & & & & & & & & \\
    & $\text{M}_{2}$ & 16 & 98 & 17 & 100 & 14 & 101 & -2 & 3 \\
    & $\text{S}_{2}$ & 6 & 94 & 4 & 78 & 4 & 87 & -3 & -7 \\
    & $\text{K}_{1}$ & 2 & 267 & 1 & 254 & 0 & 162 & -1 & -105 \\
    & $\text{O}_{1}$ & 2 & 291 & 3 & 269 & 2 & 264 & 0 & -26 \\
    \midrule
    \textbf{Oslo} & & & & & & & & & \\
    & $\text{M}_{2}$ & 17 & 100 & 18 & 102 & 13 & 143 & -4 & 43 \\
    & $\text{S}_{2}$ & 7 & 97 & 4 & 80 & 3 & 132 & -3 & 34 \\
    & $\text{K}_{1}$ & 2 & 267 & 1 & 253 & 0 & 181 & -1 & -86 \\
    & $\text{O}_{1}$ & 2 & 291 & 3 & 270 & 2 & 284 & 0 & -7 \\
    \bottomrule
  \end{tabular}}
\end{table}

To evaluate the model's performance during the winter, which is considered the season that has the highest variability and amplitudes of residual water level, a model run covering the first two weeks of January 2025 was performed.

Referring again to Table \ref{tab:rmse_corr_tides}, we note a very similar comparison to the NMA tidal predictions for this period. $\text{RMSE}_{\text{tides}} = 5$ cm at Viker, increases slightly at Oscarsborg, and then again jumps to 10 cm at Oslo. Again, $\rho_{\text{tides}}$ is high for the outer gauges, and significantly lower for the Oslo gauge, with a $24\%$ drop in correlation. This discrepancy in the inner fjord highlights the difficulty of modeling tides through the narrow Dr\o{}bak Sound.

In this and the September 2022 case, the residuals' correlation is very low at Viker, and increases slightly at Oscarsborg and Oslo. The water level in the Oslofjord is largely dependent on the weather situation, and without the weather effect at the open boundary, the tidal ADCIRC-OF model's residuals have very low correlation with the observed surge.

Here we observe that the $\text{RMSE}_{\text{resid}}$ values are approximately equal throughout the fjord at all three stations. The observed residual is again very large compared to the tidal predictions, and it peaks at about 60 cm, as shown in Figure \ref{fig:jan25}. The modeled residual is very small, with a few spikes not exceeding 10 cm, demonstrating that ADCIRC-OF lacks a crucial component of storm surge for this time period as well. 

The reconstructed tidal constituents used to compute these residuals are shown in Table \ref{tab:constituents_jan25}. It is clearly shown that the modeled tides at the Viker and Oscarsborg stations have an almost perfect fit for the phase and amplitudes for $M_2$ and $S_2$, while the $K_1$ and $O_1$, whose amplitudes are only a few centimeters, show almost perfect fit for amplitude, but large offset for the phase. However, since these are small, this is not critical, and they are also more sensitive to small errors in the TWL, that can translate into a large phase error. For Oslo we find that the error in amplitudes are, although larger than the stations further south, also very small. However, here we see large phase errors, which suggests that the increased RMSE for the Oslo station is not because of errors in the amplitude, but due to phase error. This could most likely be improved by adjusting the bottom depth of the inner parts of the fjord. 

\subsubsection{July 2025}

\begin{figure}
    \centering
    \includegraphics[width=\linewidth]{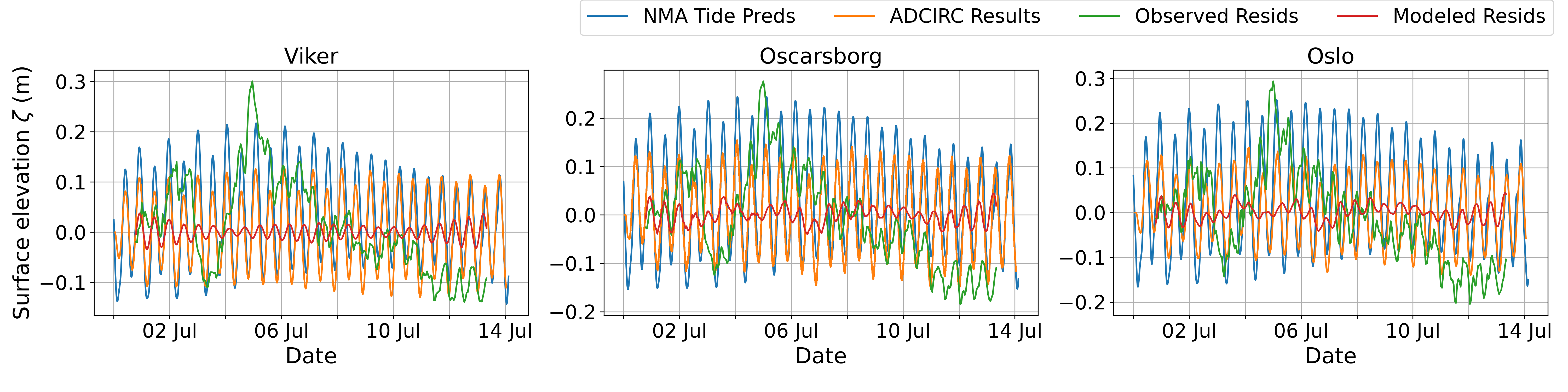}
    \caption{\textbf{Tidal ADCIRC-OF model results for July 2025}. Water surface elevation $\zeta$ given by ADCIRC-OF and NMA tidal predictions, as well as the modeled and observed residuals between TWL and tides.}
    \label{fig:jul25}
\end{figure}

\begin{table}[htbp]
  \centering
  \small
  \caption{\textbf{Reconstructed tidal constituents for ADCIRC-OF results (with major 8 forcing) and observations for July 2025}.
  Only constituents satisfying the Rayleigh criterion (Eq. \eqref{eq:rayleigh}) over the 14-day period.
  $a$ denotes amplitude (cm), and $G$ is Greenwich phase lag ($^\circ$).}
  \label{tab:constituents_jul25}
  \resizebox{\textwidth}{!}{
  \begin{tabular}{l|l|rr|rr|rr|rr}
    \toprule
    \textbf{Station} & \textbf{Const.} & \multicolumn{2}{c|}{\textbf{Observed}} & \multicolumn{2}{c|}{\textbf{Astronomical}} & \multicolumn{2}{c|}{\textbf{ADCIRC-OF}} & \multicolumn{2}{c}{\textbf{Model Error}} \\
    & & \multicolumn{2}{c|}{\textbf{(De-surged)}} & \multicolumn{2}{c|}{\textbf{Predictions}} & \multicolumn{2}{c|}{\textbf{Model}} & \multicolumn{2}{c}{\textbf{(vs Obs)}} \\
    \cmidrule(lr){3-4} \cmidrule(lr){5-6} \cmidrule(lr){7-8} \cmidrule(lr){9-10}
    & & $a$ & $G$ & $a$ & $G$ & $a$ & $G$ & $\Delta a$ & $\Delta G$ \\
    \midrule
    \textbf{Viker} & & & & & & & & & \\
    & $\text{M}_{2}$ & 13 & 123 & 11 & 123 & 9 & 100 & -3 & -23 \\
    & $\text{S}_{2}$ & 2 & 34 & 2 & 44 & 3 & 59 & 1 & 24 \\
    & $\text{K}_{1}$ & 1 & 146 & 1 & 130 & 0 & 179 & -1 & 33 \\
    & $\text{O}_{1}$ & 2 & 282 & 2 & 277 & 2 & 267 & 0 & -15 \\
    \midrule
    \textbf{Oscarsborg} & & & & & & & & & \\
    & $\text{M}_{2}$ & 16 & 136 & 14 & 139 & 10 & 108 & -6 & -28 \\
    & $\text{S}_{2}$ & 2 & 50 & 3 & 65 & 3 & 68 & 1 & 18 \\
    & $\text{K}_{1}$ & 2 & 138 & 1 & 128 & 0 & 88 & -1 & -50 \\
    & $\text{O}_{1}$ & 2 & 283 & 2 & 280 & 2 & 272 & 0 & -11 \\
    \midrule
    \textbf{Oslo} & & & & & & & & & \\
    & $\text{M}_{2}$ & 17 & 140 & 14 & 147 & 9 & 151 & -7 & 11 \\
    & $\text{S}_{2}$ & 3 & 56 & 3 & 73 & 3 & 111 & 1 & 54 \\
    & $\text{K}_{1}$ & 2 & 136 & 2 & 126 & 1 & 112 & -2 & -24 \\
    & $\text{O}_{1}$ & 2 & 283 & 2 & 281 & 2 & 292 & -1 & 9 \\
    \bottomrule
  \end{tabular}}
\end{table}

The most recent dates modeled were the first two weeks of July 2025, and the water surface elevations and residuals are shown in Figure \ref{fig:jul25}. This period contained the lowest observed residuals of all the simulations, and a $\text{RMSE}_{\text{resid}}$ of about 10 centimeters, again slightly increasing towards Oslo. These comparatively low observed residuals are still much higher than the modeled residuals, which peak at about 5 cm.

For this period, $\text{RMSE}_{\text{tides}}$ and $\rho_{\text{tides}}$ do not worsen much between the Inner and Outer fjords, as ADCIRC-OF reproduces the observed tides well throughout the mesh. The improvement seen here may be attributed to the smaller observed residuals for this period. The residual correlations $\rho_{\text{resid}}$ are also all very low during this period.

The reconstructed tidal constituents used to compute these residuals are shown in Table \ref{tab:constituents_jul25}. The errors increase further into the fjord, and $\text{M}_{2}$ peaks at a $-7$ cm amplitude error at Oslo. The phase errors also increase further north, up to a maximum of $54^{\circ}$.

\subsubsection{An extreme weather event: Storm Hans}
In August 2023, Storm Hans caused major flooding in southern Norway. While western and northern Norway regularly receive large amounts of rainfall from storms in the Norwegian Sea, the east is generally more sheltered from westerly winds and precipitation. Hans was a peculiar storm in that it arose from the collision of two low pressure systems over eastern Europe, before hitting Norway from the southeast. The storm's unusual approach meant that many parts of the country experienced 100-year precipitation levels \cite{hansmet}. MET Norway issued its highest risk warning for rainfall, and several weather stations received over 100 mm of rain over three days, with the highest precipitation values over 150 mm.
The storm hit the region north of the Oslofjord with very strong winds and rainfall, and its devastation required thousands of people to evacuate. Even after the extreme rainfall, the flood warning persisted because of high water levels in many rivers, which caused additional flooding, landslides, and destruction.

Unlike the other time periods modeled, Storm Hans was simulated with a smaller timestep of 0.1 seconds. This is due to the more intense winds experienced during this storm, which caused some numerical oscillations in a few parts of the mesh. Additionally, the storm only lasted for a few days, so we could decrease the simulation time along with the timestep, completing the three-day simulation in just over an hour.

\begin{figure}
    \centering
    \includegraphics[width=\linewidth]{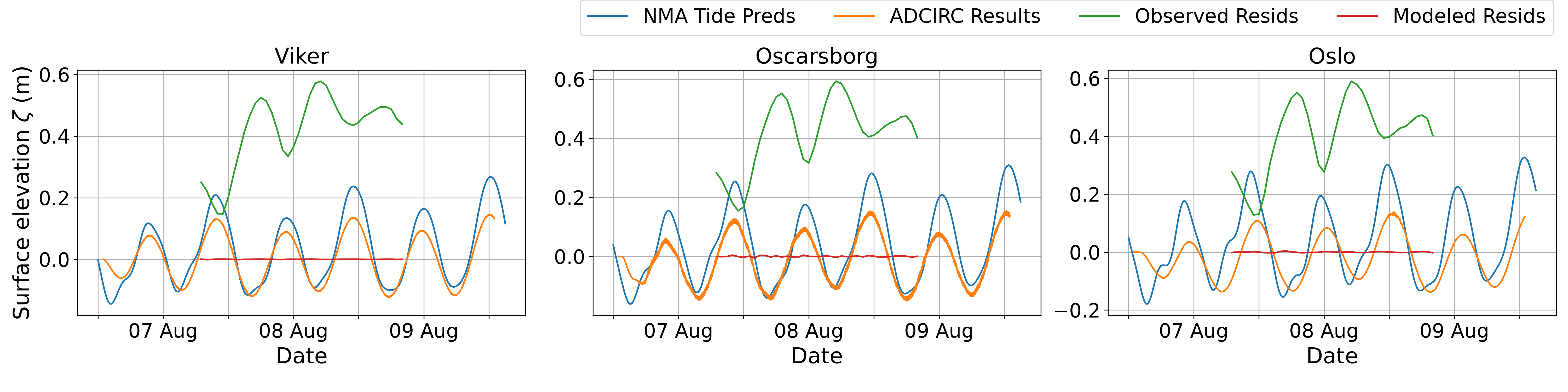}
    \caption{\textbf{Tidal ADCIRC-OF model results for Storm Hans}. Water surface elevation $\zeta$ given by ADCIRC-OF and NMA tidal predictions, as well as the modeled and observed residuals between TWL and tides.}
    \label{fig:hans_tides}
\end{figure}

\begin{table}[htbp]
  \centering
  \small
  \caption{\textbf{Reconstructed tidal constituents for ADCIRC-OF results (with major 8 forcing) and observations for Storm Hans}.
  Only constituents satisfying the Rayleigh criterion (Eq. \eqref{eq:rayleigh}) over the 3-day period.
  $a$ denotes amplitude (cm), and $G$ is Greenwich phase lag ($^\circ$).}
  \label{tab:constituents_hans}
  \resizebox{\textwidth}{!}{
  \begin{tabular}{l|l|rr|rr|rr|rr}
    \toprule
    \textbf{Station} & \textbf{Const.} & \multicolumn{2}{c|}{\textbf{Observed}} & \multicolumn{2}{c|}{\textbf{Astronomical}} & \multicolumn{2}{c|}{\textbf{ADCIRC-OF}} & \multicolumn{2}{c}{\textbf{Model Error}} \\
    & & \multicolumn{2}{c|}{\textbf{(De-surged)}} & \multicolumn{2}{c|}{\textbf{Predictions}} & \multicolumn{2}{c|}{\textbf{Model}} & \multicolumn{2}{c}{\textbf{(vs Obs)}} \\
    \cmidrule(lr){3-4} \cmidrule(lr){5-6} \cmidrule(lr){7-8} \cmidrule(lr){9-10}
    & & $a$ & $G$ & $a$ & $G$ & $a$ & $G$ & $\Delta a$ & $\Delta G$ \\
    \midrule
    \textbf{Viker} & & & & & & & & & \\
    & $\text{M}_{2}$ & 7 & 83 & 15 & 114 & 11 & 109 & 4 & 26 \\
    & $\text{K}_{1}$ & 4 & 165 & 4 & 163 & 2 & 160 & -1 & -5 \\
    \midrule
    \textbf{Oscarsborg} & & & & & & & & & \\
    & $\text{M}_{2}$ & 9 & 98 & 17 & 130 & 12 & 119 & 3 & 21 \\
    & $\text{K}_{1}$ & 4 & 162 & 5 & 164 & 3 & 148 & -2 & -14 \\
    \midrule
    \textbf{Oslo} & & & & & & & & & \\
    & $\text{M}_{2}$ & 9 & 102 & 18 & 136 & 11 & 161 & 2 & 59 \\
    & $\text{K}_{1}$ & 4 & 167 & 5 & 163 & 2 & 163 & -2 & -4 \\
    \bottomrule
  \end{tabular}}
\end{table}

ADCIRC-OF's results are compared to the predicted tides and residuals in Figure \ref{fig:hans_tides}. In this case, we see that while the ADCIRC-OF results resemble the predicted tides in part, the observed residuals are very large, peaking at about 60 cm. Meanwhile, the modeled residuals are very small, indicating that the Oslofjord mesh allows for accurate tidal modeling even in the presence of heavy atmospheric forcings. At the same time, the large discrepancy between the residuals confirms that weather effects are lacking from the tidal boundary condition.

The results in Table \ref{tab:rmse_corr_tides} show that the $\text{RMSE}_{\text{resid}}$ is much higher than that of the previous periods, at 42--43 cm for each gauge. However, we see similar patterns between ADCIRC-OF and NMA tidal predictions: small $\text{RMSE}_{\text{tides}}$ and high $\rho_{\text{tides}}$ at the mouth of the fjord, which both worsen slightly as we look further north. Again, the $\text{RMSE}_{\text{tides}}$ doubles between Viker and Oslo. 

Because of the much shorter time span of this period, we can only reconstruct two tidal constituents, which are shown in Table \ref{tab:constituents_hans}. We again observe that the maximum amplitude error is about 4 centimeters for this case, but unusually, this peak is found at the Viker gauge. Because of the strong winds accompanying this storm, some deviations from the first three cases can be expected, but the low error in these two reconstructed constituents shows the skill of reproducing the tidal signal despite the additional extreme weather effects.

\subsection{Incorporating Additional Tidal Constituents} \label{sec:tides}

The previous results showcase ADCIRC-OF's ability to accurately reproduce tidal signals in the fjord using the major 8 tidal constituents, with a small computational footprint. In order to further analyze ADCIRC-OF's tidal modeling capabilities, we examine the use of different tidal constituent sets to force the model.
Incorporating these extra parameters into a large ADCIRC model has not been attempted much in the literature, because adding smaller and smaller tidal terms has diminishing returns, and the major 8 constituents are usually sufficient \cite{militello}. Thus, this relatively small scale mesh provides a viable testing ground for these methods, and we may examine the effect and cost of the additional data used.
Varying tidal constituent sets have been used with ADCIRC in tidal studies, in order to evaluate and improve tidal databases. One such study uses the same set of 15 constituents to help evaluate the EC2015 ADCIRC database, which provides all 37 standard tidal constituents, but its performance is only evaluated using the major 8 constituents \cite{szpilka}. 

The two 14-day time periods from January and July 2025 were simulated using three different selections of forcing tidal constituents. The typical major 8 constituents, as documented in section \ref{results_main}, were used in the first model. The second model used the nine constituents used by the former MET FjordOs model (Section \ref{sec:fjordos}). The third model uses all fifteen available constituents from TPXO-9. These parameters are outlined in Table \ref{tab:constituents}.

\subsubsection{Computational Resources}

These larger sets of tidal constituents have not been used much at a large scale, in part because the extra fidelity gained comes with an increased computational cost. We present the computation times for each tidal constituent set in Table \ref{tab:timings}.

\begin{table}[]
    \centering
    \small
    \begin{tabular}{c|cc}
        \toprule
        \textbf{Tidal} & \multicolumn{2}{|l}{\textbf{CPU hours per day simulated}} \\
        \textbf{Constituent Set} & $\qquad$ January 2025 & July 2025 \\
        \midrule
        Major 8 & $\qquad$ 3.26 & 3.15 \\
        FjordOs & $\qquad$ 3.19 & 3.07 \\
        All & $\qquad$ 3.32 & 3.25 \\
        \bottomrule
    \end{tabular}
    \caption{\textbf{CPU hours per day simulated, for each forcing set of tidal constituents}.}
    \label{tab:timings}
\end{table}

In both cases, the FjordOs selection of nine constituents resulted in the fastest model, albeit by a small amount. Compared with only using the major 8 constituents, we observe speedups of 2.1 and 2.6\% in January and July, respectively. Using all seven additional parameters resulted in a slowdown of 1.7\% and 3.3\% compared to the major 8 runs. This is an expected result, but it shows that, at least for a relatively small mesh, the additional constituents do not cause a very large increase in required computational resources.

We note that the timing differences depend largely on the length and resolution of the tidal boundary nodestring, which is generally much coarser than the nearshore parts of ADCIRC meshes. Hence, for example, if the Oslofjord mesh is refined further, the differences in computation times are expected to decrease.

\subsubsection{Evaluation of Additional Tidal Constituents}


\begin{table}
    \centering
    \scriptsize
    \begin{tabular}{c|c|c|c|c|c}
        \toprule
        \textbf{Gauge} & \textbf{Tidal} & \multicolumn{2}{l|}{\textbf{Jan 2025}} & \multicolumn{2}{l}{\textbf{Jul 2025}} \\
        & \textbf{Constituent Set} & $\text{RMSE}_{\text{tides}}$ & $\rho_{\text{tides}}$ & $\text{RMSE}_{\text{tides}}$ & $\rho_{\text{tides}}$ \\
        \midrule
        & Major 8 & 4.6 & 0.941 & 4.9 & 0.891 \\
        Viker & FjordOs & 4.4 & 0.950 & 4.8 & 0.891 \\
        & All & 4.6 & 0.959 & 4.3 & 0.911 \\
        \midrule
        & Major 8 & 5.6 & 0.933 & 6.4 & 0.856 \\
        Oscarsborg & FjordOs & 5.4 & 0.943 & 6.3 & 0.856 \\
        & All & 5.8 & 0.943  & 5.8 & 0.877 \\
        \midrule
        & Major 8 & 10.0 & 0.689 & 6.4 & 0.891 \\
        Oslo & FjordOs & 9.8 & 0.700 & 6.3 & 0.893 \\
        & All & 10.2 & 0.686 & 5.9 & 0.902 \\
        \bottomrule
    \end{tabular}
    \caption{\textbf{Error analysis of ADCIRC-OF models with different tidal constituent sets used for forcing}. RMSE is shown in centimeters.}
    \label{tab:rmse_tides}
\end{table}

The water level for ADCIRC-OF with each set of tidal constituents is shown, together with the observed gauge data, in \ref{appendix:tidal}. We again calculate the RMSE and correlation between ADCIRC-OF's computed TWL and the NMA tidal predictions. These values are shown in Table \ref{tab:rmse_tides} for each of the tidal constituent sets. The three models perform very similarly, requiring an extra decimal place to examine the small changes caused by the different forcing. While the additional tidal constituents resulted in no decrease in $\text{RMSE}_{\text{tide}}$ in the January case, from the July simulation we do observe an 8--14\% decrease in $\text{RMSE}_{\text{tides}}$ with the full fifteen-parameter set. On average, and particularly for the July case, the fifteen-constituent set produces a slightly higher $\rho_{\text{tides}}$ value, confirming that these additional constituents slightly reduce the error of the simulated tidal signal.

\begin{figure}
    \centering
    \begin{subfigure}{0.32\textwidth}
        \includegraphics[width=\textwidth]{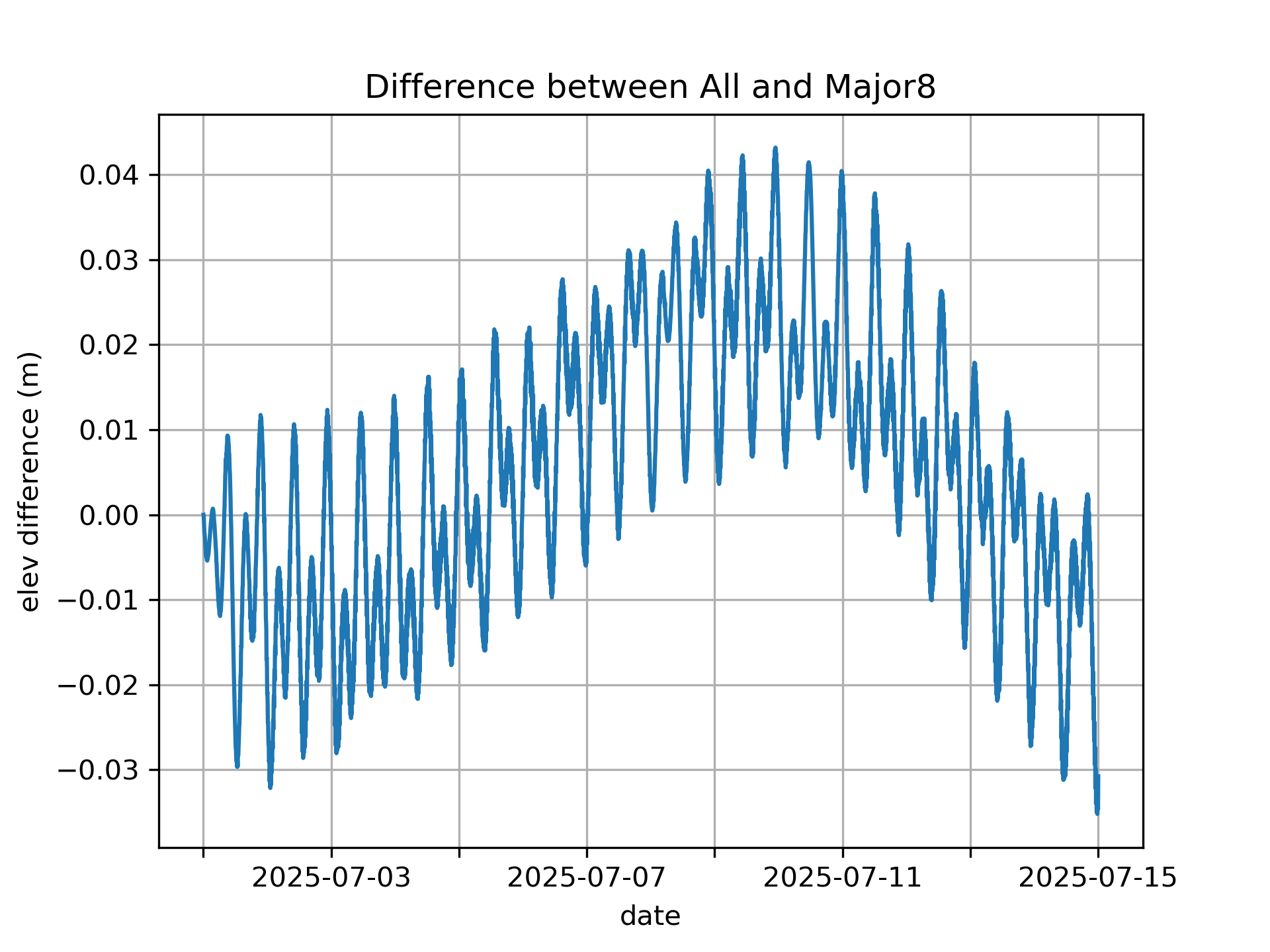}
        \caption{Viker}
        \label{fig:jul25_viker_all_maj8}
    \end{subfigure}
    \begin{subfigure}{0.32\textwidth}
        \includegraphics[width=\textwidth]{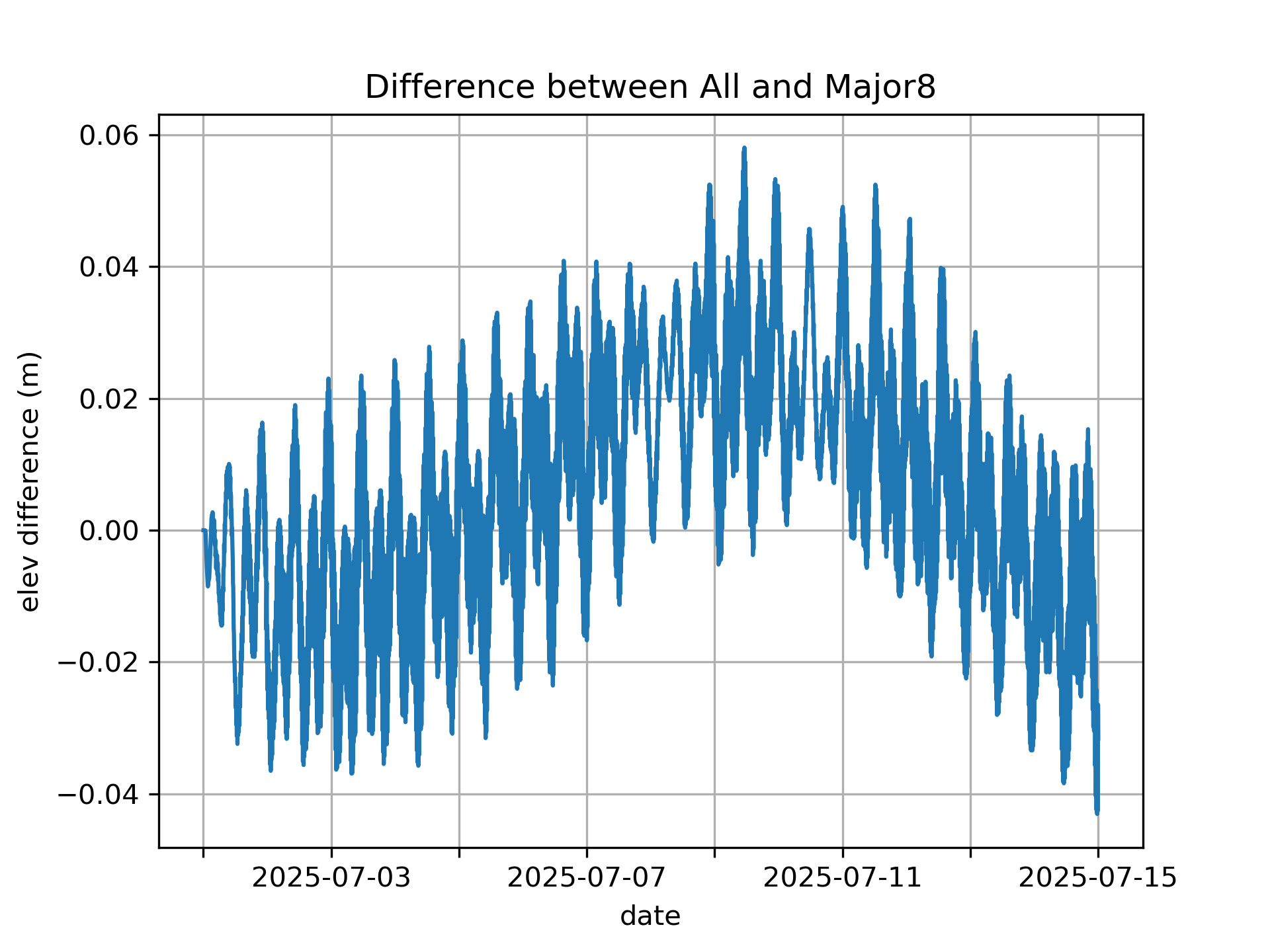}
        \caption{Oscarsborg}
        \label{fig:jul25_oscarsborg__all_maj8}
    \end{subfigure}
    \begin{subfigure}{0.32\textwidth}
        \includegraphics[width=\textwidth]{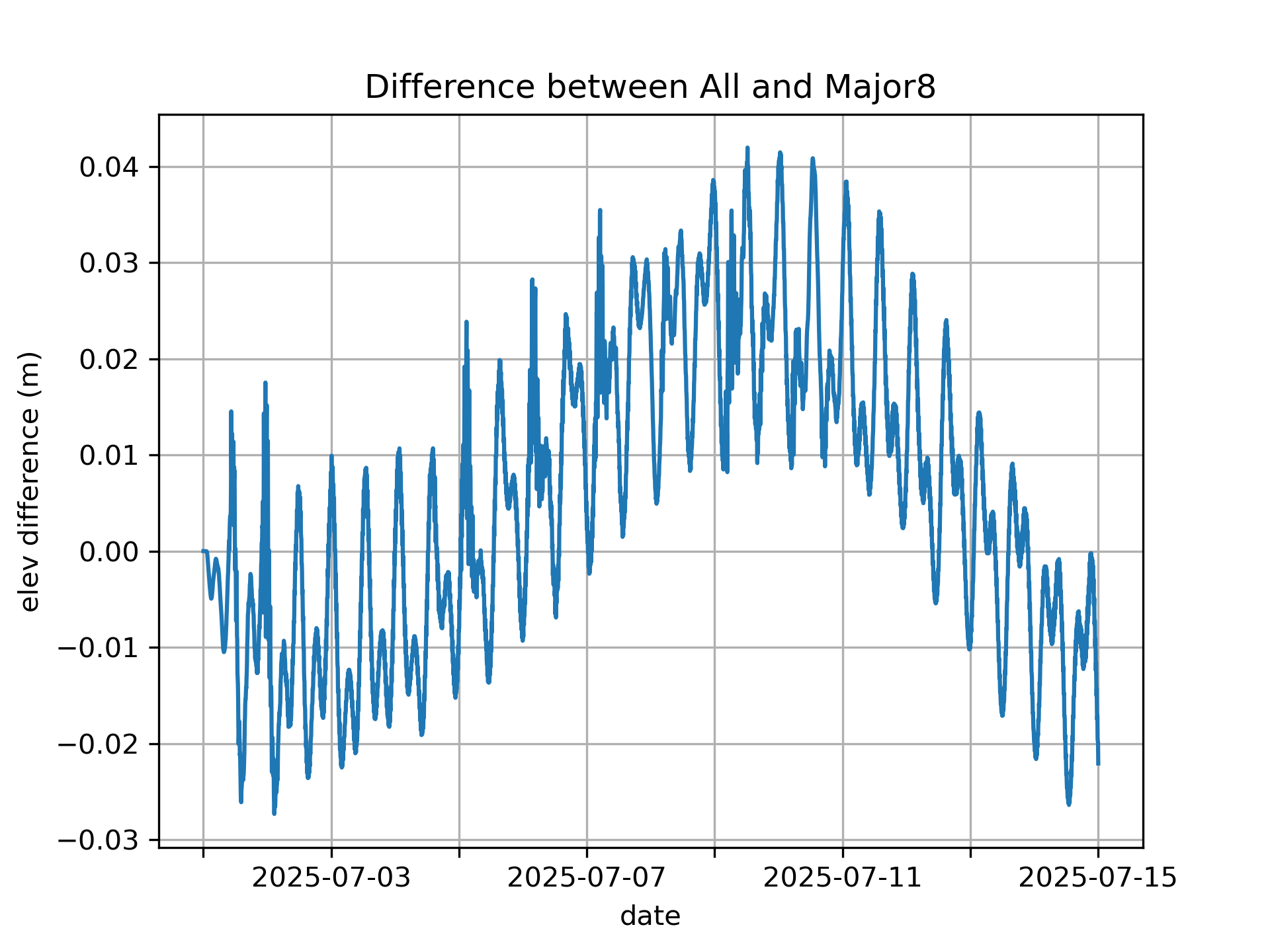}
        \caption{Oslo}
        \label{fig:jul25_oslo_all_maj8}
    \end{subfigure}
    \caption{\textbf{Example of change in water surface elevation due to additional tidal constituents}. Shown here is the difference in ADCIRC-OF's simulated TWL for July 2025, between the fifteen-constituent set and the major 8 constituents.}
    \label{fig:jul25_all_maj8}
\end{figure}

Nevertheless, the three ADCIRC-OF model runs produce very similar results. To illustrate this, we present an example of the difference between the fifteen-constituent and major-8 ADCIRC-OF models in Figure \ref{fig:jul25_all_maj8}. The maximum difference due to the five additional tidal constituents is about six centimeters. Further difference plots that compare the results of the varying sets tidal constituents are shown in \ref{appendix:diffs}.

In order to more thoroughly examine these additional constituents, much longer time spans would enable reconstructions of more of the constituents. In order to distinguish the major 8, for example, a simulated time span of 100 days is required, using the same Rayleigh number of 0.5. Distinguishing each of the fifteen constituents requires a 171-day time span. 

\section{Full Realistic Forcing} \label{sec:full_forcing}

Finally, we explore a realistic forcing scenario, and compare the results with other models and available gauge data. Section \ref{sec:results} showed that while ADCIRC-OF reproduces tidal signals very effectively, its residuals diverge strongly from the observed residuals. This demonstrates how the atmospheric effects 
that generate the local residuals inside the model domain only account for a small part of the actual observed residuals. We suspect that weather effects outside the fjord, in the Skagerrak and North Sea area, largely influence the water level in the Oslofjord \citep{KRISTENSEN2024102406}. These effects are not present in the small ADCIRC-OF mesh domain. In order to model the storm surge, and hence the TWL, more precisely, we introduce a different approach to the model, replacing the tidal input with realistic forcing elevation data along the entire open boundary. This sea surface elevation data comes from the aforementioned STOFS-2D-Global model \cite{stofsdata}. Using the relatively coarse STOFS-2D-Global model will allow us to better capture weather effects from outside the fjord, and use them to force the high-resolution Oslofjord mesh.

To apply the water surface elevation boundary condition from STOFS, ADCIRC-OF's boundary nodes were matched with a corresponding node in the STOFS-2D-Global mesh. The nearest nodes were selected by minimizing the geodesic distance to the ADCIRC-OF boundary node. Then, STOFS's simulated water surface elevation for our model periods was extracted along the boundary and turned into an input for ADCIRC-OF.

\begin{figure}
    \centering
    \includegraphics[width=0.7\linewidth]{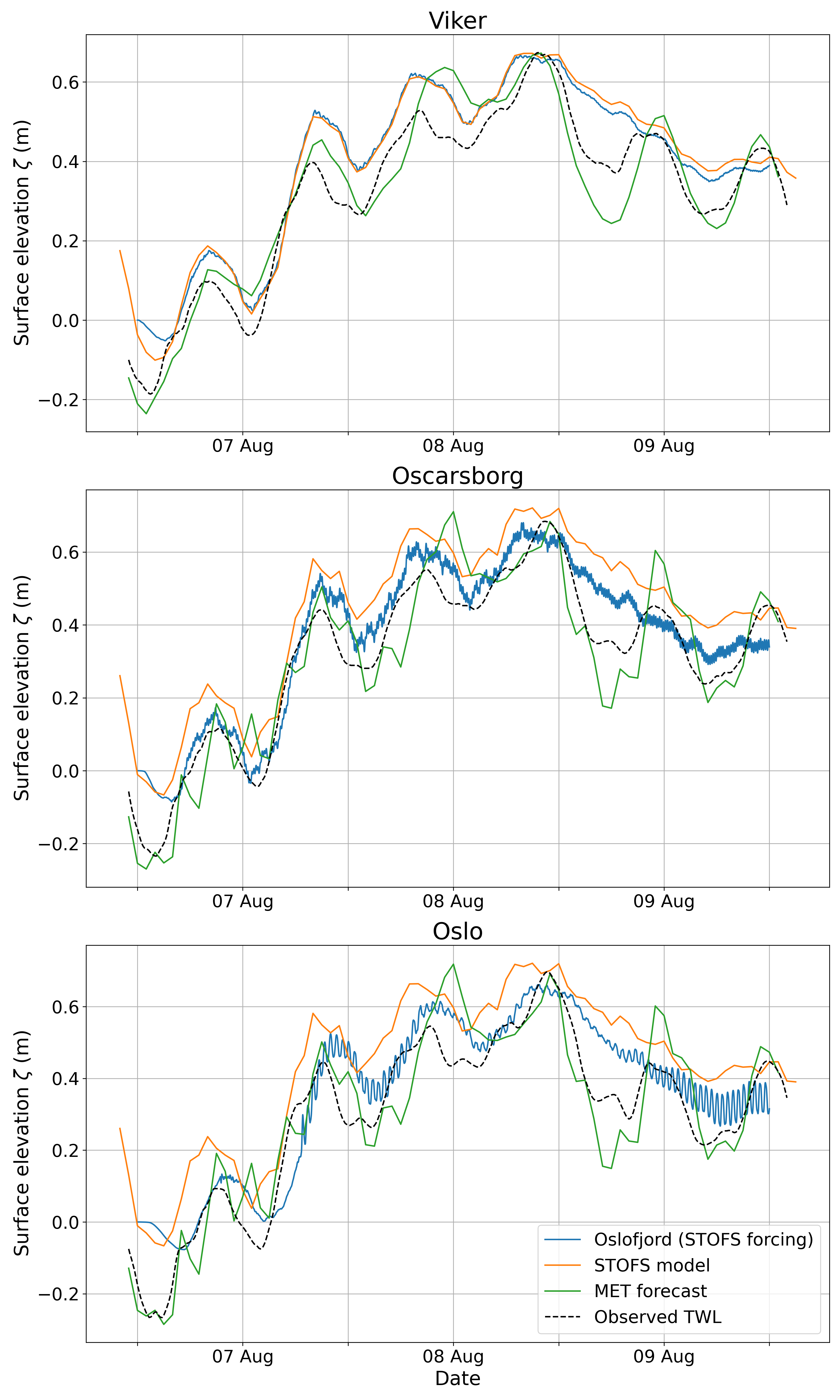}
    \caption{\textbf{Comparison of ADCIRC-OF with three existing models}. Shown is the water surface elevation $\zeta$ during Storm Hans given by ADCIRC-OF, STOFS-2D-Global, and the MET Norway forecast given on 2023-08-06, as well as the observed TWL at each gauge. Error analysis from this simulation are shown in Table \ref{tab:rmse_stofs}.}
    \label{fig:stofs_hans}
\end{figure}

In particular, the tidal ADCIRC-OF model of Storm Hans strongly diverged from the observed gauge data. We employ this STOFS forcing with Storm Hans to analyze the high-resolution model with such a severe weather event. The resulting water surface elevations portray a remarkably more accurate model for the storm. The results are presented in Figure \ref{fig:stofs_hans}, along with outputs from the MET Norway operational forecast model (Section \ref{sec:forecasting}) and STOFS's own outputs \cite{stofsdata}. With ADCIRC-OF, we attempt to refine the modeling capability of the STOFS model in the Oslofjord, and we validate against the existing operational ROMS model from MET Norway. Note that, because the STOFS mesh does not penetrate as far north as Oscarsborg or Oslo, the nearest nodes are taken for comparison in Figure \ref{fig:stofs_hans}.

While the Viker plot is smooth, we observe some high frequency oscillations further upstream. These oscillations are likely a combination of numerical artifacts and basin-scale oscillations, particularly through the narrow channels into the inner fjord. These types of oscillations are often seen in both models and observations, and are correlated with high energy input, as is present in this case with the high winds from Hans.

\begin{figure}
    \centering
    \begin{subfigure}{0.48\textwidth}
        \includegraphics[width=\linewidth]{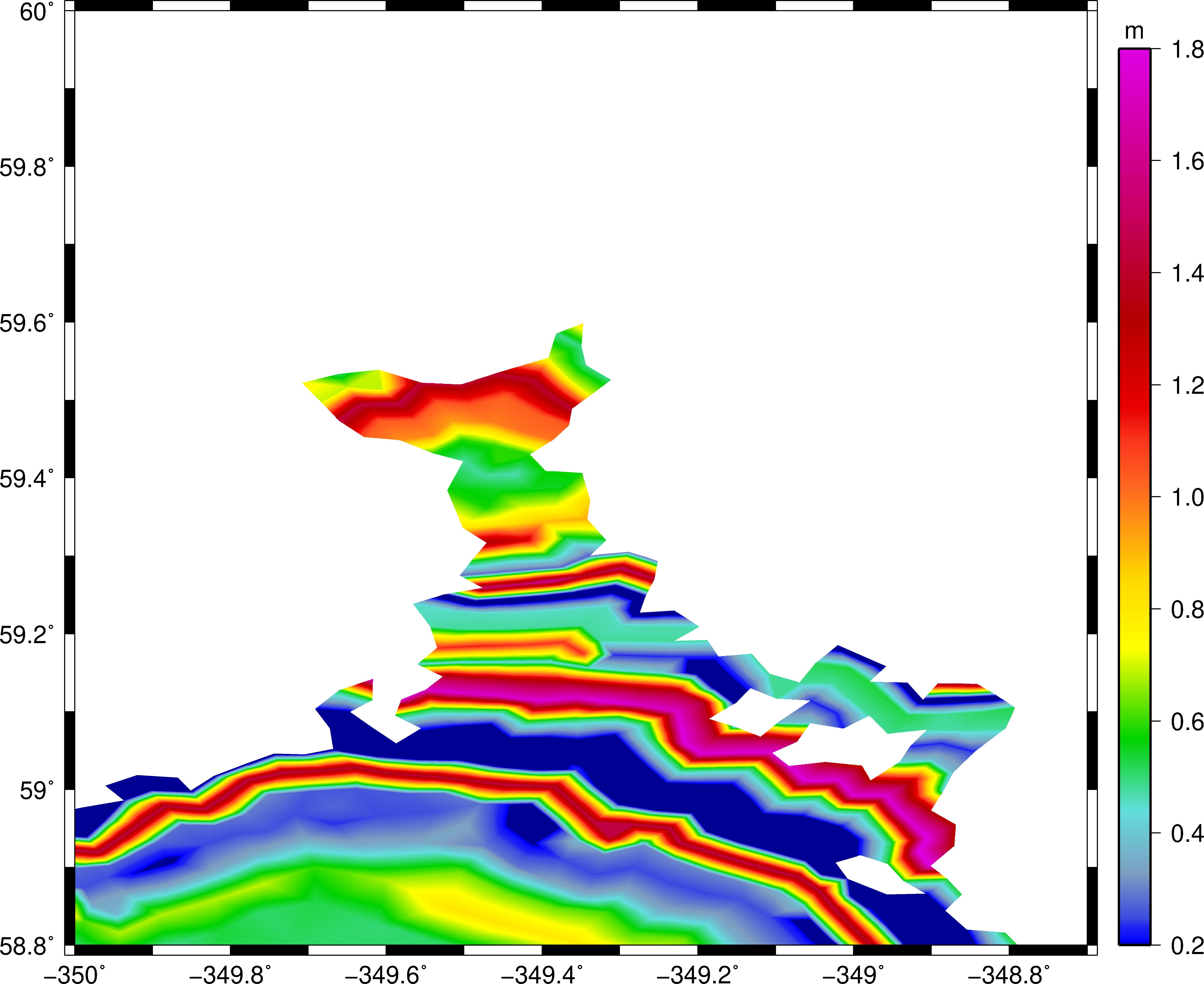}
        \subcaption{STOFS-2D-Global}
    \end{subfigure}
    \begin{subfigure}{0.48\textwidth}
        \includegraphics[width=\linewidth]{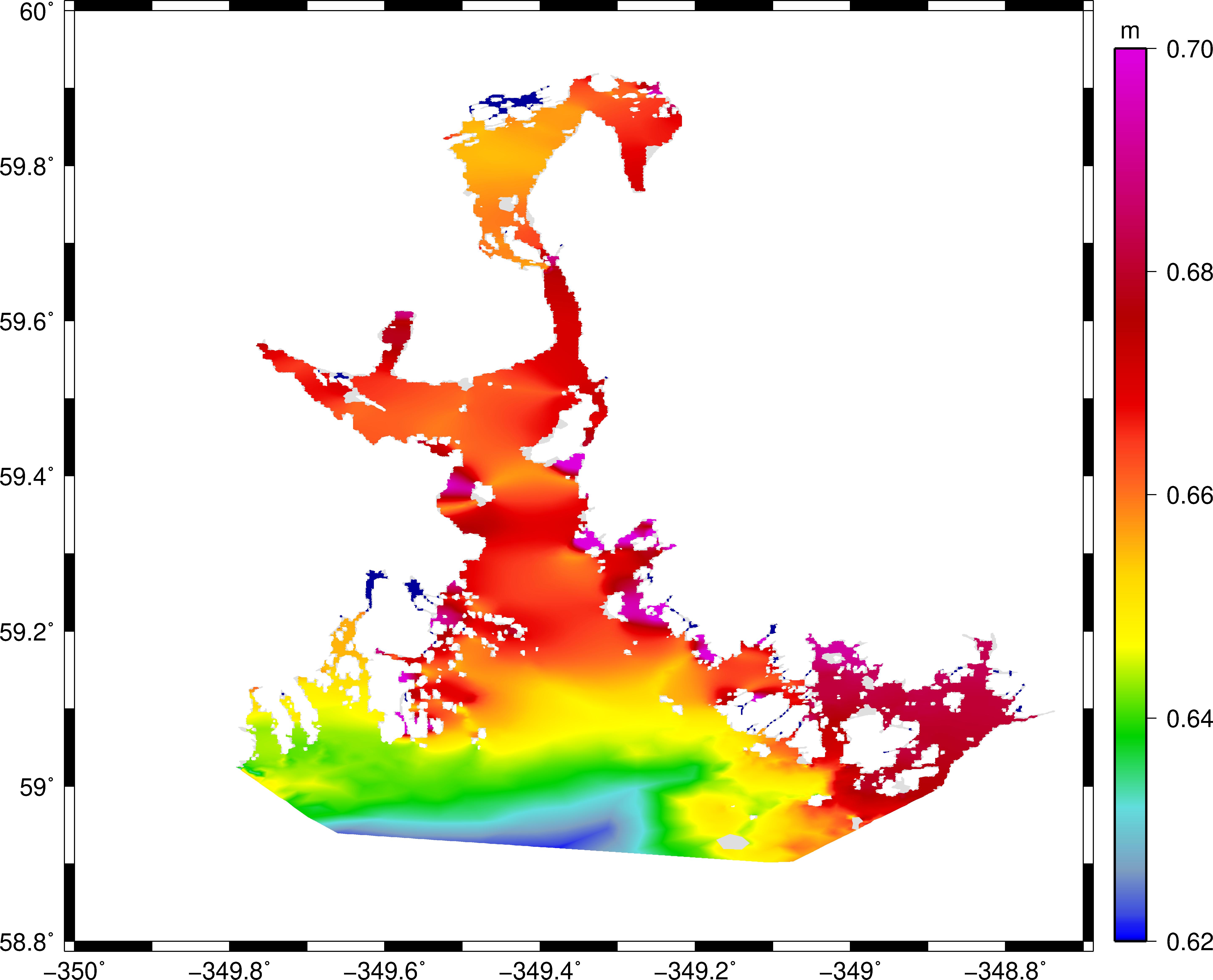}
        \subcaption{ADCIRC-OF}
    \end{subfigure}
    \caption{\textbf{Maximum elevations achieved by each model during Storm Hans}. Note the significant difference in scale.}
    \label{fig:max_elevations}
\end{figure}

In order to further illustrate the difference in resolution between the ADCIRC-OF and STOFS-2D-Global models, the peak simulated TWL is shown for every node on each model's native mesh in Figure \ref{fig:max_elevations}. Almost all of the max elevations from the high-resolution model fall between 0.62 and 0.7 meters, whereas the STOFS model has a wide range, from about 0.2 to 1.8 meters. We observe a ripple-like pattern going into the fjord, which is likely caused by some numerical oscillations with this coarse resolution part of the mesh.

We also observe some areas of the high-resolution mesh with lower peak values than expected, e.g., near Sandvika and T\o{}nsberg. These areas are only accessible via very narrow channels, and it appears that they do not receive the same influx of surge as the surrounding areas, warranting further investigation in the future.

\begin{table}
    \centering
    \scriptsize
    \begin{tabular}{c|c|c|c|c|c|c}
        \toprule
        \textbf{Gauge} & \multicolumn{2}{l|}{\textbf{Oslofjord with STOFS Forcing}} & \multicolumn{2}{l|}{\textbf{STOFS Model}} & \multicolumn{2}{l}{\textbf{Residuals}} \\
        & $\text{RMSE}_{\text{TWL}}$ & $\rho_{\text{TWL}}$ & $\text{RMSE}_{\text{TWL}}$ & $\rho_{\text{TWL}}$ & $\text{RMSE}_{\text{resid}}$ & $\rho_{\text{resid}}$ \\
        \midrule
        Viker & 9 & 0.96 & 9 & 0.97 & 11 & 0.78 \\
        Oscarsborg & 8 & 0.95 & 13 & 0.97 & 10 & 0.73 \\
        Oslo & 11 & 0.92 & 15 & 0.96 & 11 & 0.67 \\
        \bottomrule
    \end{tabular}
    \caption{\textbf{RMSE (cm) and correlation of the two models of Storm Hans}. Time series from this simulation are shown in Figure \ref{fig:stofs_hans}.}
    \label{tab:rmse_stofs}
\end{table}

The RMSE and correlation of the STOFS-forced ADCIRC-OF model are shown compared to the observed TWL in Table \ref{tab:rmse_stofs}, along with the same data for the STOFS model itself.  The RMSE of the two models are similar at the Viker gauge, at just over 9 cm. This makes sense intuitively, since ADCIRC-OF uses STOFS elevations at the boundary near Viker. Further into the fjord, however, we see the benefits of the high resolution model, as its RMSE is significantly smaller at the Oscarsborg and Oslo gauges.

Each model's correlations are very high, but the high-resolution model's correlation at the Oslo gauge is the lowest. This is attributed to the aforementioned high-frequency oscillations, which could be mitigated with smaller timestepping or a post-processing low-pass filter.

Finally, we calculate the RMSE and correlation between the observed and modeled residuals. Here, because of the large non-tidal signal present in the ADCIRC-OF results, we apply the Doodson $X_{0}$ to those as well as the gauge data, then reconstruct the two tidal constituents distinguishable per our Rayleigh criterion, $\text{M}_{2}$ and $\text{K}_{1}$. We find $\text{RMSE}_{\text{resid}}$ values just over 10 centimeters at each gauge, whereas they were over 40 cm in the tidal ADCIRC-OF model. The residual correlations $\rho_{\text{resid}}$ are fairly high, at 67--78\%, decreasing slightly further into the fjord. The very short timespan of this simulation, further shortened by the Doodson filter, prevents adequate reconstruction of tidal constituents, and we speculate that these correlations will improve with a longer simulation.

The low $\text{RMSE}_{\text{TWL}}$ and high $\rho_{\text{TWL}}$ show that the ADCIRC-OF model performs well at simulating the TWL. The residual analysis, while suboptimal given the short timespan of the simulation, also shows strong modeling capability for non-tidal weather effects. This test case helps to solidify the Oslofjord mesh's validity for a variety of weather events, and warrants further investigation into this type of forcing input for future models.

\section{Conclusions} \label{sec:conclusions}

In this work, we present a new finite element based numerical model of the Oslofjord, and analyze its performance in a variety of different scenarios. It is shown that the model is able to propagate tides in the fjord with high precision. Moreover, when forced with realistic water levels that include weather effects at the open boundary, it can reproduce the observed TWL with finer detail than the existing global mesh. The model requires minimal computational resources, so that it could be run on a laptop if needed. The new mesh also enables further extensive studies of the Oslofjord using ADCIRC or other finite element models, e.g. SWEMniCS~\cite{dawson2024swemnics} or DG-SWEM~\cite{wichitrnithed2024discontinuous}.

Using harmonic analysis, we reconstruct 2--4 tidal constituents per simulated period in order to analyze the model's ability to propagate tidal forcing through the narrow channels of the Oslofjord mesh. The reconstructed constituents have small amplitude errors, and the tidal ADCIRC-OF model has very high correlation with the NMA tidal predictions, indicating a good reproduction of tidal signal. In general, however, the correlations were slightly lower at the Oslo gauge in the inner fjord, beyond Dr\o{}bak Sound.
We also analyze the effects of using additional tidal constituents on the tidal model, finding that additional constituents may slightly improve the tidal correlations with minimal additional computation time. When the model is forced with output from STOFS-2D-Global, we observe much greater correlation between the residuals, and show that ADCIRC-OF can produce similar TWL predictions to both STOFS and MET Norway's forecast model. Compared to the STOFS model, ADCIRC-OF produces a much higher-resolution solution.

Future iterations of the model could be enhanced by more accurate data, particularly higher resolution wind data from MET Norway. Adding river input and rainfall on the grid would also help with future compound flooding studies near river mouths in the fjord, particularly in the spring when there is a large amount of snowmelt. 
In order to further improve the mesh itself, finer bathymetry data could be incorporated, along with topography to help extend the mesh further inland. Using parameter estimation to develop a spatially varying bottom friction map could also produce better results than the constant friction factor used here.

Other possibilities include developing a model based on the adaptive hydraulics suite~\cite{hammack2008modeling} that uses a ``ship'' as transient forcing to simulate the mini-tsunamis that have been observed in the fjord, as well as probabilistic studies with variable storm tracks and intensities.

Our results indicate that the major limitations are likely associated with the external boundary conditions and the accuracy of the bathymetry, but the sensitivity of the model to the level of detail in the atmospheric forcing is also an interesting topic for a follow-up study. In order to more effectively capture external weather effects without using output from an existing model, a mesh including much of the North and Norwegian Seas is desired. This would require an additional sea ice model, and could allow for results approximating those of Figure \ref{fig:stofs_hans} with only tidal and atmospheric forcing. It is nevertheless clear that the model is a flexible and computationally efficient tool for predicting water level changes in complex fjord environments with high relevance for operational forecasting.

\section*{Acknowledgements}
The authors would like to gratefully acknowledge the use of the ``DMS23001,'' ``DMS21031,'' and ``ADCIRC'' allocations on the Frontera supercomputer at the Texas Advanced Computing Center at the University of Texas at Austin.




\appendix

\newpage

\section{Mesh element size}\label{appendix:size}

\begin{figure}[H]
    \centering
    \includegraphics[width=\linewidth]{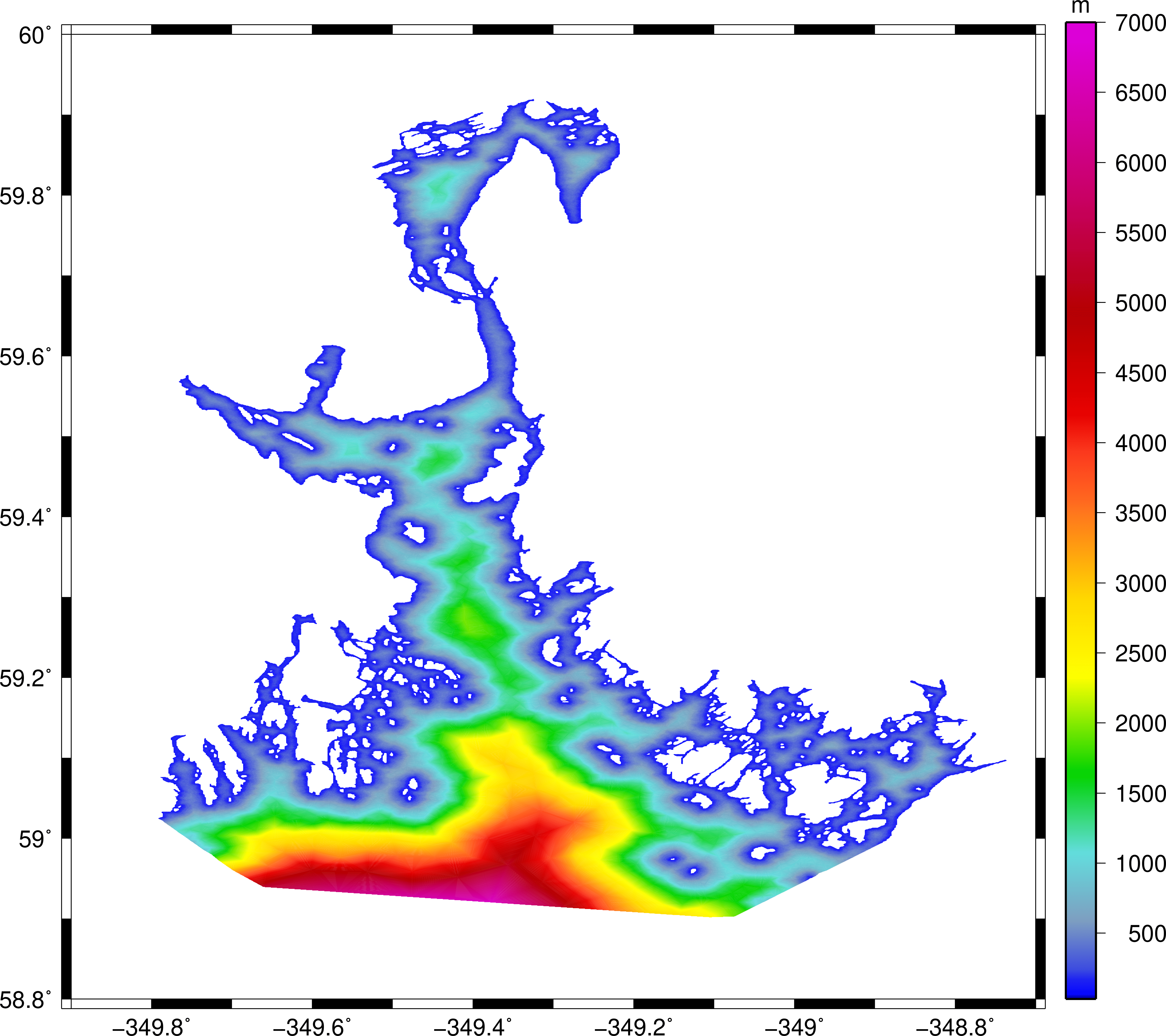}
    \caption{Element size of the Oslofjord mesh.}
    \label{fig:gridsize}
\end{figure}

\newpage

\section{Tidal constituent comparisons}\label{appendix:tidal}

\begin{figure}[H]
    \centering
    \begin{subfigure}{0.5\textwidth}
        \includegraphics[width=\textwidth]{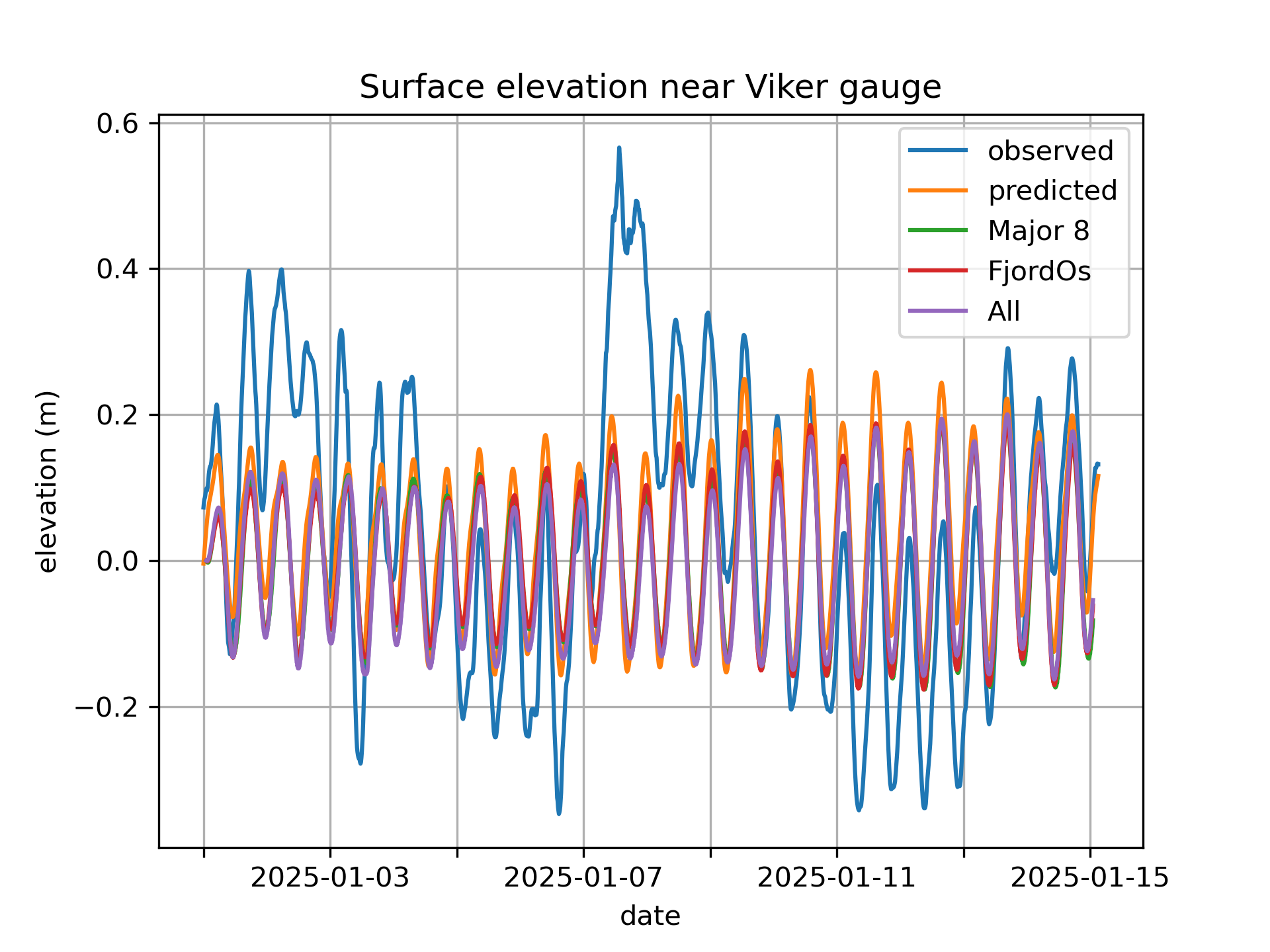}
        \caption{Viker}
        \label{fig:jan25_viker_tides}
    \end{subfigure}
    
    \begin{subfigure}{0.5\textwidth}
        \includegraphics[width=\textwidth]{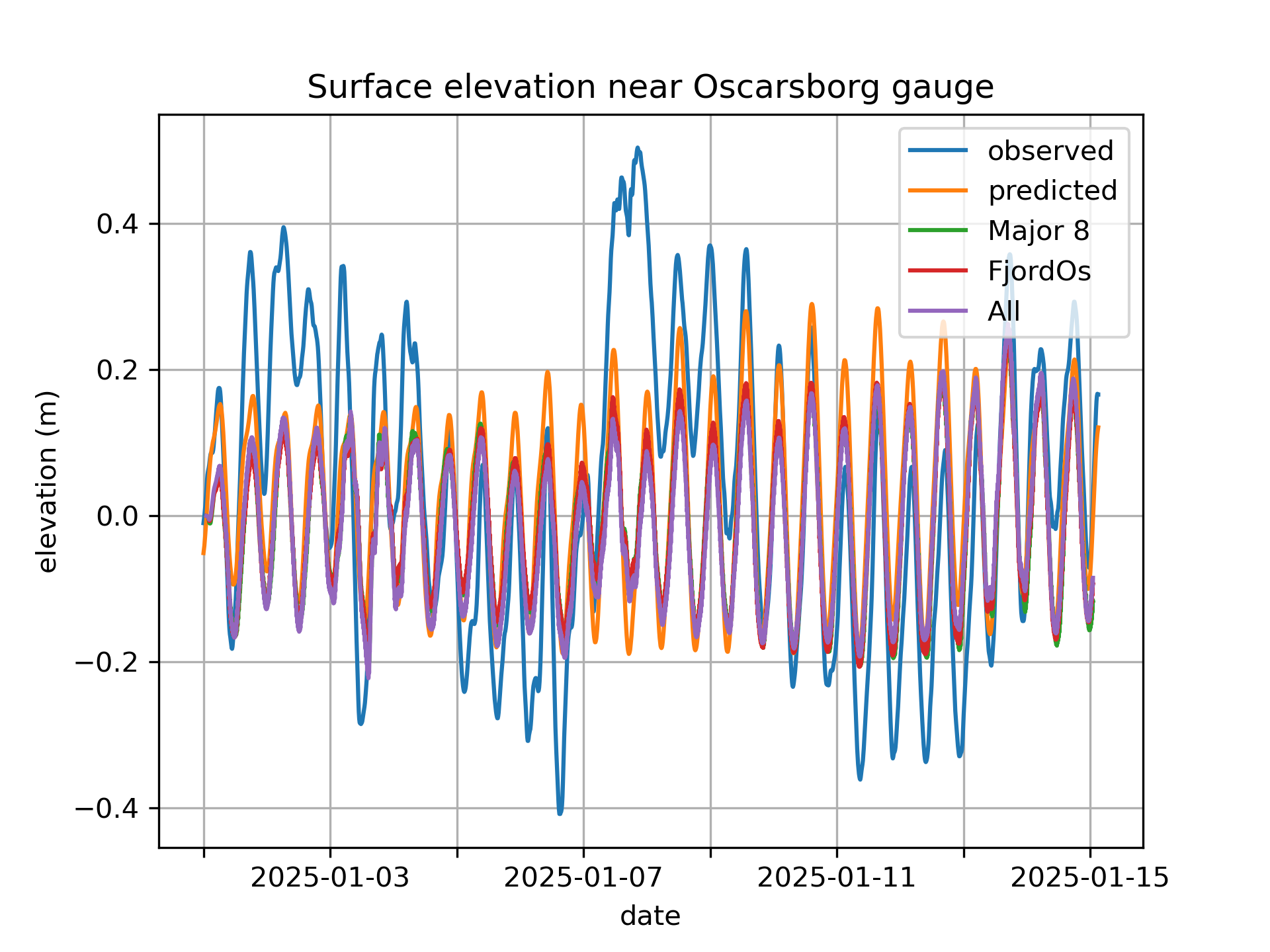}
        \caption{Oscarsborg}
        \label{fig:jan25_oscarsborg_tides}
    \end{subfigure}
    
    \begin{subfigure}{0.5\textwidth}
        \includegraphics[width=\textwidth]{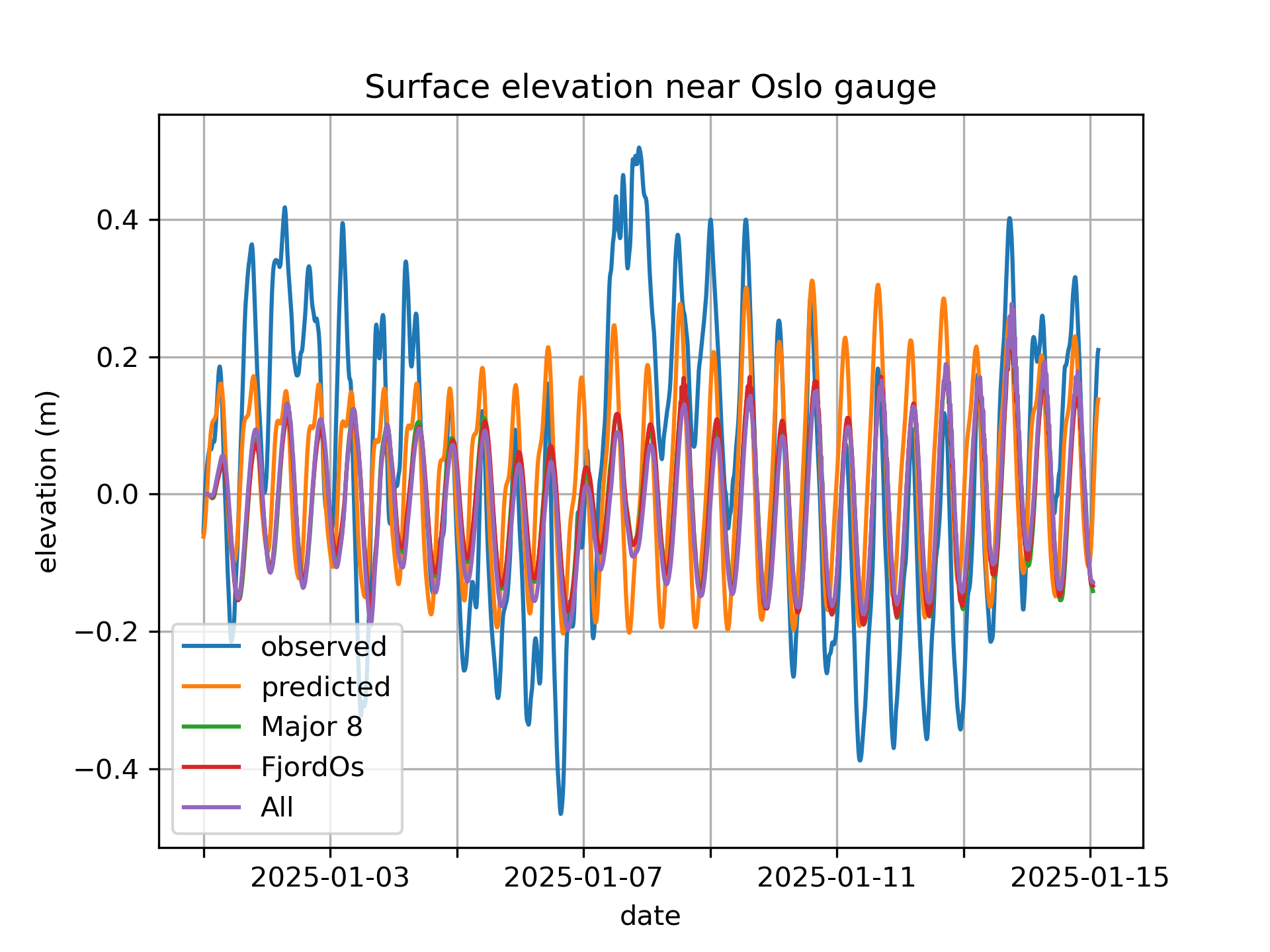}
        \caption{Oslo}
        \label{fig:jan25_oslo_tides}
    \end{subfigure}
    \caption{Tidal constituent comparisons, January 2025.}
    \label{fig:jan25tide_comparisons}
\end{figure}

\pagebreak

\begin{figure}[H]
    \centering
    \begin{subfigure}{0.5\textwidth}
        \includegraphics[width=\textwidth]{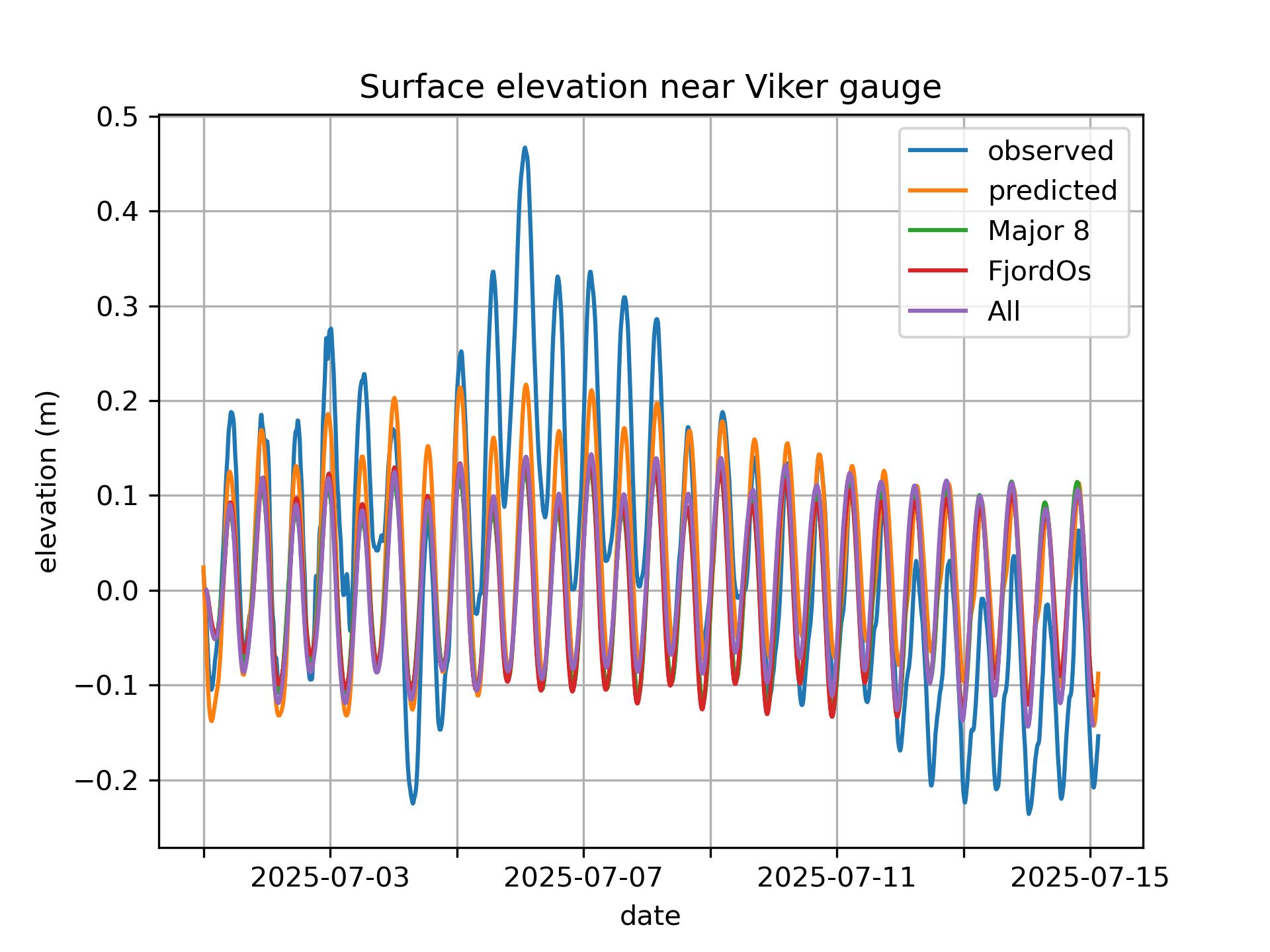}
        \caption{Viker}
        \label{fig:jul25_viker}
    \end{subfigure}
    
    \begin{subfigure}{0.5\textwidth}
        \includegraphics[width=\textwidth]{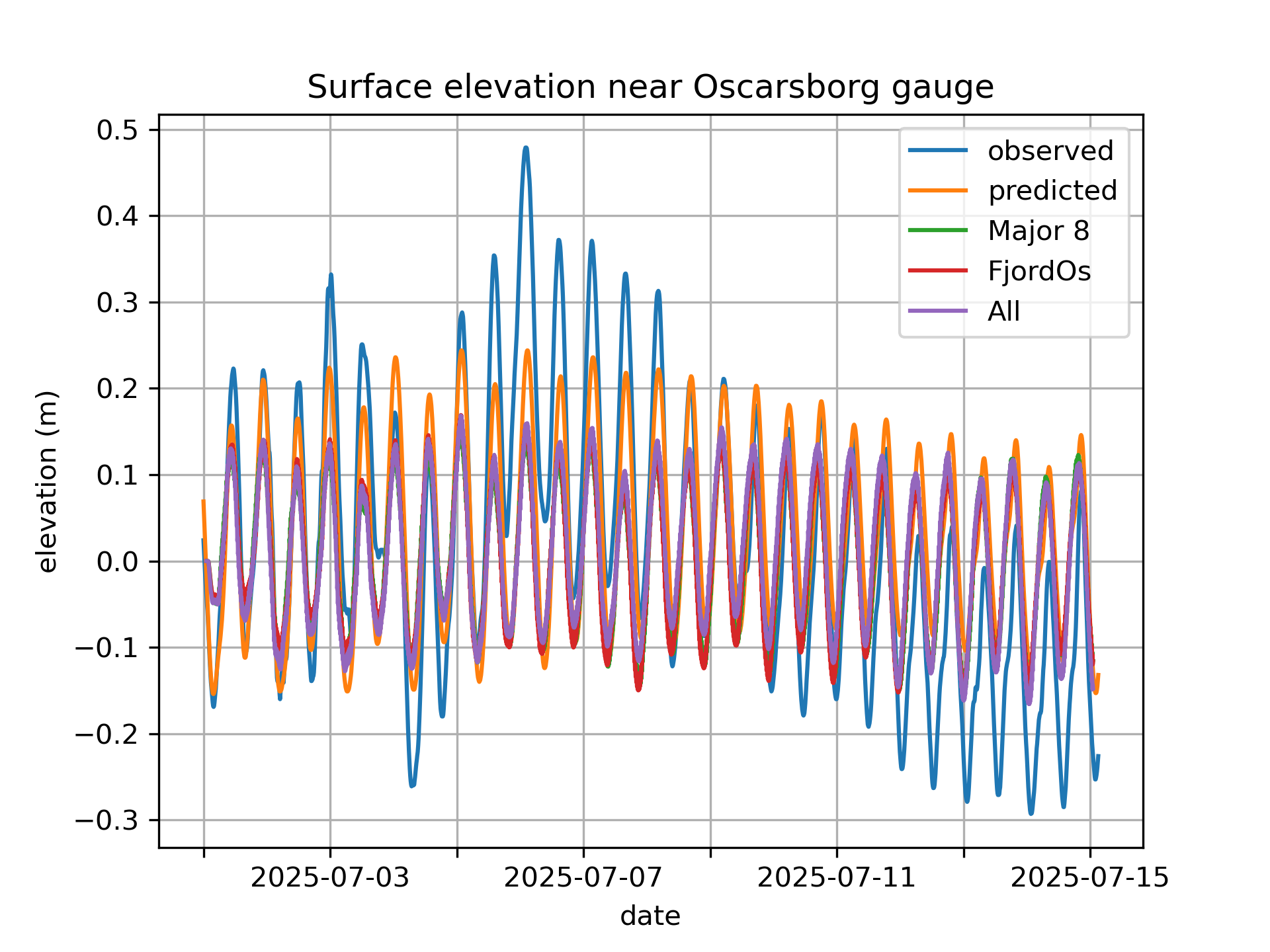}
        \caption{Oscarsborg}
        \label{fig:jul25_oscarsborg}
    \end{subfigure}
    
    \begin{subfigure}{0.5\textwidth}
        \includegraphics[width=\textwidth]{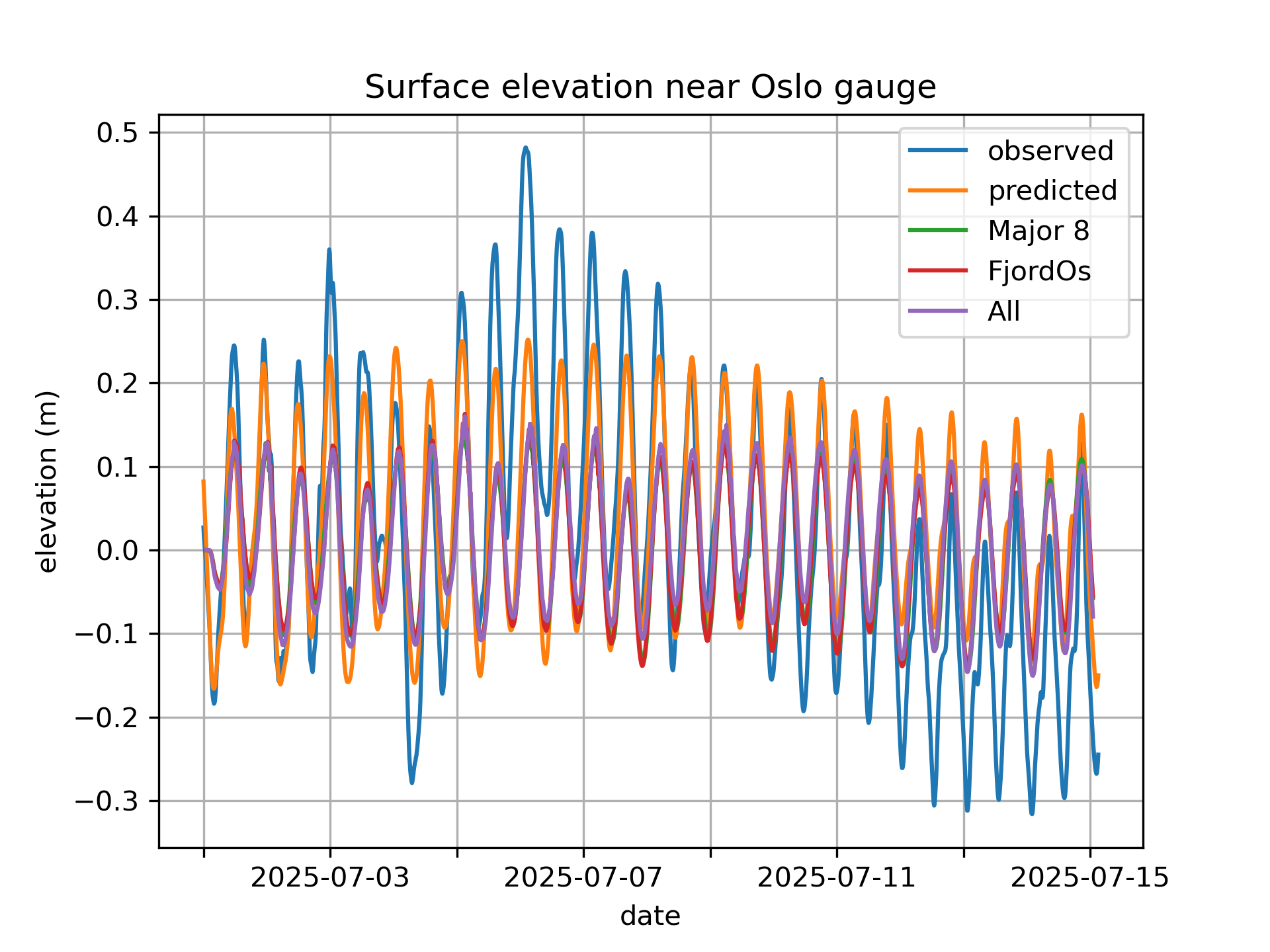}
        \caption{Oslo}
        \label{fig:jul25_oslo}
    \end{subfigure}
    \caption{Tidal constituent comparisons, July 2025.}
    \label{fig:jul25tide_comparisons}
\end{figure}

\pagebreak

\section{Differences between various sets of tidal constituents}\label{appendix:diffs}

\begin{figure}[H]
    \centering
    \begin{subfigure}{0.4\textwidth}
        \includegraphics[width=\textwidth]{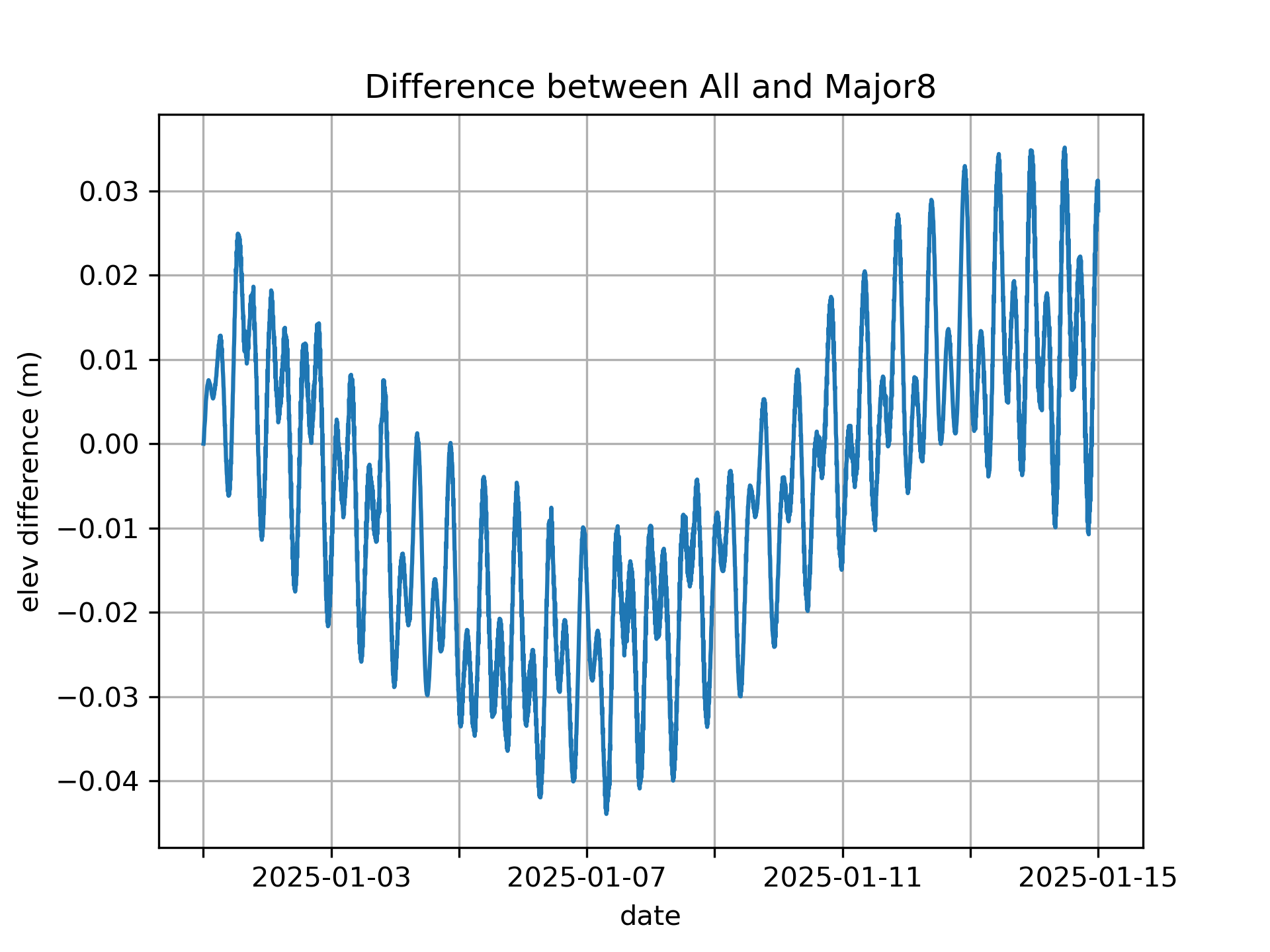}
        \caption{Viker}
        \label{fig:jan25_viker_all_maj8}
    \end{subfigure}
    
    \begin{subfigure}{0.4\textwidth}
        \includegraphics[width=\textwidth]{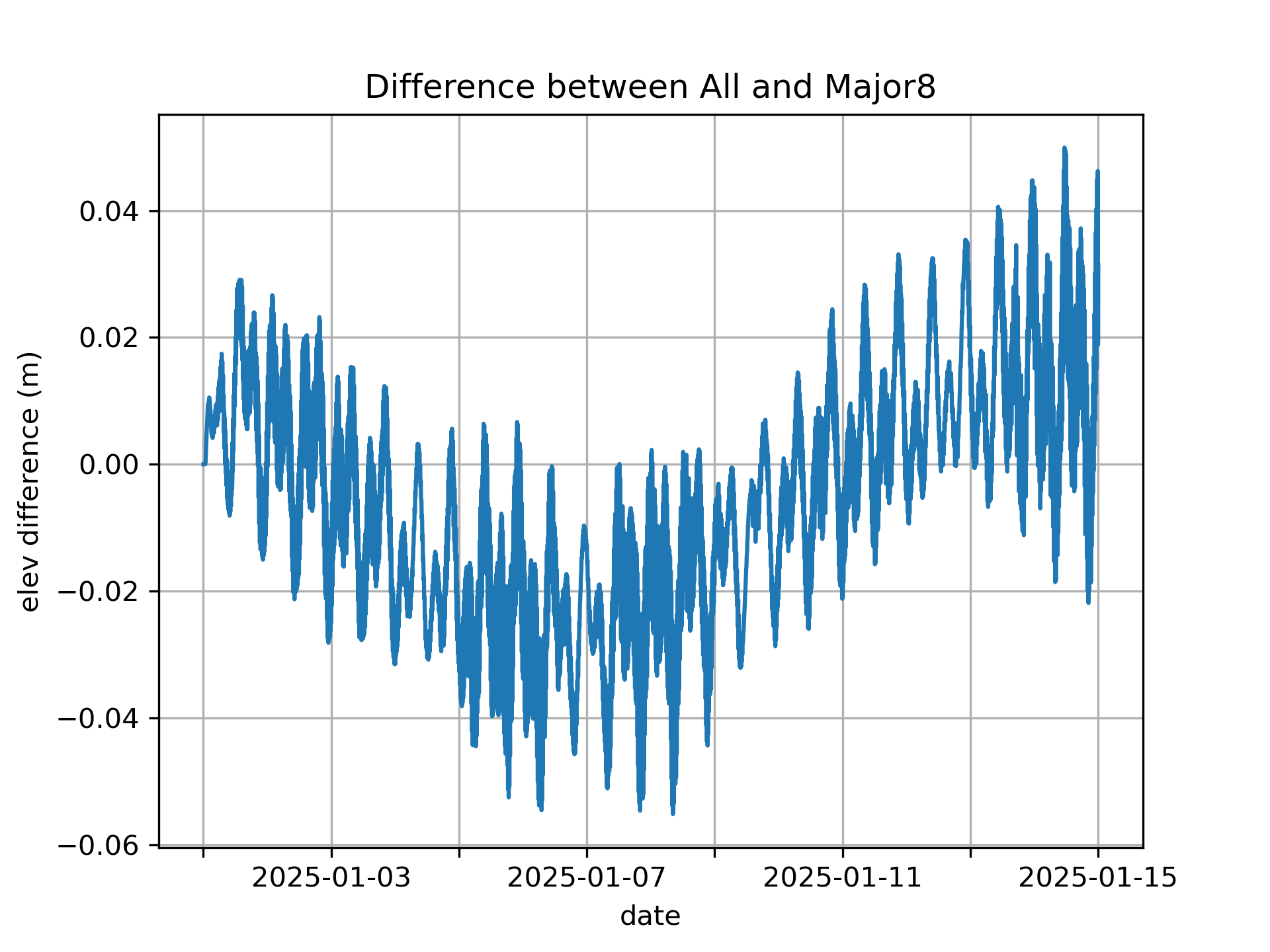}
        \caption{Oscarsborg}
        \label{fig:jan25_oscarsborg__all_maj8}
    \end{subfigure}
    
    \begin{subfigure}{0.4\textwidth}
        \includegraphics[width=\textwidth]{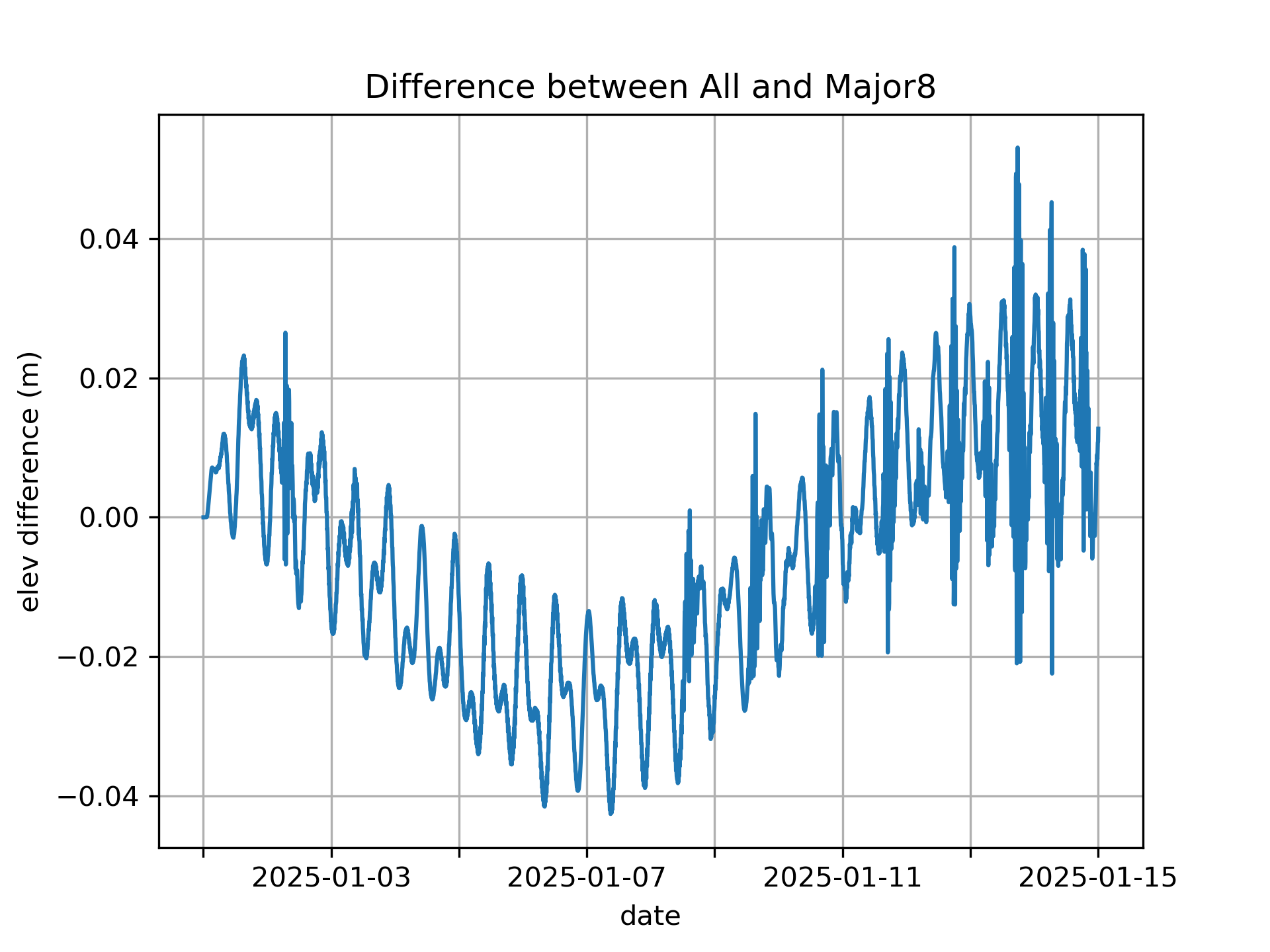}
        \caption{Oslo}
        \label{fig:jan25_oslo_all_maj8}
    \end{subfigure}
    \caption{Difference between ADCIRC-OF surface elevations with the 15-constituent set of tidal constituents vs. the major 8 constituents, for January 2025.}
    \label{fig:jan25_all_maj8}
\end{figure}

\pagebreak

\begin{figure}[H]
    \centering
    \begin{subfigure}{0.4\textwidth}
        \includegraphics[width=\textwidth]{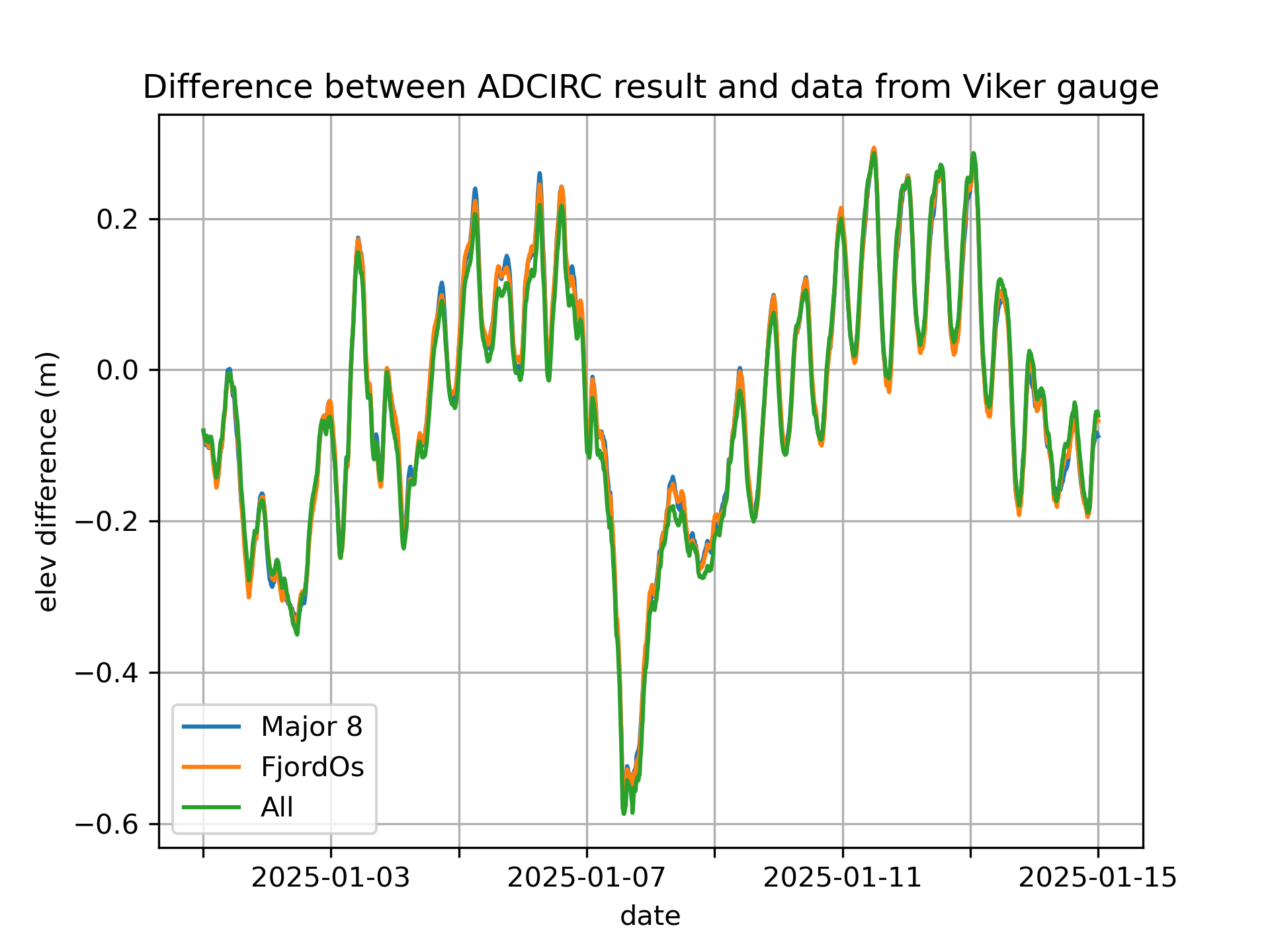}
        \caption{Viker}
        \label{fig:jan25_viker_diff}
    \end{subfigure}
    
    \begin{subfigure}{0.4\textwidth}
        \includegraphics[width=\textwidth]{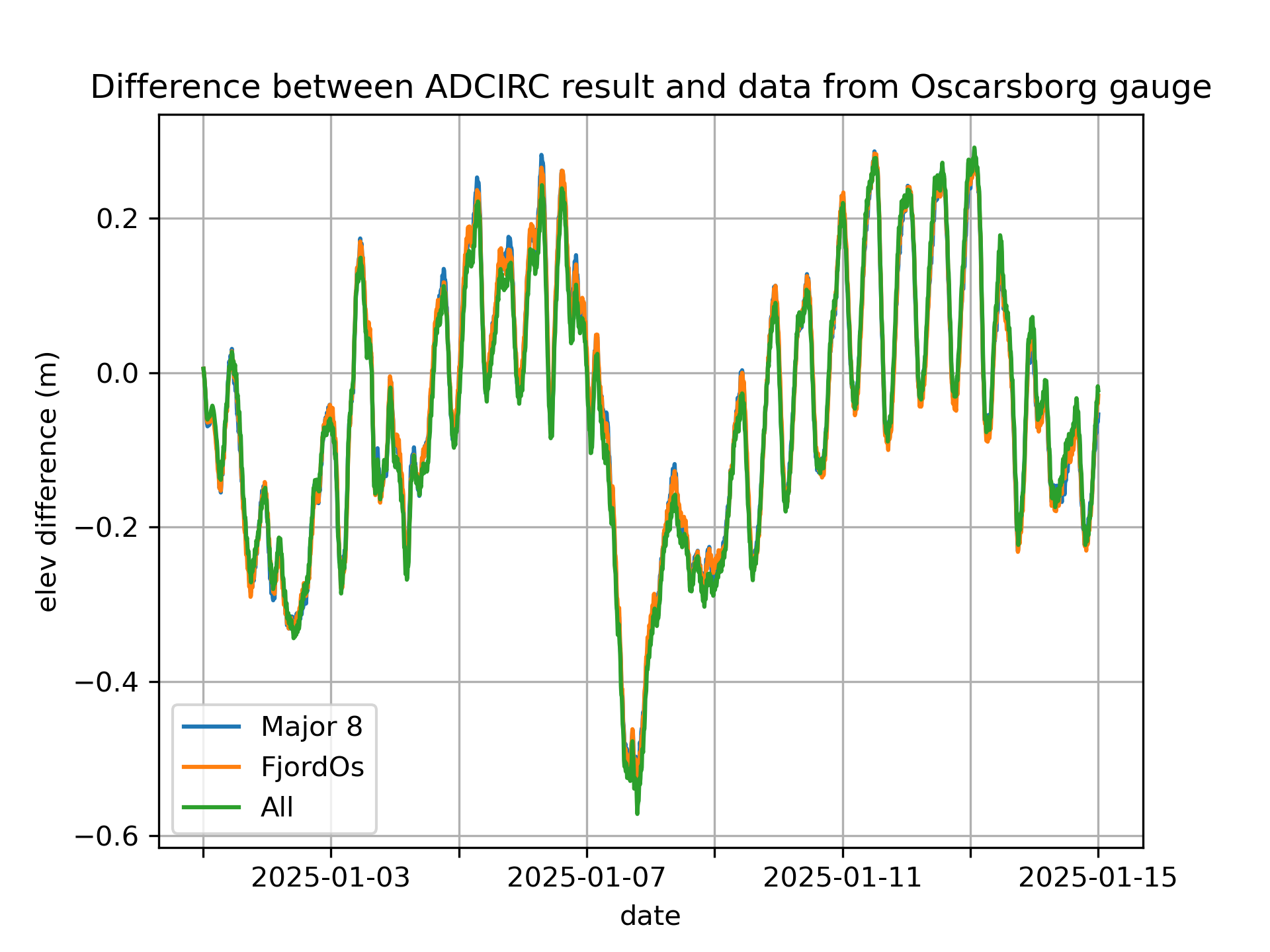}
        \caption{Oscarsborg}
        \label{fig:jan25_oscarsborg_diff}
    \end{subfigure}
    
    \begin{subfigure}{0.4\textwidth}
        \includegraphics[width=\textwidth]{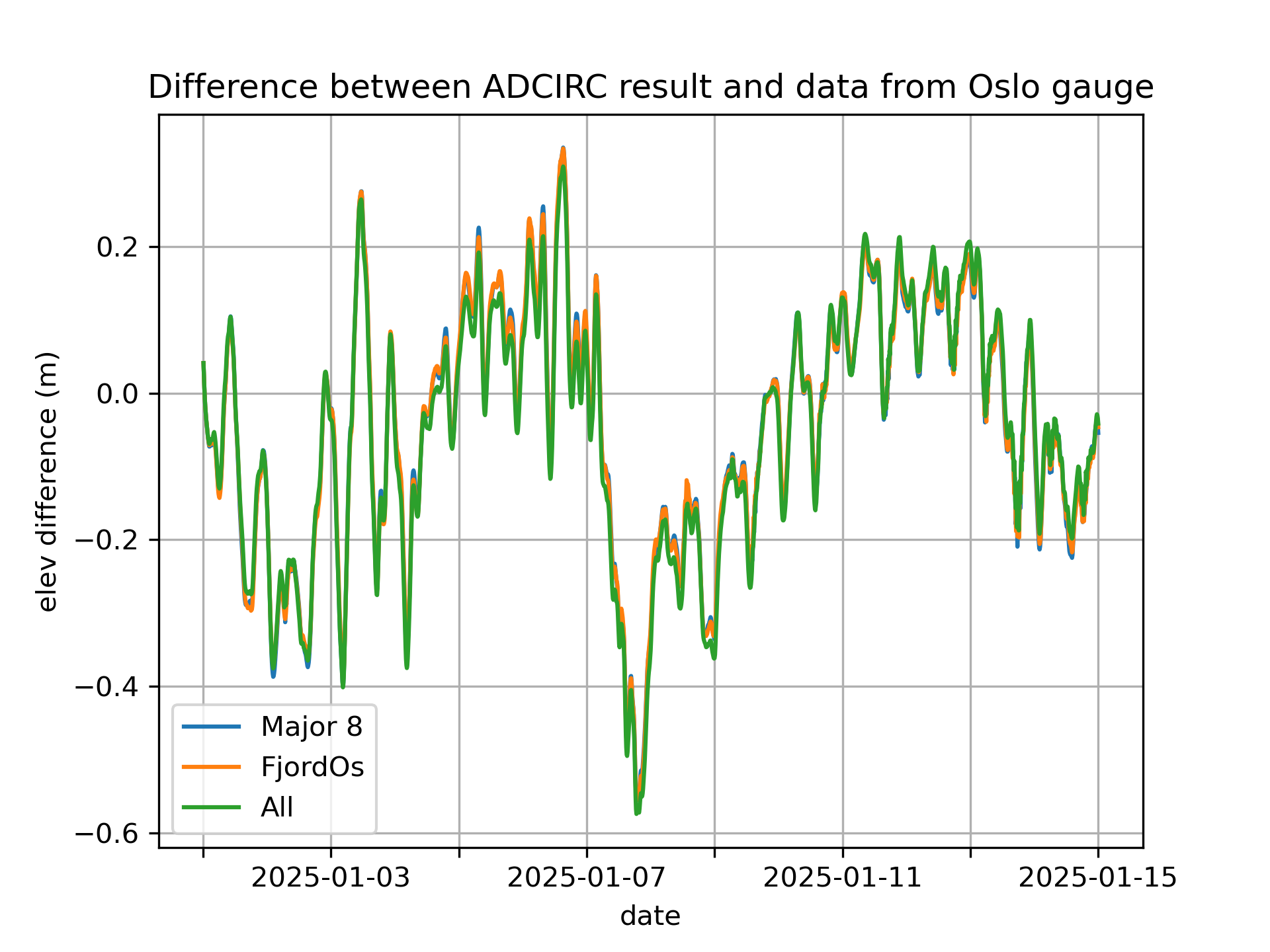}
        \caption{Oslo}
        \label{fig:jan25_oslo_diff}
    \end{subfigure}
    \caption{Difference between ADCIRC-OF results and observed gauge data for January 2025.}
    \label{fig:jan25_tides_diff}
\end{figure}

\pagebreak

\begin{figure}[H]
    \centering
    \begin{subfigure}{0.4\textwidth}
        \includegraphics[width=\textwidth]{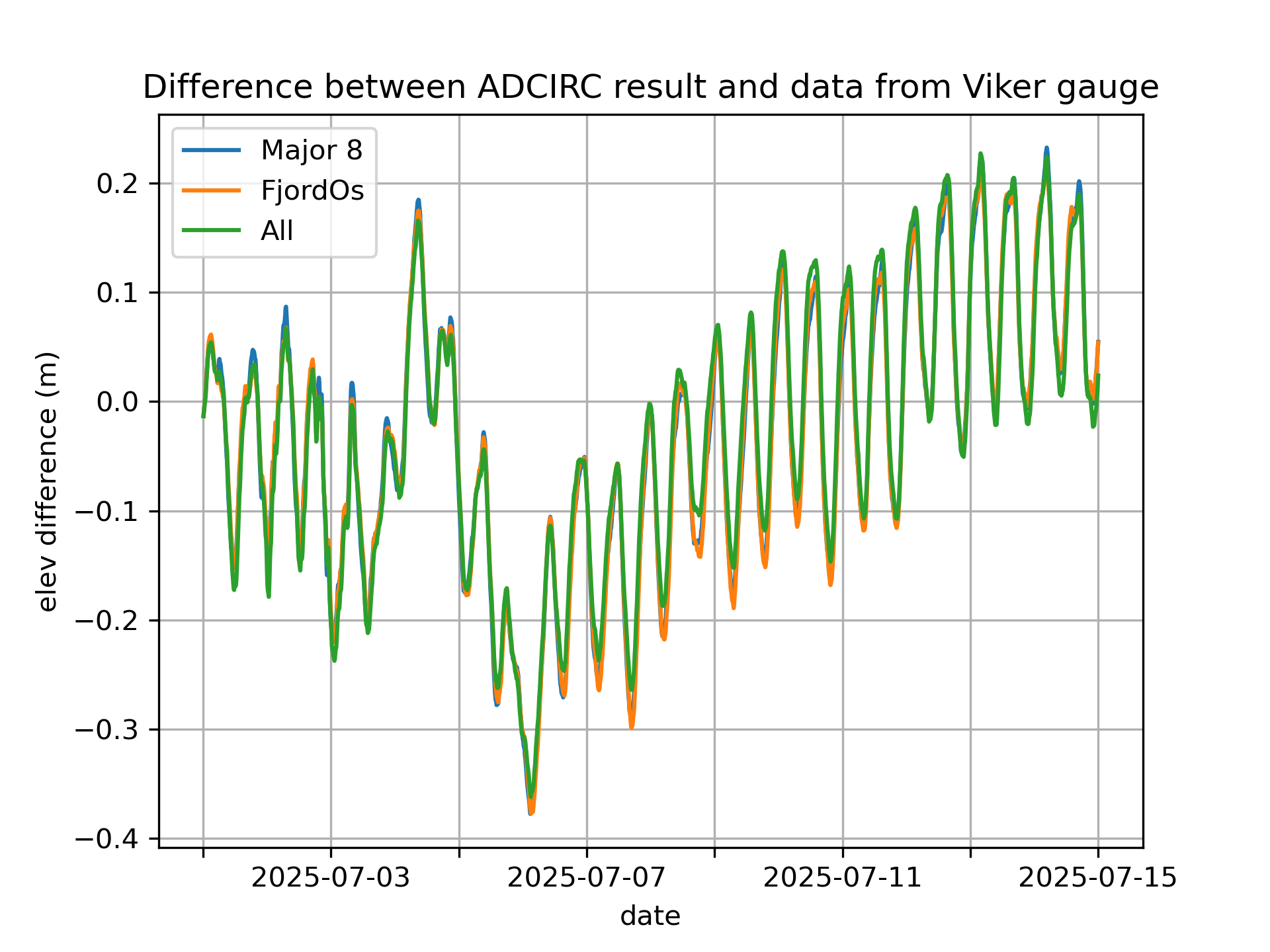}
        \caption{Viker}
        \label{fig:jul25_viker_diff}
    \end{subfigure}
    
    \begin{subfigure}{0.4\textwidth}
        \includegraphics[width=\textwidth]{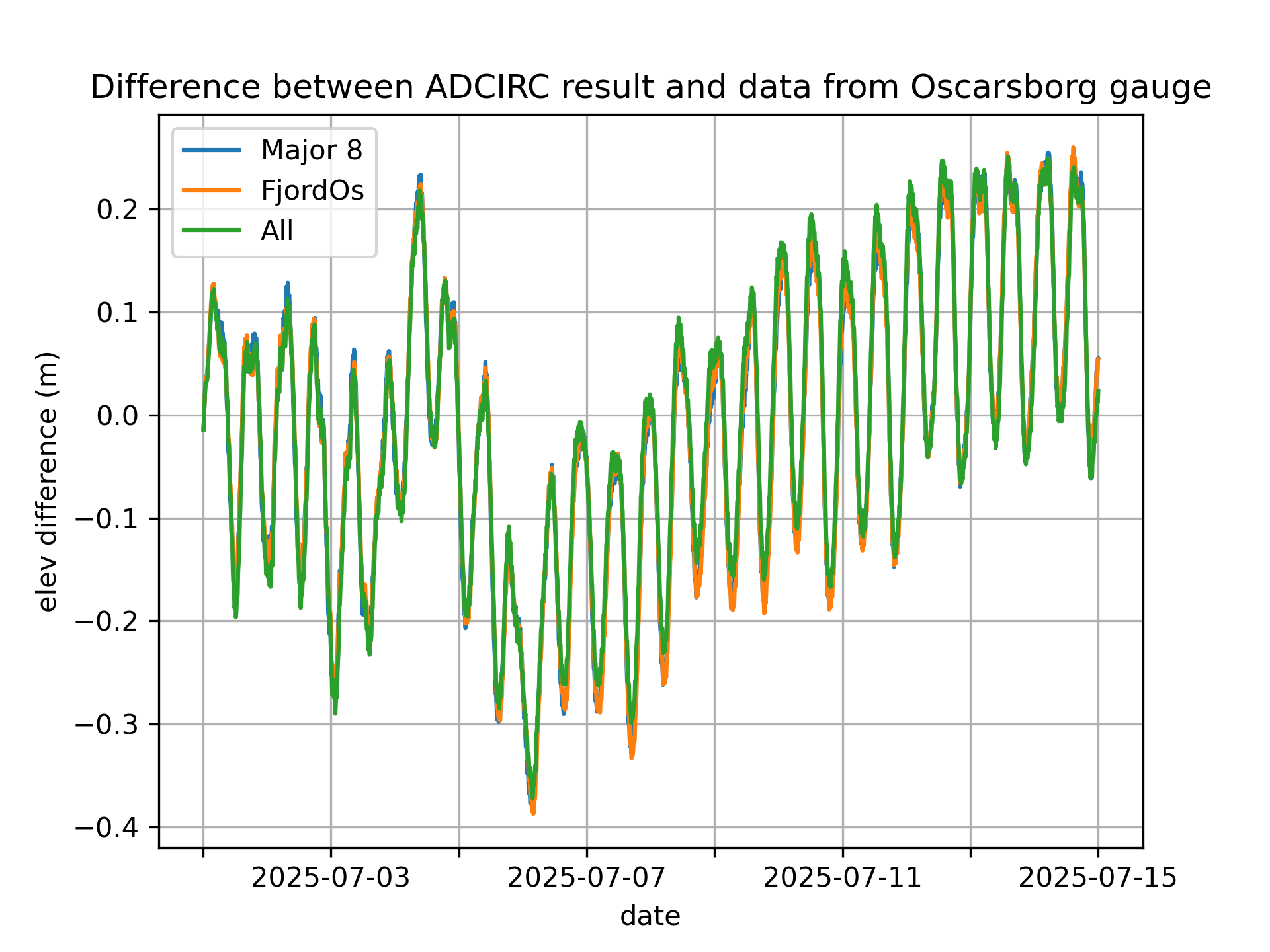}
        \caption{Oscarsborg}
        \label{fig:jul25_oscarsborg_diff}
    \end{subfigure}
    
    \begin{subfigure}{0.4\textwidth}
        \includegraphics[width=\textwidth]{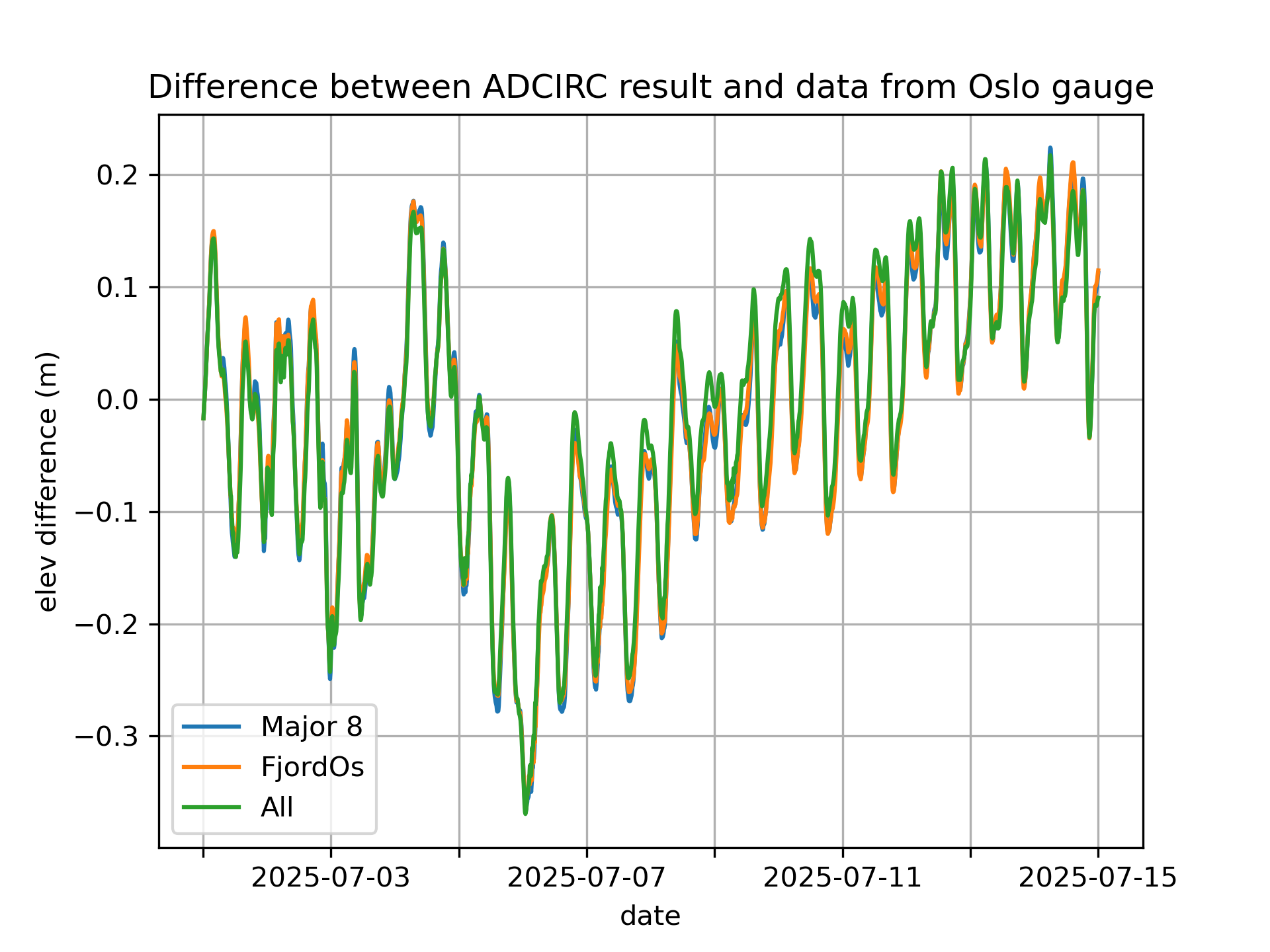}
        \caption{Oslo}
        \label{fig:jul25_oslo_diff}
    \end{subfigure}
    \caption{Difference between ADCIRC-OF results and observed gauge data for July 2025.}
    \label{fig:jul25_tides_diff}
\end{figure}

\pagebreak

\begin{figure}[H]
    \centering
    \begin{subfigure}{0.4\textwidth}
        \includegraphics[width=\textwidth]{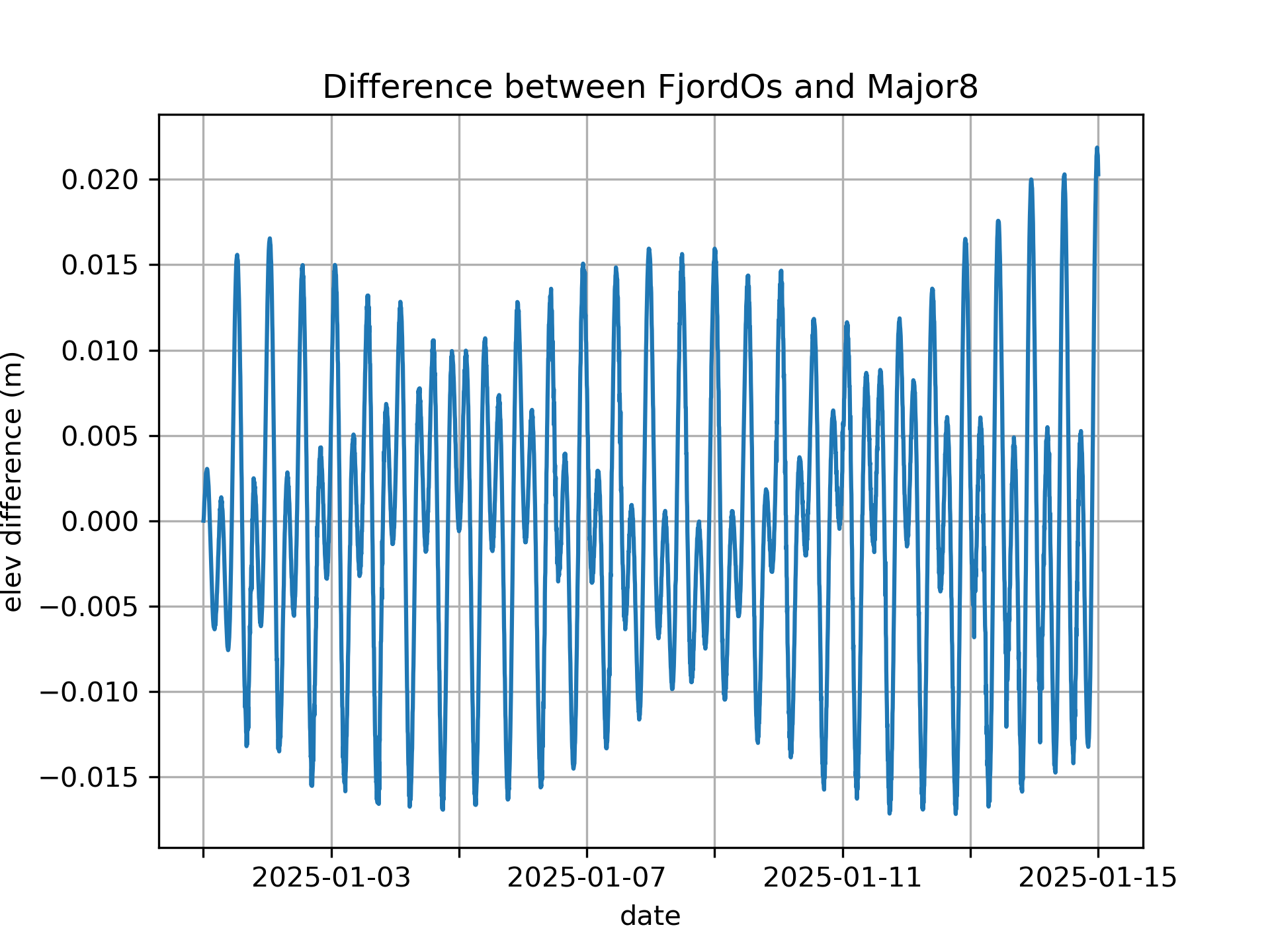}
        \caption{Viker}
        \label{fig:jan25_viker_diff_nine_maj8}
    \end{subfigure}
    
    \begin{subfigure}{0.4\textwidth}
        \includegraphics[width=\textwidth]{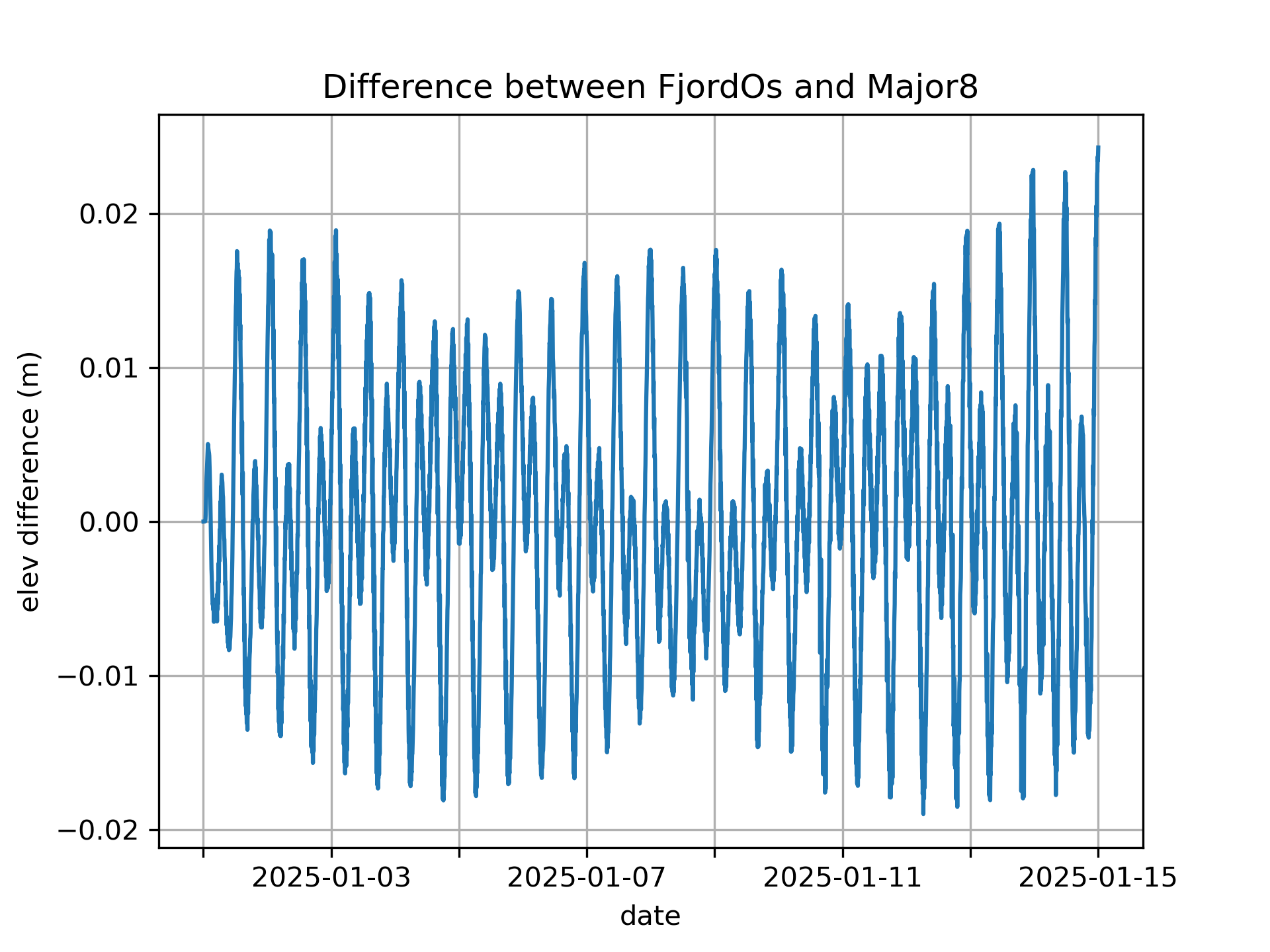}
        \caption{Oscarsborg}
        \label{fig:jan25_oscarsborg_diff_nine_maj8}
    \end{subfigure}
    
    \begin{subfigure}{0.4\textwidth}
        \includegraphics[width=\textwidth]{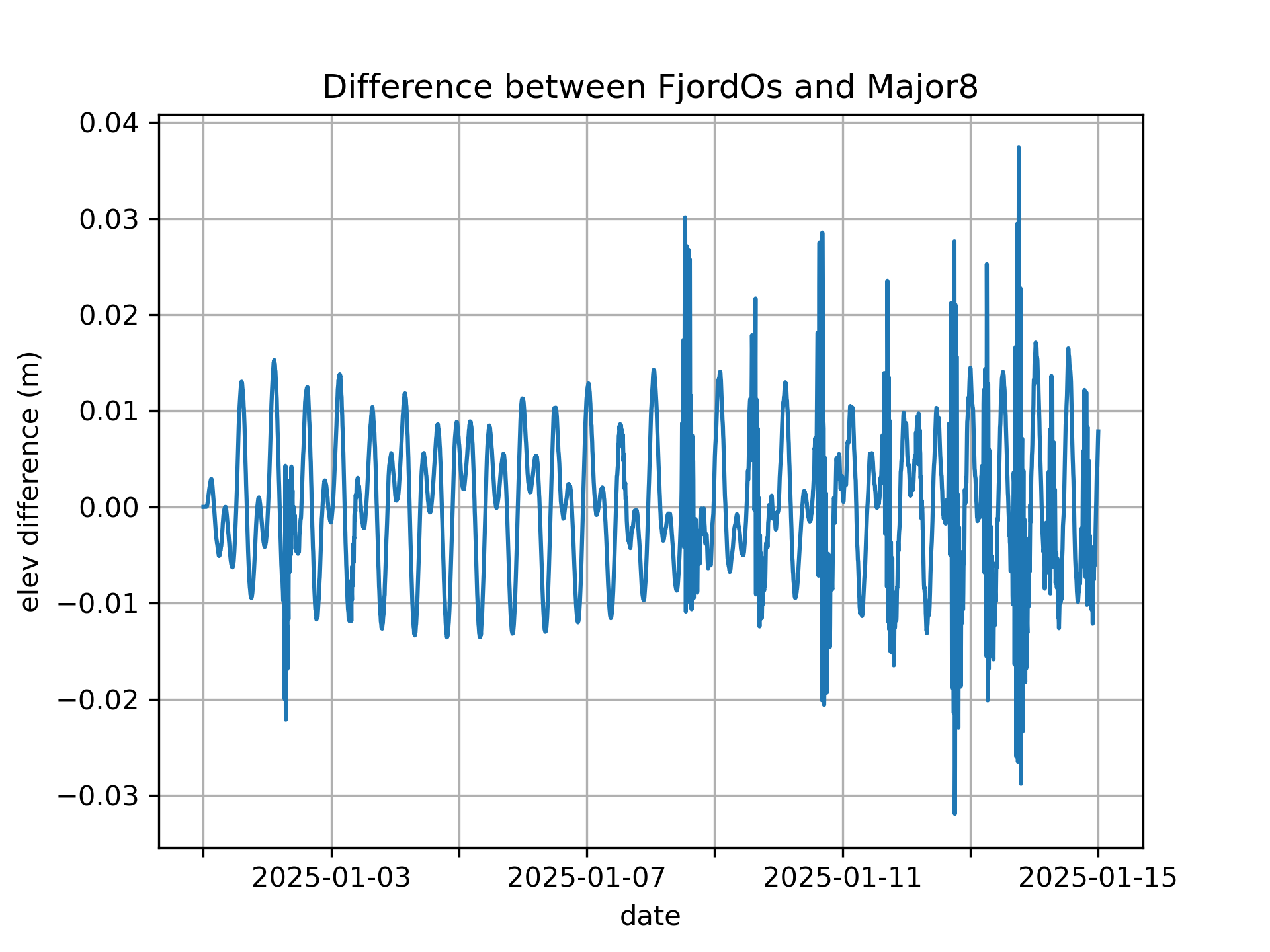}
        \caption{Oslo}
        \label{fig:jan25_oslo_diff_nine_maj8}
    \end{subfigure}
    \caption{Difference between ADCIRC-OF surface elevations with the FjordOs set of tidal constituents vs. the major 8 constituents, for January 2025.}
    \label{fig:jan25_tides_diff_nine_maj8}
\end{figure}

\pagebreak

\begin{figure}[H]
    \centering
    \begin{subfigure}{0.4\textwidth}
        \includegraphics[width=\textwidth]{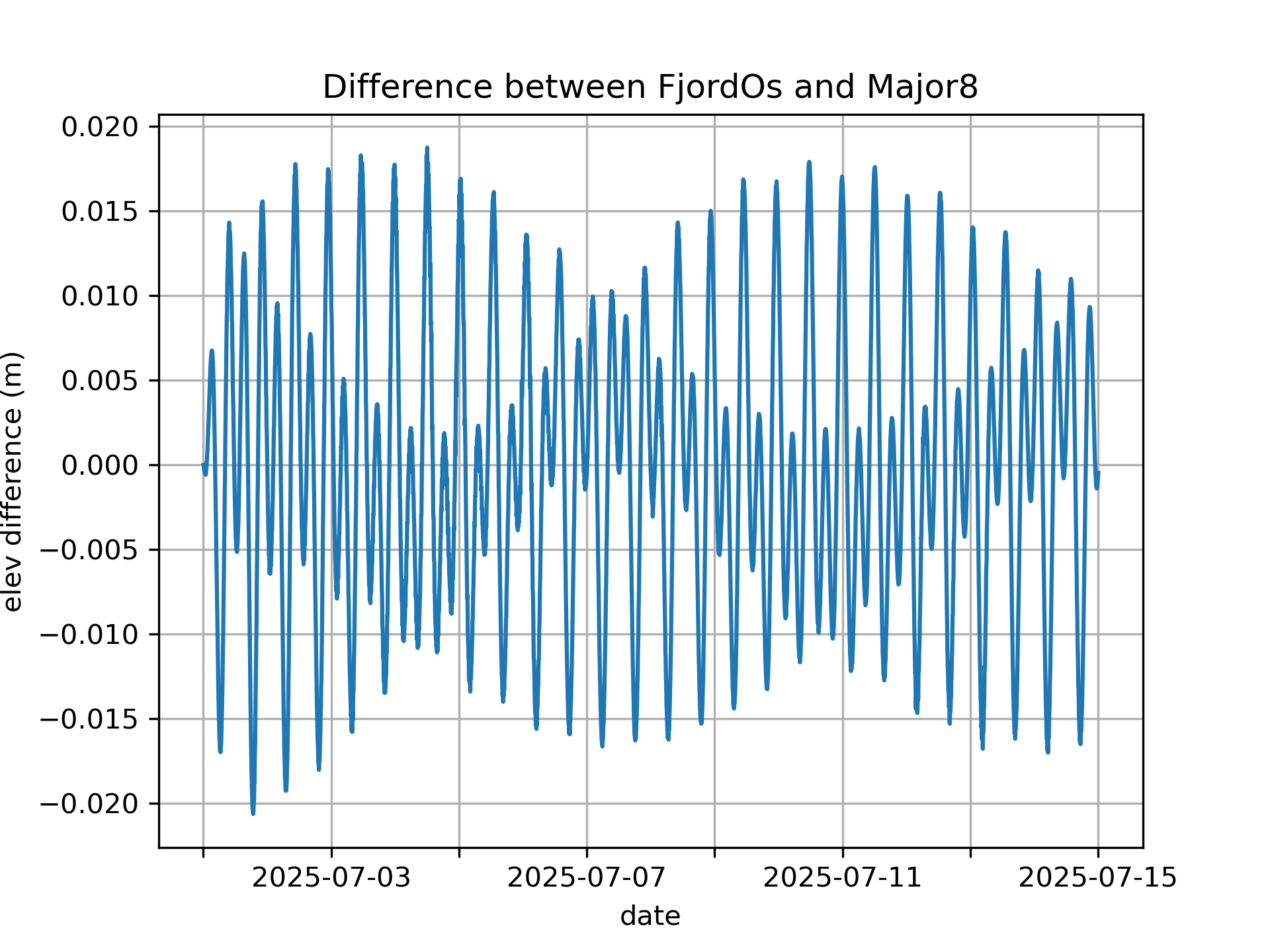}
        \caption{Viker}
        \label{fig:jul25_viker_diff_nine_maj8}
    \end{subfigure}
    
    \begin{subfigure}{0.4\textwidth}
        \includegraphics[width=\textwidth]{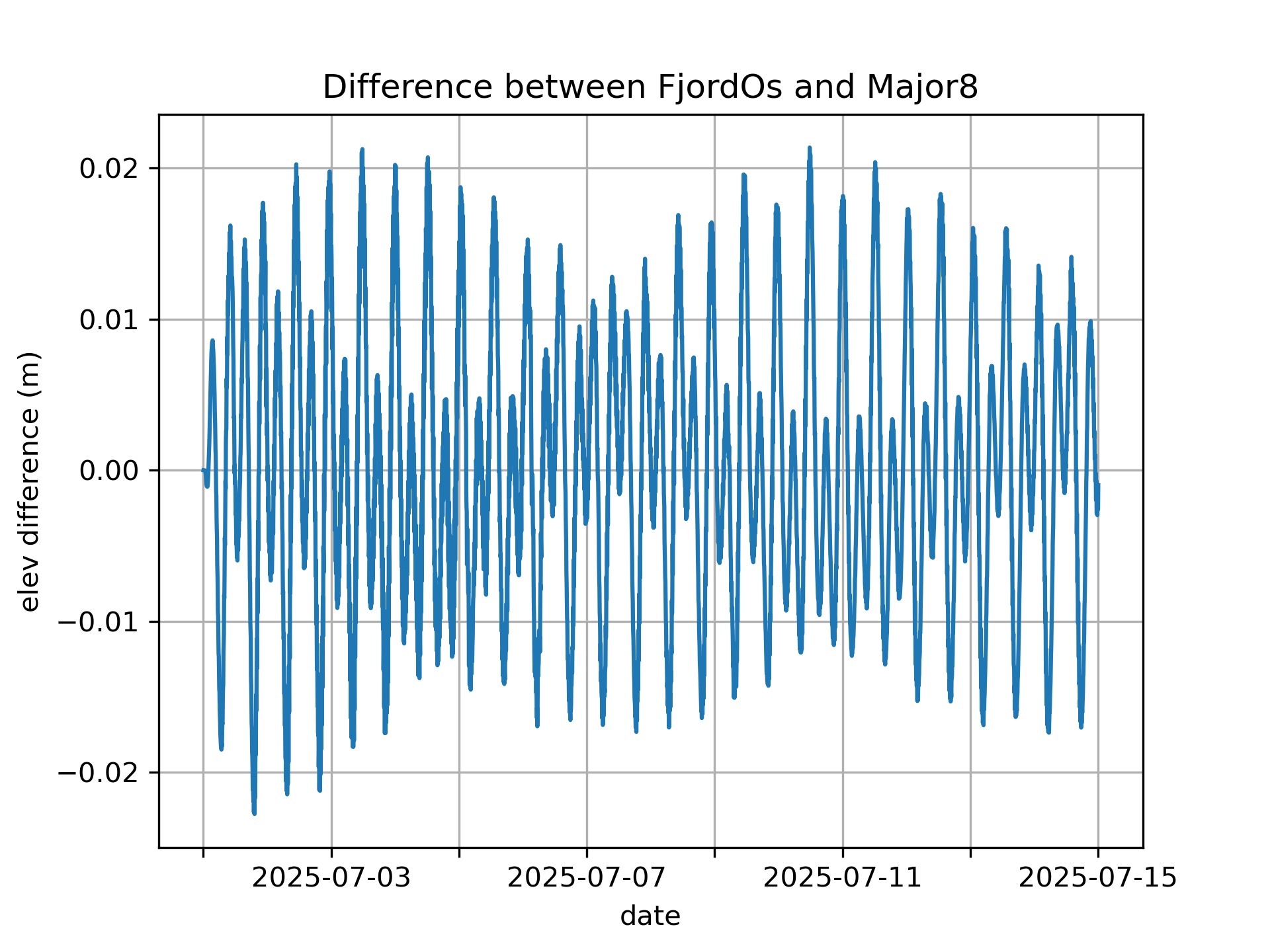}
        \caption{Oscarsborg}
        \label{fig:jul25_oscarsborg_diff_nine_maj8}
    \end{subfigure}
    
    \begin{subfigure}{0.4\textwidth}
        \includegraphics[width=\textwidth]{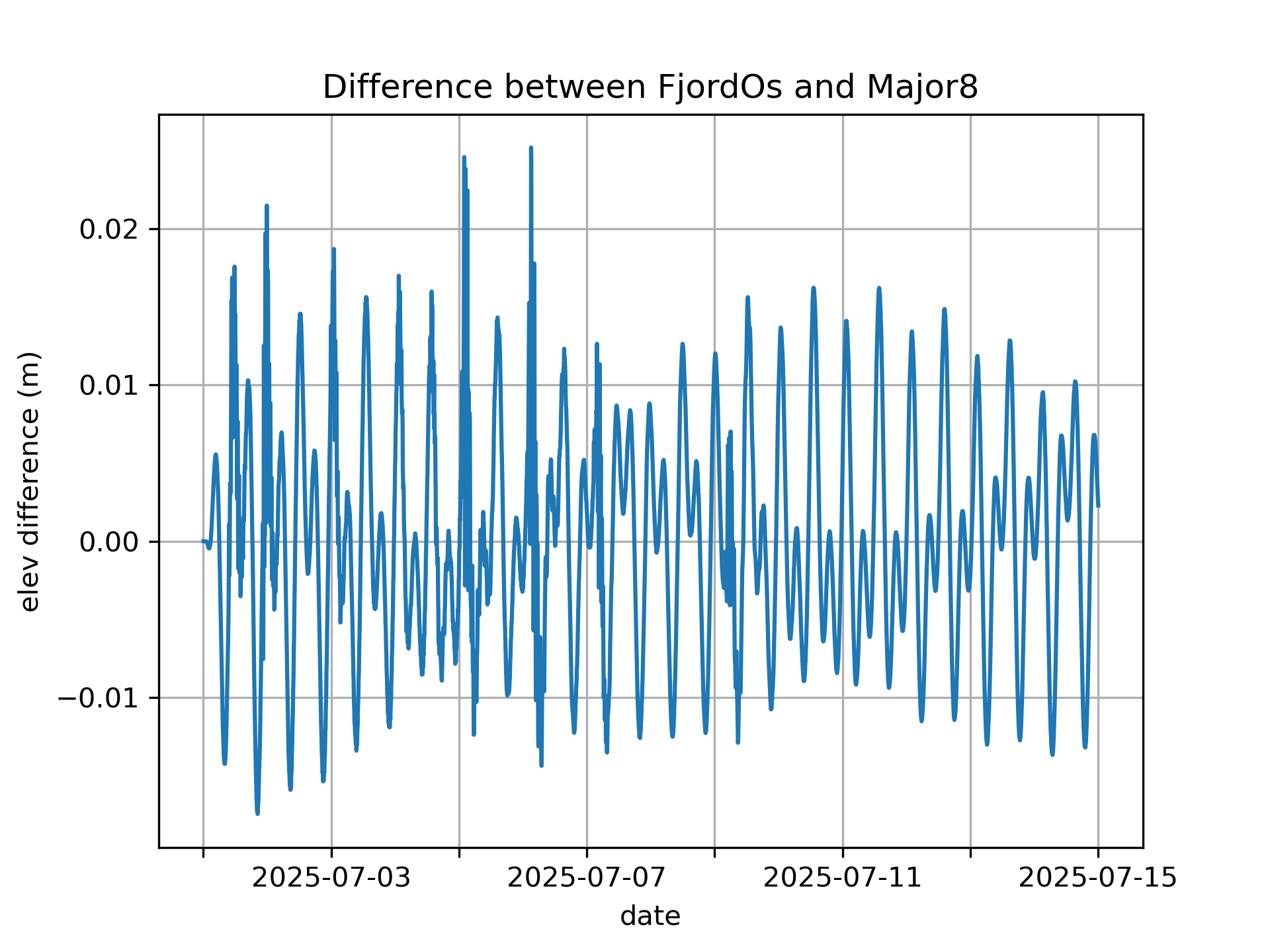}
        \caption{Oslo}
        \label{fig:jul25_oslo_diff_nine_maj8}
    \end{subfigure}
    \caption{Difference between ADCIRC-OF surface elevations with the FjordOs set of tidal constituents vs. the major 8 constituents, July 2025.}
    \label{fig:jul25_tides_diff_nine_maj8}
\end{figure}

\pagebreak

\begin{figure}[H]
    \centering
    \begin{subfigure}{0.4\textwidth}
        \includegraphics[width=\textwidth]{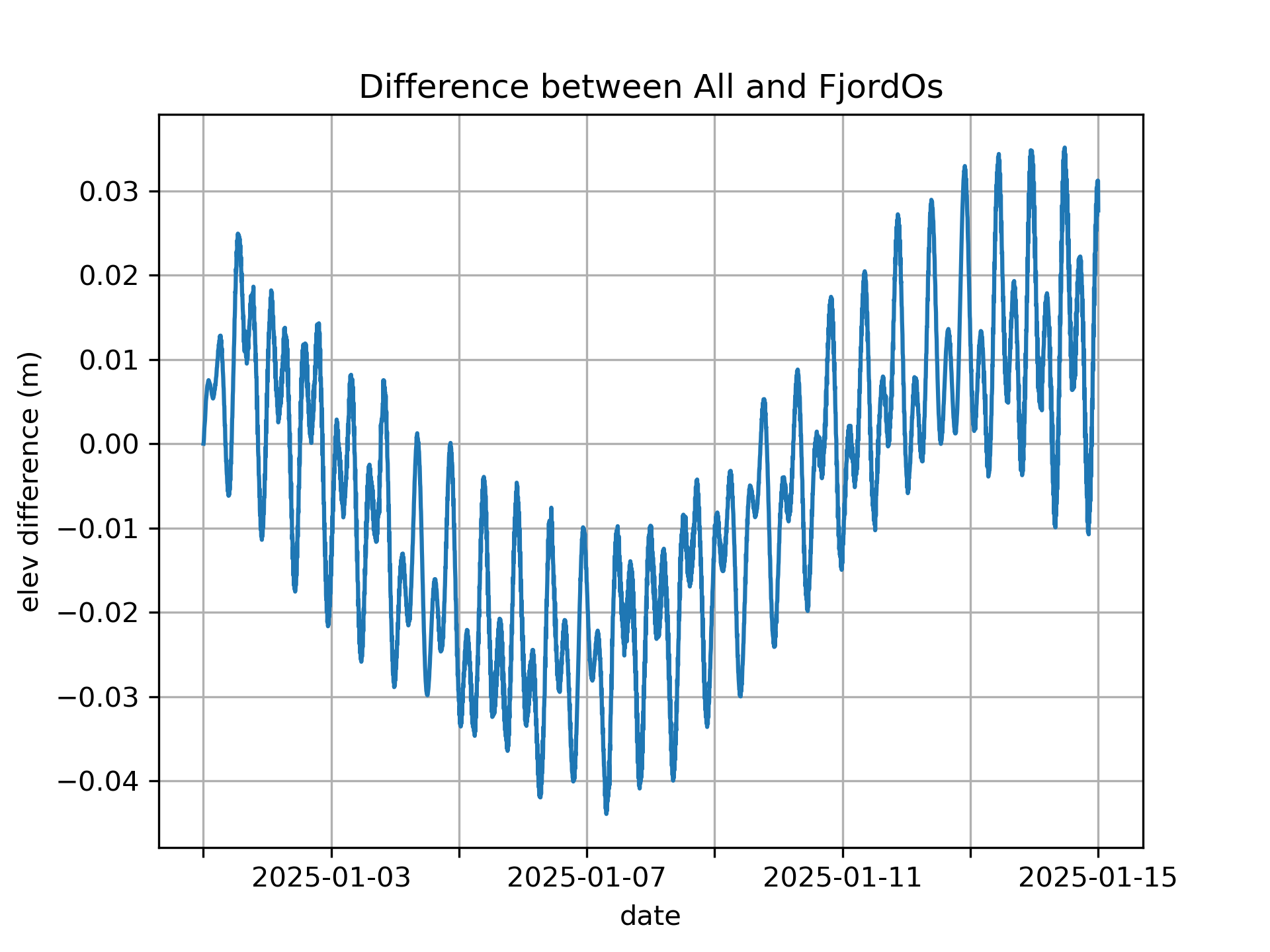}
        \caption{Viker}
        \label{fig:jan25_viker_diff_all_nine}
    \end{subfigure}
    
    \begin{subfigure}{0.4\textwidth}
        \includegraphics[width=\textwidth]{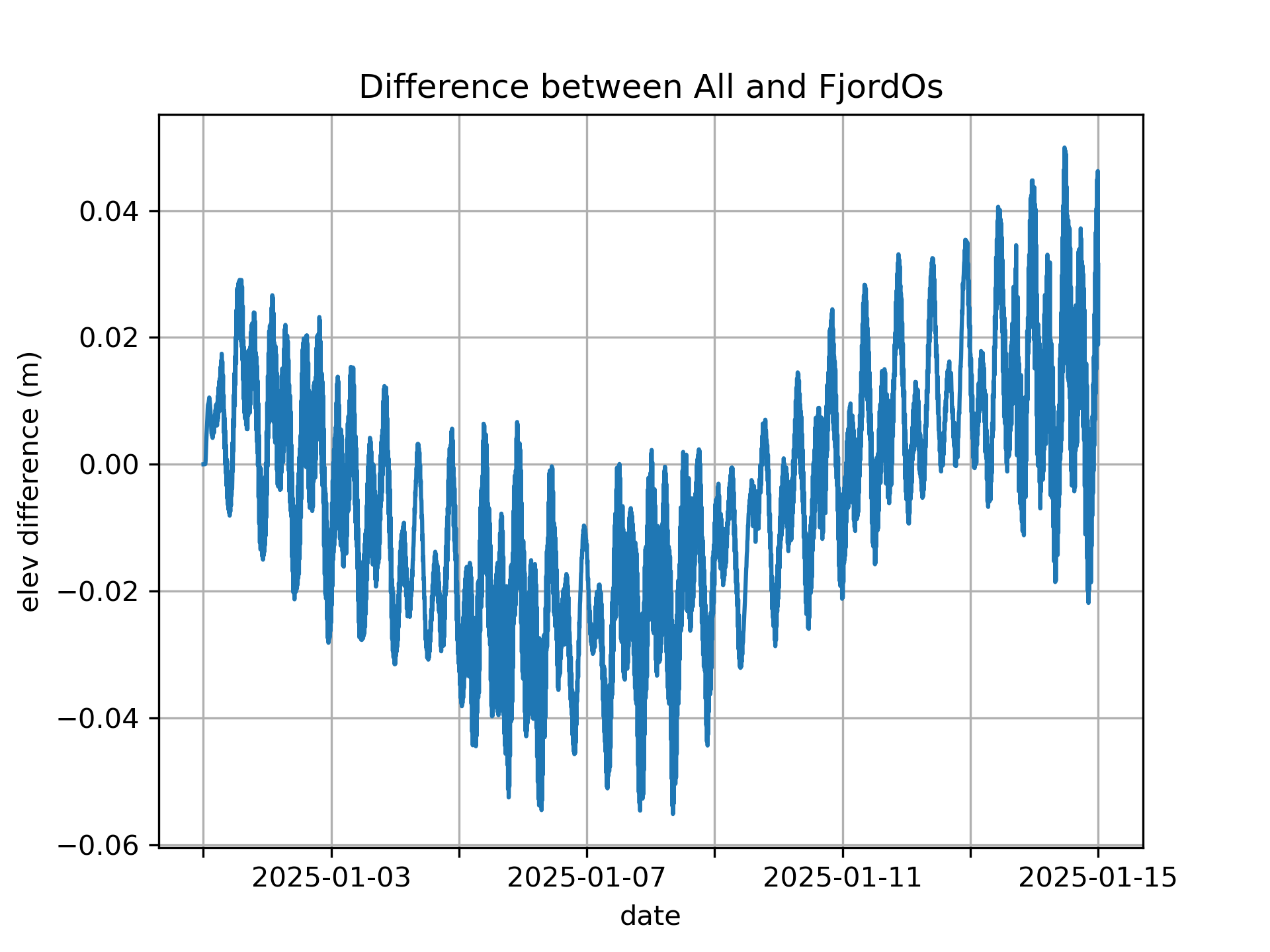}
        \caption{Oscarsborg}
        \label{fig:jan25_oscarsborg_diff_all_nine}
    \end{subfigure}
    
    \begin{subfigure}{0.4\textwidth}
        \includegraphics[width=\textwidth]{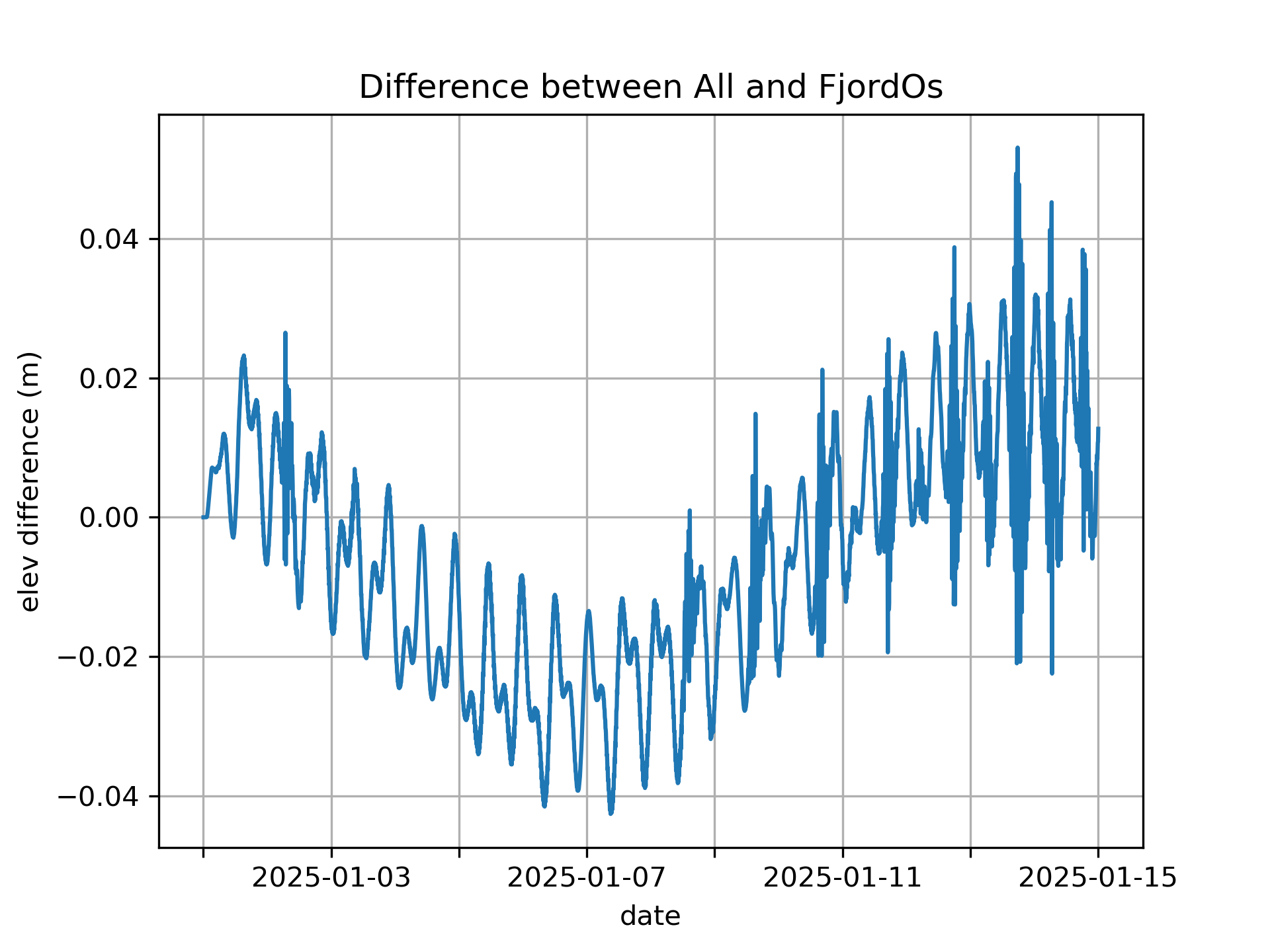}
        \caption{Oslo}
        \label{fig:jan25_oslo_diff_all_nine}
    \end{subfigure}
    \caption{Difference between ADCIRC-OF surface elevations with the 15-constituent model vs. the FjordOs constituents, for January 2025.}
    \label{fig:jan25_tides_diff_all_nine}
\end{figure}

\pagebreak

\begin{figure}[H]
    \centering
    \begin{subfigure}{0.4\textwidth}
        \includegraphics[width=\textwidth]{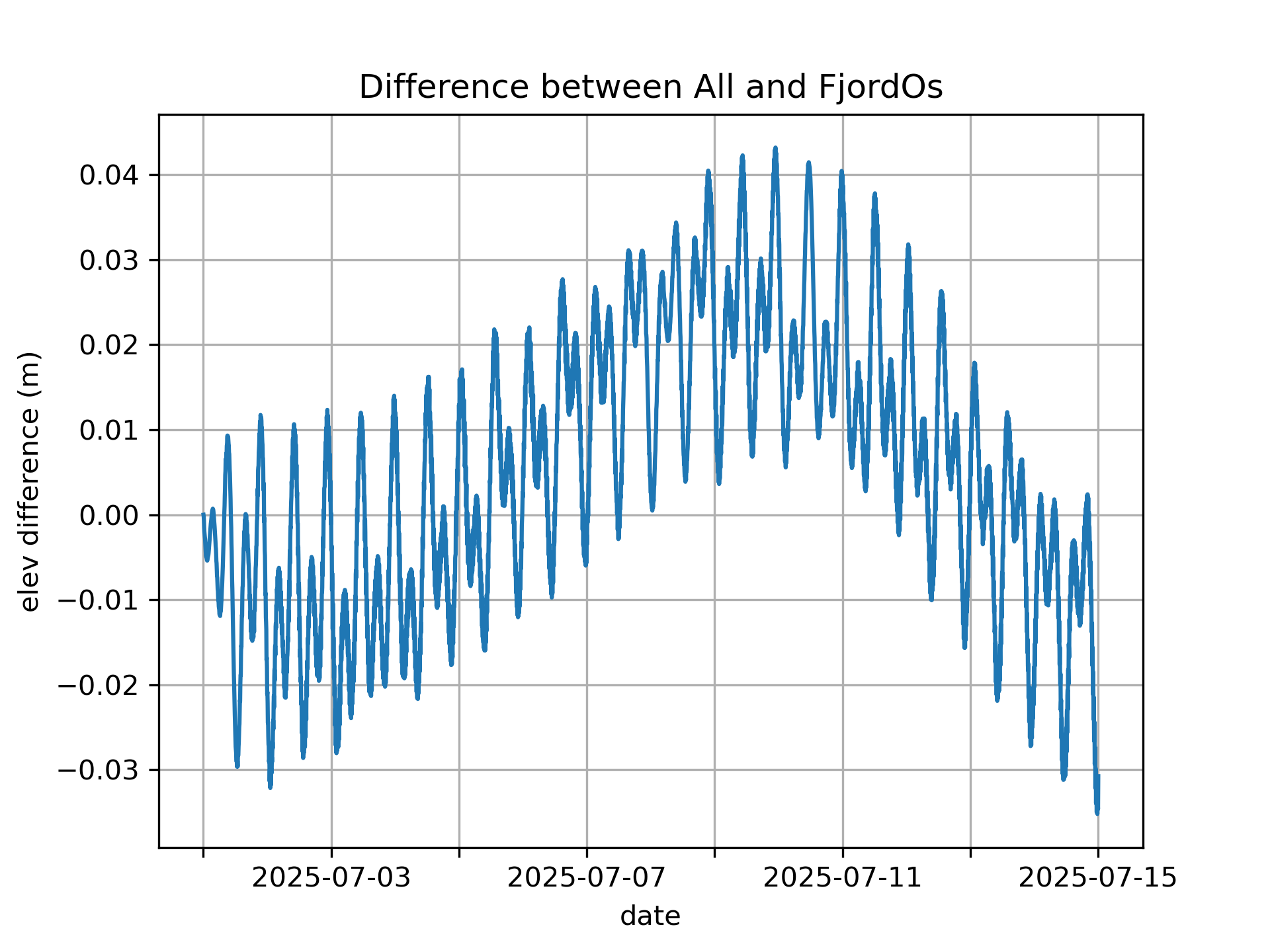}
        \caption{Viker}
        \label{fig:jul25_viker_diff_all_nine}
    \end{subfigure}
    
    \begin{subfigure}{0.4\textwidth}
        \includegraphics[width=\textwidth]{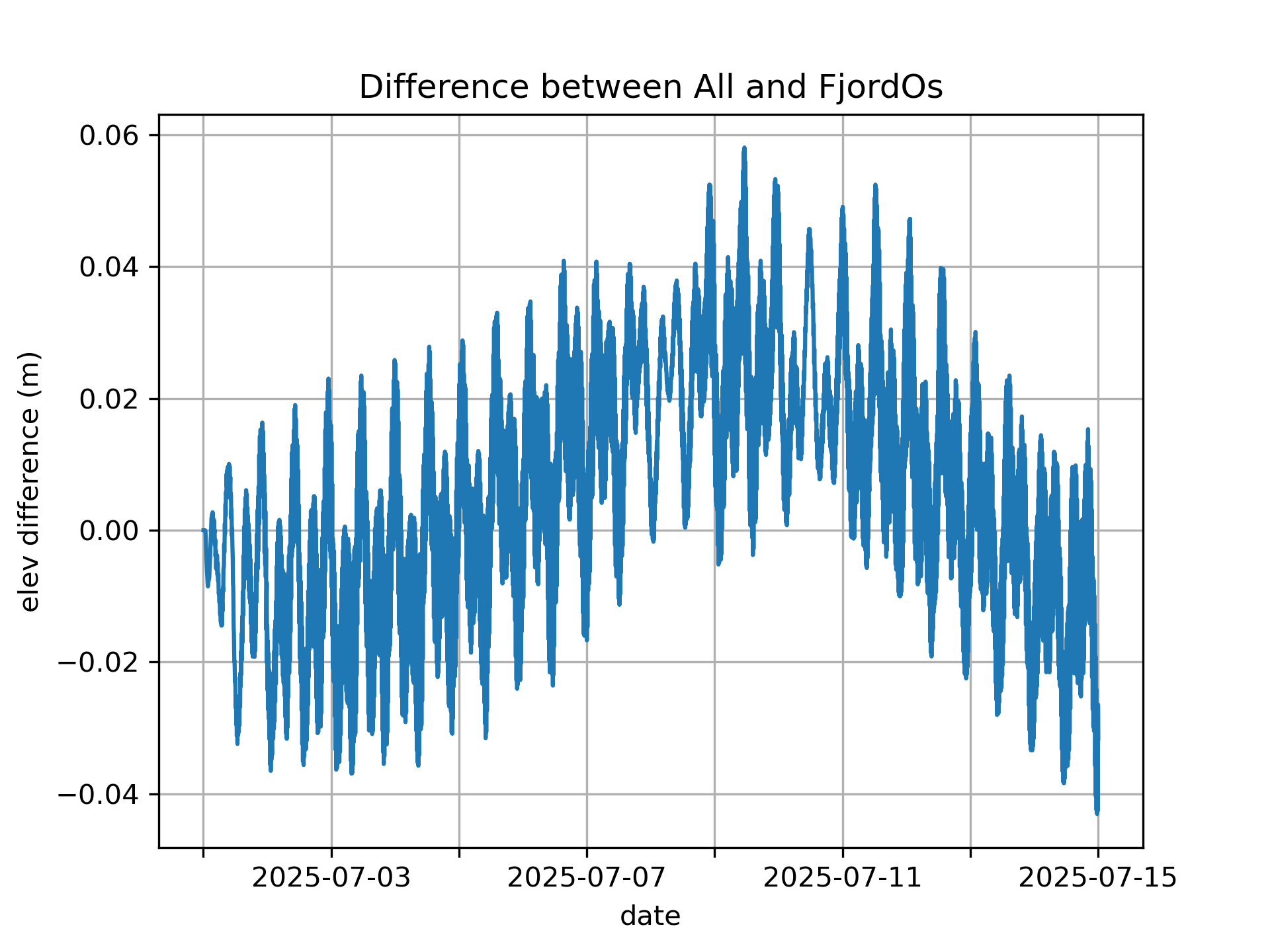}
        \caption{Oscarsborg}
        \label{fig:jul25_oscarsborg_diff_all_nine}
    \end{subfigure}
    
    \begin{subfigure}{0.4\textwidth}
        \includegraphics[width=\textwidth]{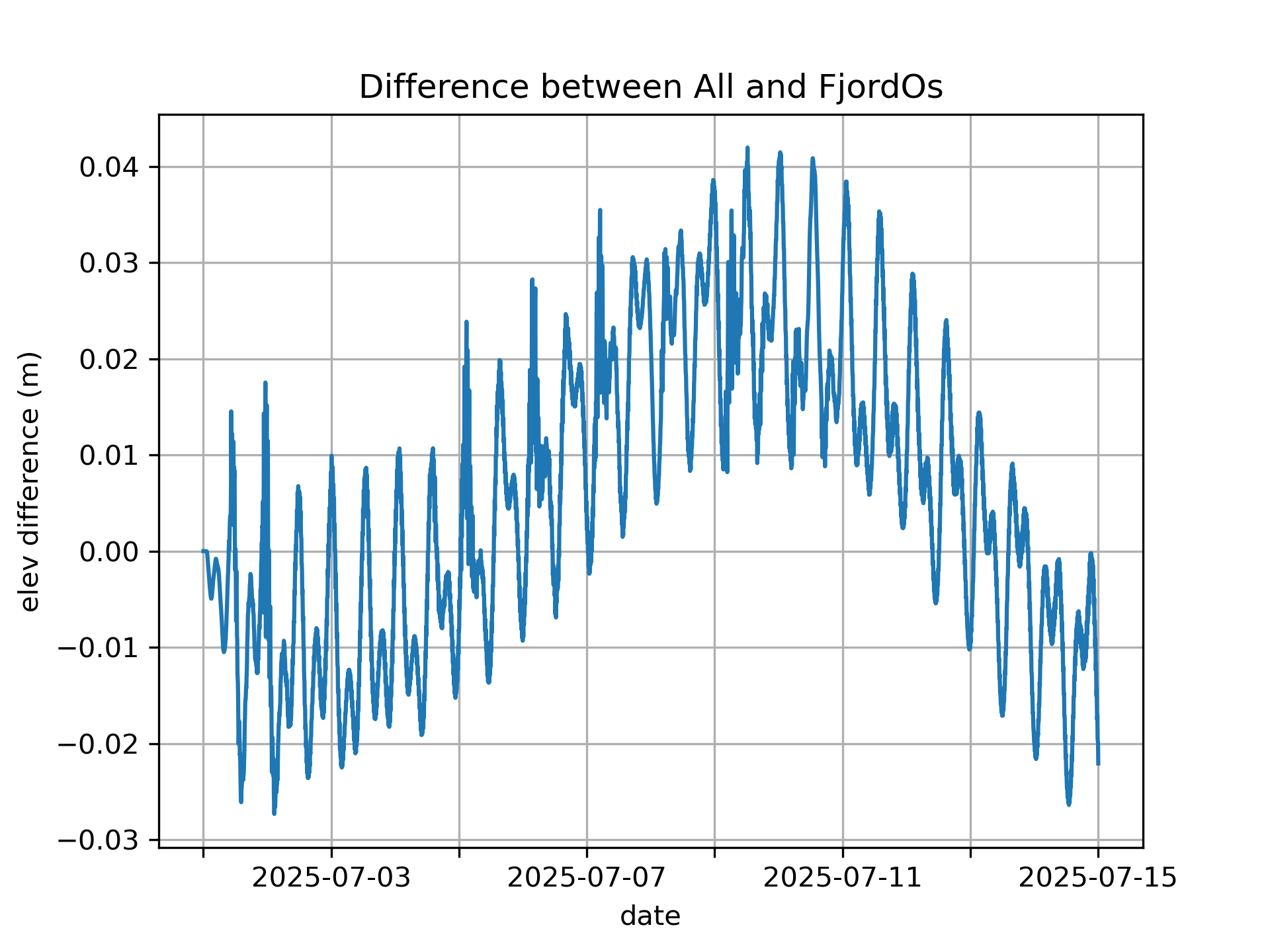}
        \caption{Oslo}
        \label{fig:jul25_oslo_diff_all_nine}
    \end{subfigure}
    \caption{Difference between ADCIRC-OF surface elevations with the 15-constituent model vs. the FjordOs constituents, July 2025.}
    \label{fig:jul25_tides_diff_all_nine}
\end{figure}

\newpage 
\bibliographystyle{elsarticle-num} 
\bibliography{references}

@ARTICLE{Webb09,
  author  = {Webb, K. E. et al.},
  title   = {Benthic ecology of pockmarks in the Inner {Oslofjord}, {Norway}},
  journal = {Mar. Ecol. Prog. Ser.}, 
  volume  = {387},
  year    = {2009},
  pages   = {15-25}
}

@techreport{fjordos,
  author      = {R\o{}ed, L. P. et al.},
  title       = {A high-resolution, curvilinear {ROMS} model for the {Oslofjord}},
  institution = {Norwegian Meterological Institute},
  year        = {2016}
}

@book{tan1992shallow,
  title={Shallow water hydrodynamics: Mathematical theory and numerical solution for a two-dimensional system of shallow-water equations},
  author={Tan, {Wei-Yan}},
  year={1992},
  publisher={Elsevier}
}

@ARTICLE{oceanmesh,
  author  = {Roberts, K. J. et al.},
  title   = {OceanMesh2D 1.0: MATLAB-based software for two-dimensional unstructured mesh generation in coastal ocean modeling},
  journal = {Geosci. Model Dev. (GMD)}, 
  volume  = {12},
  number  = {5},
  year    = {2019},
  pages   = {1847–1868},
  url     = {https://doi.org/10.5194/gmd-12-1847-2019}
}

@ARTICLE{gshhg,
  author  = {Wessel, P. and Smith, W. H. F.},
  title   = {A global, self-consistent, hierarchical, high-resolution shoreline database},
  journal = {J. of Geophys. Res.}, 
  volume  = {101},
  number  = {B4},
  year    = {1996},
  pages   = {8741–8743},
  url     = {doi:10.1029/96JB00104}
}

@misc{geonorge,
  author = {Statens Kartverk},
  title = {Dybdedata - terrengmodeller 50 meters grid},
  year = {2018},
  note = {data retrieved from Geonorge, 
          \url{https://kartkatalog.geonorge.no/metadata/bbd687d0-d34f-4d95-9e60-27e330e0f76e}},
}

@ARTICLE{srtm,
  author  = {Tozer, B. et al.},
  title   = {Global bathymetry and topography at 15 arc seconds: SRTM15+},
  journal = {Accept. Earth and Space Sci.}, 
  volume  = {6},
  number  = {10},
  year    = {2019},
  pages   = {1847-1864},
  url     = {https://doi.org/10.1029/2019EA000658}
}

@INPROCEEDINGS{metget,
       author = {{Cobell}, Zach et al.},
        title = "{MetGet: Cloud-Based Meteorological Forcing for Ocean Circulation}",
    booktitle = {102nd Annual AMS Meeting 2022},
         year = 2022,
       series = {American Meteorological Society Meeting Abstracts},
       volume = {102},
        month = jan,
          eid = {J9.3},
        pages = {J9.3},
       adsurl = {https://ui.adsabs.harvard.edu/abs/2022AMS...10299147C}
}

@ARTICLE{tpxo,
    title = {Efficient inverse modeling of barotropic ocean tides},
    author = {Egbert, G. D. and Erofeeva, S. Y.},
    journal = {J. of Atmos. and Ocean. Technol.},
    volume = {19},
    number = {2},
    year = {2002},
    pages = {183-204}
}

@article{kristensen23,
  author  = {Kristensen, N. M. et al.},
  title   = {A forecasting and warning system of storm surge events along the Norwegian coast},
  journal = {Environmental Fluid Mechanics}, 
  volume  = {23},
  year    = {2023},
  pages   = {307--329}
}

@techreport{stofs,
  author  = {Funakoshi, Y. et al.},
  title   = {NOAA's Surge and Tide Operational Forecast System 2D Global {(STOFS-2D-Global)} Version 2},
  institution = {Coast Survey Development Laboratory (U.S.), Ocean Associates, Inc., and University of Notre Dame}, 
  number  = {42},
  year    = {2025},
}

@techreport{norkyst,
    author = {Christensen, K. H. et al.},
    title = {``Norkyst'' version 3: the coastal ocean forecasting system for
{Norway}},
    year = {2025},
    institution={Norwegian Meteorological Office (Norway)}
}

@article{beiser,
  author  = {Beiser, F. et al.},
  title   = {Combining barotropic and baroclinic simplified models for drift trajectory predictions},
  journal = {J. of Oper. Oceanogr.}, 
  volume  = {17},
  number  = {3},
  year    = {2024},
  pages   = {187--206}
}

@manual{barker,
  title       = {Tidal Analysis and Prediction},
  author      = {Barker, B. B.},
  year        = {2007},
  organization = {National Oceanic and Atmospheric Administration},
  type = {NOAA Special Publication NOS CO-OPS 3}
}

@article{szpilka,
  author  = {Szpilka, C.et al.},
  title   = {Improvements for the Western North Atlantic, Caribbean and {Gulf of Mexico ADCIRC} Tidal Database (EC2015)},
  journal = {J. of Mar. Sci. and Eng.}, 
  volume  = {4},
  number  = {4},
  year    = {2016},
  pages   = {72}
}

@manual{fjord,
  title        = {The coast and fjords},
  year         = {2019},
  author       = {Institute of Marine Research},
  url          = {https://www.hi.no/en/hi/temasider/ocean-and-coast/the-ocean-coast-and-fjords/the-coast-and-fjords},
  urldate      = {2026-03-09}
}

@booklet{shipowners,
    author = {Norwegian Shipowners Association},
    title  = {Maritime Outlook 2025},
    year   = {2025}
}

@manual{sealevel,
    author = {Kartverket},
    title = {Future Sea Level Along the {Norwegian} Coast},
    year = {2024},
    url = {https://www.kartverket.no/en/at-sea/se-havniva/sea-level/future-sea-level-along-the-norwegian-coast},
    urldate = {2026-03-09}
}

@book{floods,
    author = {Roald, L. A.},
    title = {Floods in {Norway}},
    year = {2021},
    publisher = {Norwegian Water Resources and Energy Directorate}
}

@article{tsunamis,
  author = {Grue, J.},
  title = {Ship generated mini-tsunamis},
  journal = {J. of Fluid Mech.},
  year = {2017},
  volume = {816},
  pages = {142-166}
}

@book{milliman,
    author = {Milliman, J. D. and Farnsworth, K. L.},
    title = {River discharge to the coastal ocean: a global synthesis},
    publisher = {Cambridge University Press},
    year = {2011}
}

@manual{amy,
  author  = {Norsk naturskadepool},
  title   = {Trolig naturskader for rundt 1,8 milliarder kroner etter "Amy"},
  year    = {2025},
  url     = {https://www.naturskade.no/nyheter/2025/trolig-naturskader-for-rundt-18-milliarder-kroner-etter-amy},
  urldate = {2026-03-09}
}

@manual{amymet,
    author = {Fagerlid, G. O. et al.},
    title = {Ekstremv\ae{}ret “Amy”; ekstremt kraftige vindkast i S\o{}r-Norge-
3.-4. oktober 2025.},
    institution = {Meteorologisk institutt},
    year = {2025}
}

@book{dsb,
  author = {Norwegian Directorate for Civil Protection (DSB)},
  title = {Analyses of Crisis Scenarios 2019: Disasters That May Affect {Norwegian} Society},
  publisher = {ETN Grafisk},
  address   = {Skien, Norway},
  year      = {2019}, 
}

@techreport{hansmet,
    author = {Graner\o{}d, M. et al.},
    title = {Ekstremv\ae{}ret “Hans”, ekstremt mye nedb\o{}r i deler av S\o{}r-Norge 07.-09. august 2023},
    institution = {Meteorologisk institutt},
    year = {2023}
}

@article{roms,
  author  = {Haidvogel, D. B. et al.},
  title   = {Ocean forecasting in terrain-following coordinates: Formulation and skill assessment of the Regional Ocean Modeling System},
  journal = {J. of Comput. Phys.}, 
  volume  = {227},
  year    = {2008},
  pages   = {3595--3624}
}

@techreport{adcirc,
  title={{ADCIRC}: an advanced three-dimensional circulation model for shelves, coasts, and estuaries. {R}eport 1, Theory and methodology of {ADCIR-2DD1} and {ADCIRC-3DL}},
  author={Luettich, Richard Albert et al.},
  year={1992},
  publisher={Coastal Engineering Research Center (US)},
  institution={Coastal Engineering Research Center (US)}
}

@book{vreugdenhil1994numerical,
  title={Numerical methods for shallow-water flow},
  author={Vreugdenhil, Cornelis Boudewijn},
  volume={13},
  year={1994},
  publisher={Springer Science \& Business Media}
}

@book{luettich2004formulation,
  title={Formulation and numerical implementation of the {2D/3D {ADCIRC}} finite element model version 44.XX},
  author={Luettich, Richard Albert and Westerink, Joannes J},
  volume={20},
  year={2004},
  publisher={R. Luettich Chapel Hill, NC, USA}
}

@article{aunan,
    author = {Aunan, K. and Romstad, B.},
    title = {Strong Coasts and Vulnerable Communities: Potential Implications of Accelerated Sea-Level Rise for {Norway}},
    journal = {J. of Coast. Res.},
    year = {2008},
    volume = {242},
    pages = {403--409}
}

@inbook{dietrich05,
    author = {Dietrich, J. C. et al.},
    title = {Refinements in Continuous {Galerkin} Wetting and Drying Algorithms},
    booktitle = {Estuarine and Coastal Modeling},
    year = {2005},
    doi = {10.1061/40876(209)37},
    URL = {https://ascelibrary.org/doi/abs/10.1061/40876%28209%2937},
    publisher = {American Society Of Civil Engineers},
    pages = {637--656}
}

@article{luettich99,
  title={Elemental wetting and drying in the {ADCIRC} hydrodynamic model: Upgrades and documentation for {ADCIRC} version 34.XX},
  author={Luettich, R. A. and Westerink, J. J.},
  journal={Contract Report, prepared for Headquarters, US Army Corps of Engineers, Vicksburg, MS},
  year={1999}
}

@misc{gfs,
  author = {National Centers for Environmental Prediction{,} National Weather Service{,} NOAA{,} U.S. Department of Commerce},
  title = {{NCEP} {GFS} 0.25 Degree Global Forecast Grids Historical Archive},
  publisher = {NSF National Center for Atmospheric Research},
  address = {Boulder, CO},
  year = {2015},
  doi = {10.5065/D65D8PWK},
  url = {https://doi.org/10.5065/D65D8PWK}
}

@article{lynch1979wave,
  title={A wave equation model for finite element tidal computations},
  author={Lynch, Daniel R and Gray, William G},
  journal={Comp. \& Fluids},
  volume={7},
  number={3},
  pages={207--228},
  year={1979},
  publisher={Elsevier}
}

@techreport{hammack2008modeling,
  author      = {Hammack, E. Allen et al.},
  title       = {Modeling Vessel-Generated Currents and Bed Shear Stresses},
  institution = {U.S. Army Engineer Research and Development Center, Coastal and Hydraulics Laboratory},
  address     = {Vicksburg, MS},
  year        = {2008},
  month       = {June},
  number      = {ERDC/CHL TR-08-7},
}

@article{wichitrnithed2024discontinuous,
  title={A discontinuous {Galerkin} finite element model for compound flood simulations},
  author={Wichitrnithed, Chayanon et al.},
  journal={Comp. Methods in Appl. Mech. and Eng.},
  volume={420},
  pages={116707},
  year={2024},
  publisher={Elsevier}
}

@article{dawson2024swemnics,
  title={{SWEMniCS}: a software toolbox for modeling coastal ocean circulation, storm surges, inland, and compound flooding},
  author={Dawson, Clint et al.},
  journal={npj Nat. Hazards},
  volume={1},
  number={1},
  pages={44},
  year={2024},
  publisher={Nature Publishing Group UK London}
}

@manual{sehavniva,
  title        = {Se havnivå},
  url          = {https://www.kartverket.no/en/at-sea/se-havniva},
  year         = {2018},
  organization = {Norwegian Mapping Authority},
  urldate      = {2026-03-09}
}

@manual{oscarsborg_jetty,
  title        = {Analyserapport CFD-simuleringer av utdypning i sjeté},
  url          = {https://www.kystverket.no/globalassets/sjovegen/tiltak-i-farvannet/utb-prosjekt/haoya-vest/cfd-simuleringer-av-utdypning-i-sjete--dr.techn.-olav-olsen-as.pdf},
  year         = {2022},
  organization = {Dr.techn.Olav Olsen AS},
  publisher    = {Norwegian Coastal Administration},
  urldate      = {2026-03-09}
}

@manual{stofsdata,
  title        = {NOAA Global Surge and Tide Operational Forecast System 2-D {(STOFS-2D-Global)}},
  organization = {Office of Coast Survey; National Centers for Environmental Prediction},
  publisher    = {National Oceanic and Atmospheric Administration},
  year         = {2020},
  url          = {https://doi.org/10.25923/ng4h-4b85},
  urldate      = {2026-06-01}
}

@article{garratt,
  title     = {Review of Drag Coefficients over Oceans and Continents},
  author    = {Garratt, J.R.},
  journal   = {Mon. Weather Rev.},
  volume    = {105},
  number    = {7},
  pages     = {915--929},
  year      = {1977},
  publisher = {American Meteorological Society},
  url       = {https://doi.org/10.1175/1520-0493(1977)105<0915:RODCOO>2.0.CO;2}
}

@TechReport{engen:1988,
  author = {Engen, I. K.},
  title  = {{F}lommen på {S}{\o}r- og {{\O}}stlandet i oktober 1987},
  institution =  {Norwegian Water Resources and Energy Directorate},
  year = 	 {1988},
  type = 	 {NVE Publikasjon},
  number = 	 {15},
 }

@TechReport{andresen:etal:1987,
  author = {Andresen, L. and Johnsen, {\O}. and Kristoffersen, D.},
  title  = {{U}været 16. og 17. oktober (forel{\o}pig rapport)},
  institution =  {Norwegian Meteorological Institute},
  year = 	 {1987},
  type = 	 {DNMI-Rapport},
  number = 	 {35/87},
 }

@article{doodson,
 ISSN = {02643952},
 URL = {http://www.jstor.org/stable/91217},
 author = {A. T. Doodson},
 journal = {Philos. Trans. of the R. Soc. of Lond. Ser. A, Contain. Pap. of a Math. or Phys. Character},
 pages = {223--279},
 publisher = {Royal Society},
 title = {The Analysis of Tidal Observations},
 urldate = {2026-09-15},
 volume = {227},
 year = {1928}
}

@article{KRISTENSEN2024102406,
title = {{NORA-Surge}: A storm surge hindcast for the Norwegian Sea, the North Sea and the Barents Sea},
journal = {Ocean Model.},
volume = {191},
pages = {102406},
year = {2024},
issn = {1463-5003},
doi = {https://doi.org/10.1016/j.ocemod.2024.102406},
url = {https://www.sciencedirect.com/science/article/pii/S1463500324000933},
author = {Nils Melsom Kristensen et al.}
}

@techreport{militello,
    author = {Adele Militello and Alan K. Zundel},
    title = {Surface-Water Modeling System Tidal
Constituents Toolbox for ADCIRC},
    institution = {U.S. Army Corps of Engineers},
    year = {1999},
    type = {Coastal Engineering Technical Note IV-21}
}

\end{document}